\documentclass{llncs}
\usepackage{amsfonts,mathrsfs,rotate}

\usepackage{xcolor}
\usepackage{listings}
\usepackage{algpseudocode}

\usepackage{setspace} 
\usepackage[english]{babel}
\usepackage[nottoc]{tocbibind}
\usepackage{epsfig}
\usepackage[linesnumbered,ruled,vlined]{algorithm2e}
\SetCommentSty{mycommfont}
\SetKwInput{KwInput}{Input}     \SetKwInput{KwOutput}{Output}  
\usepackage{graphicx}
\usepackage{multirow}
\usepackage{subcaption}
\usepackage{chemarrow}
\usepackage[a4paper, total={7in, 9in}]{geometry}
\usepackage{amssymb}
\usepackage{float}
\usepackage{cite}
\usepackage{caption}
\usepackage{lineno,hyperref}
\usepackage{enumerate}
\usepackage{makecell}
\usepackage{color}
\usepackage{array}
\newcolumntype{?}{!{\vrule width 2pt}}
\usepackage{tikz}
\usepackage{MnSymbol}

\begin{document}
\title{A Theory of L-shaped Floor-plans}
\author{Raveena, Krishnendra Shekhawat}
\institute{Department of Mathematics, BITS Pilani, Pilani Campus, India-333031}
\maketitle

\begin{abstract}
\begin{figure}[]
\includegraphics[width=13cm,height=7cm]{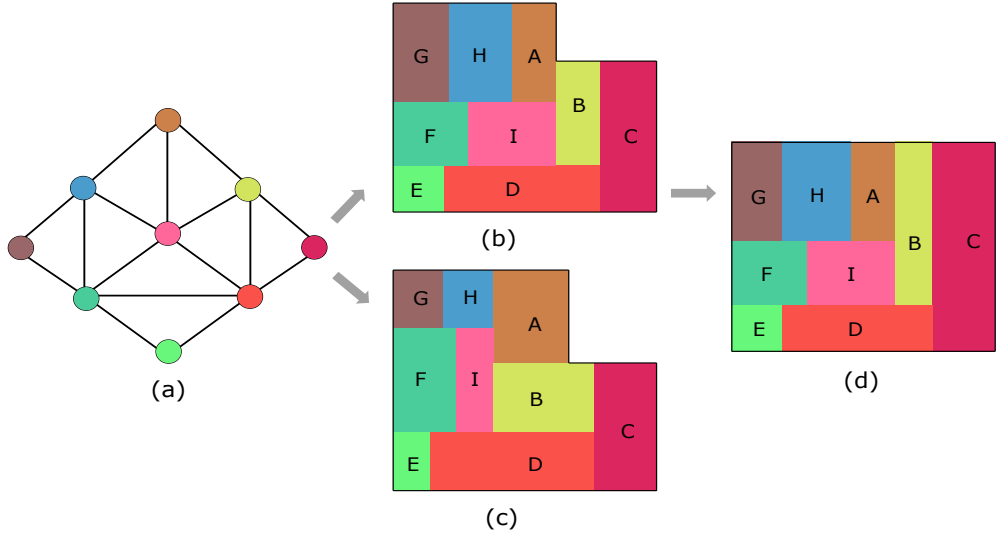}
\centering
\caption{(a) A planar graph \textit{G}, (b - c) Two topologically distinct floor-plans with \textit{L}-shaped boundary, (d) A rectangular floor-plan obtained from (b) by stretching modules $B$ and $C$ in upward direction without disturbing any adjacency while the floor-plan in (c) is non-trivial because it can not be transformed into a rectangular floor-plan by stretching modules and preserving given adjacencies}
\label{drawing--7}
\end{figure}

Existing graph theoretic approaches are mainly restricted to floor-plans with rectangular boundary. In this paper, we introduce floor-plans with $L$-shaped boundary (boundary with only one concave corner).
To ensure the $L$-shaped boundary, we introduce the concept of non-triviality of a floor-plan. A floor-plan with a rectilinear boundary with at least one concave corner is non-trivial if the number of concave corners can not be reduced, without affecting the modules adjacencies within it. 
Further, we present necessary and sufficient conditions for the existence of a non-trivial $L$-shaped floor-plan corresponding to a \textit{properly triangulated planar graph} (PTPG) \textit{G}. Also, we develop an $O(n^2)$ algorithm for its construction, if it exists.

\end{abstract}

\keywords{\rm adjacency, algorithm, graph theory, concave corner, non-rectangular floor-plan, $L$-shaped floor-plan}

\section{Introduction}

Representing planar graphs as floor-plans (layouts) have been studied for many decades due to its applications in various fields as in architectural design, in VLSI circuit designs, in cartography etc. In such a representation, vertices correspond to rectangular or rectilinear polygons (modules), while edges correspond to the common line segments shared by two modules and the adjacency graph itself is the \textit{dual graph} of the floor-plan. Most of prior work on the problem of floor-planning was restricted to rectangular boundaries  \cite{kozminski1988rectangular,bhasker1987linear,he1993finding,alam2013computing}. However, in recent years, researchers have focused on generating floor plans with non-rectangular boundaries and have proposed a few approaches, as in \cite{wu2018based, wu2019data-driven, wang2020generating,rahbar2021architectural}. 
These approaches may not ensure the specific shape (\textit{L}-shaped, \textit{T}-shaped etc.)  of the boundary of a floor-plan and may not be generalized for any given graph. Furthermore, for any given graph, the existence conditions of floor-plans with the prescribed outer boundary and the non-triviality of the floor-plan have not been addressed yet. In contrast, for the construction of floor-plans with rectangular boundary, there exist well-known graph algorithms as well as necessary and sufficient conditions for the existence of such floor-plans corresponding to any graph. Further, with the recent advancement in technology and design, it is mathematically demanding to generate floor-plans with rectilinear boundaries (e.g., $L$-shaped, $T$-shaped, $Z$-shaped) while satisfying the provided  adjacency requirements. These floor-plans with rectilinear boundaries are called \textit{non-rectangular floor-plans} (NRFPs).

We begin the study by considering the outer boundary to be rectilinear with one concave corner ($L$-shaped boundary) and modules to be rectangles. Such a floor-plan is known as $L$-shaped floor-plan. The boundary of an $L$-shaped floor-plan can sometimes be converted into a rectangle without modifying the module adjacencies by expanding or reducing the outer boundary walls of some modules. Two topologically distinct $L$-shaped floor-plans are depicted in Figures \ref{drawing--7}(b) and \ref{drawing--7}(c) corresponding to a planar graph $G$ (Figure \ref{drawing--7}(a)). Here a rectangular floor-plan (Figure \ref{drawing--7}(d)) is generated from Figure \ref{drawing--7}(b) by expanding one of the outer walls of modules $B$ and $C$, without affecting any of the adjacencies of $G$. Such floor-plans with non-rectangular boundaries (Figure \ref{drawing--7}(b)) are insignificant since, ultimately, these are rectangular floor-plans and we already have several efficient algorithms for producing such layouts. Furthermore, Figure \ref{drawing--7}(c) is the required output because none of the above operations (expanding or reducing the outer walls) can convert it into a rectangular floor-plan.
To ensure that the boundary of an $L$-shaped floor-plan is not transferable into a rectangle, we define the property of \textit{non-triviality} of a floor-plan. A floor-plan with a rectilinear boundary with at least one concave corner is \textit{non-trivial} if the number of concave corners can not be reduced, without modifying the module adjacencies. We aim to obtain a non-trivial $L$-shaped floor-plan correspond to a PTPG $G$, if it exists. In Figure \ref{drawing--7}(c), a non-trivial $L$-shaped floor-plan is obtained for the graph shown in Figure \ref{drawing--7}(a).

\subsection{Related Work} 
\begin{figure}[]
\includegraphics[width=6.3cm,height=3.7cm]{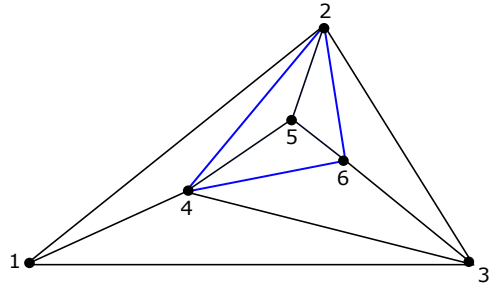}
\centering
\caption{A PTG with complex triangle $(2, 4, 6)$}
\label{draw13}
\end{figure}

We begin with some existing graph terminologies and then we survey the literature on the generation of floor-plans for a particular class of graphs.
\begin{definition}{\rm 
A planar graph \textit{G} is a  \textit{planar triangulated graph} (PTG) \cite{bhasker1986linear} if all of its interior and exterior faces are triangles. In a PTG, a cycle of length three consisting of at least one vertex inside it is known as \textit{complex triangle} or \textit{separating triangle} \cite{he1993finding} (see Figure \ref{draw13}).}
\end{definition}

\begin{figure}[]
 \centering
 \begin{subfigure}[b]{0.24\textwidth}
  \includegraphics[width=.64\textwidth]{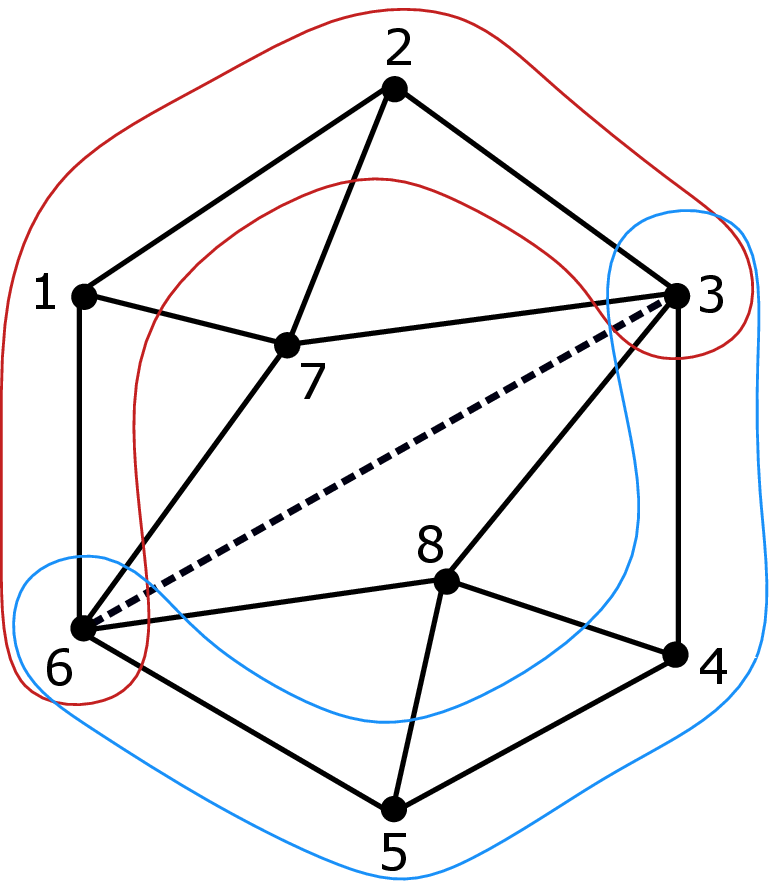}
  \caption{A PTPG}
  \label{drawing23}
 \end{subfigure}%
 \begin{subfigure}[b]{0.28\textwidth}
  \includegraphics[width=.75\textwidth]{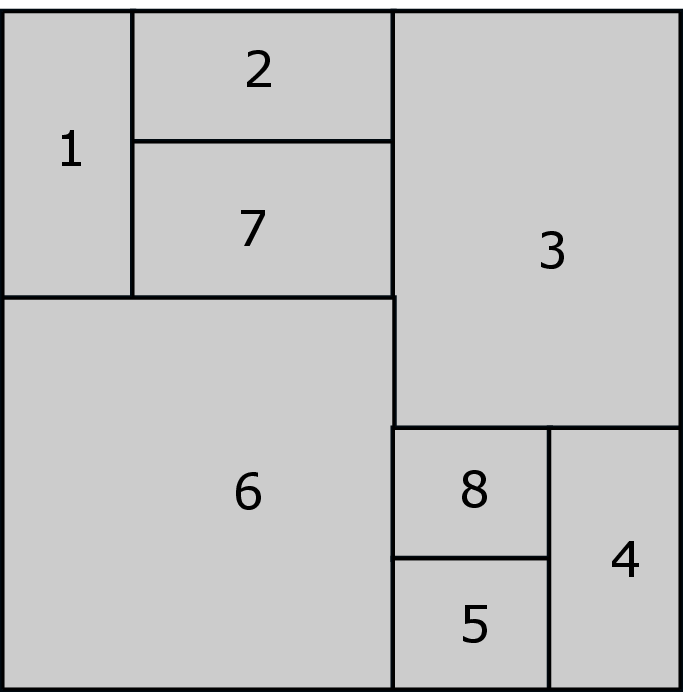}
  \caption{RFP}
  \label{drawing-3}
 \end{subfigure}
 \begin{subfigure}[b]{0.30\textwidth}
  \includegraphics[width=.70\textwidth]{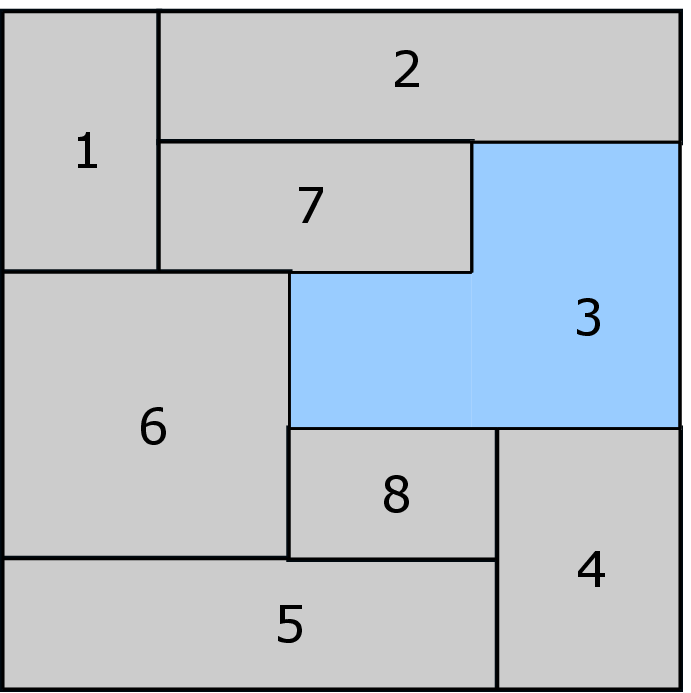}
  \caption{OFP}
  \label{drawing-4}
 \end{subfigure}
 \caption{}
\end{figure}

\begin{definition}{\rm 
A \textit{properly triangulated planar graph} (PTPG) \cite{bhasker1986linear} is a connected planar graph that satisfies the following properties: (refer to Figure \ref{drawing23})
\begin{enumerate}[i.]
\item Every face (except the exterior) is a triangle,
\item It should not have complex triangles.
\end{enumerate}

In this paper, a PTPG refers to a bi-connected properly triangulated planar graph.}
\end{definition}
\begin{definition}{\rm 
A \textit{corner implying path} (CIP) \cite{kozminski1988rectangular} in a planar bi-connected graph $G$ is a path, containing $n$ consecutive vertices $u_1, u_2,\dots, u_n$ on the outer boundary of $G$ such that the end vertices $u_1$ and $u_n$ are adjacent and none of the other non-consecutive vertices are adjacent. 
In Figure \ref{drawing23}, PTPG $G$ has two CIPs namely, $\{6, 1, 2, 3\}$ and $\{3, 4, 5, 6\}$.

An interior edge $e$ between two outer vertices $u$ and $v$ of a graph $G$ is called a \textit{shortcut} \cite{kozminski1988rectangular}, if there exist another path between $u$ and $v$ made up of outer vertices. 
In Figure \ref{drawing23}, edge $(6, 3)$ is a shortcut.}
\end{definition}

\begin{figure}[]
\includegraphics[width=9.8cm,height=5.1cm]{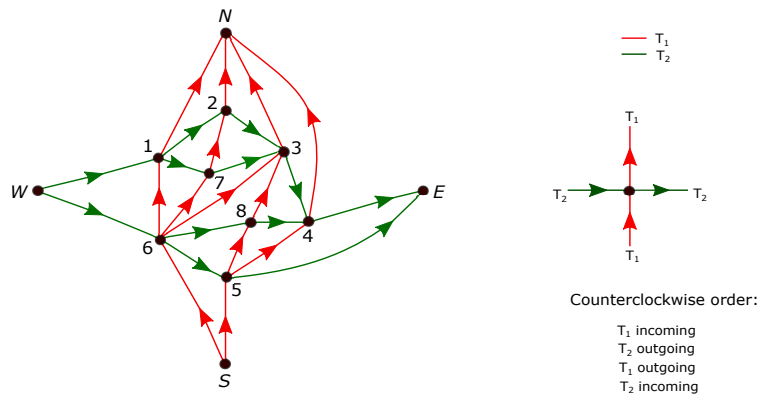}
\centering
\caption{REL of the PTPG in Figure \ref{drawing23}}
 \label{drawing-5}
\end{figure}

\begin{definition}{\rm 
A \textit{regular edge labeling} (REL) \cite{he1993finding} of a graph \textit{G} is obtained by partitioning the set of edges into two subsets $T_1$ and $T_2$ while assigning direction to each edge 
such that the edges incident to each vertex $v$ in counterclockwise order follows a regular pattern:
\begin{enumerate}[i.]
\item a set of edges in $T_1$, leaving $v$
    
\item a set of edges in $T_2$, entering $v$
    
\item a set of edges in $T_1$, entering $v$
    
\item a set of edges in $T_2$, leaving $v$
    
\end{enumerate}

In REL of a graph $G$ (see Figure \ref{drawing-5}) there exists four exterior vertices $N$, $W$, $S$, $E$ in counterclockwise order, whereas all the edges incident to vertex $N$ belongs to set $T_1$ and entering $N$, edges incident to $W$ belongs to set $T_2$ and leaving $W$, edges incident to $S$ belongs to set $T_1$ and leaving $S$ and edges incident to $E$ belongs to set $T_2$ and entering $E$.
}
\end{definition}

\begin{figure}[H]
\centering
\begin{subfigure}[b]{0.40\textwidth} \includegraphics[width=.90\textwidth]{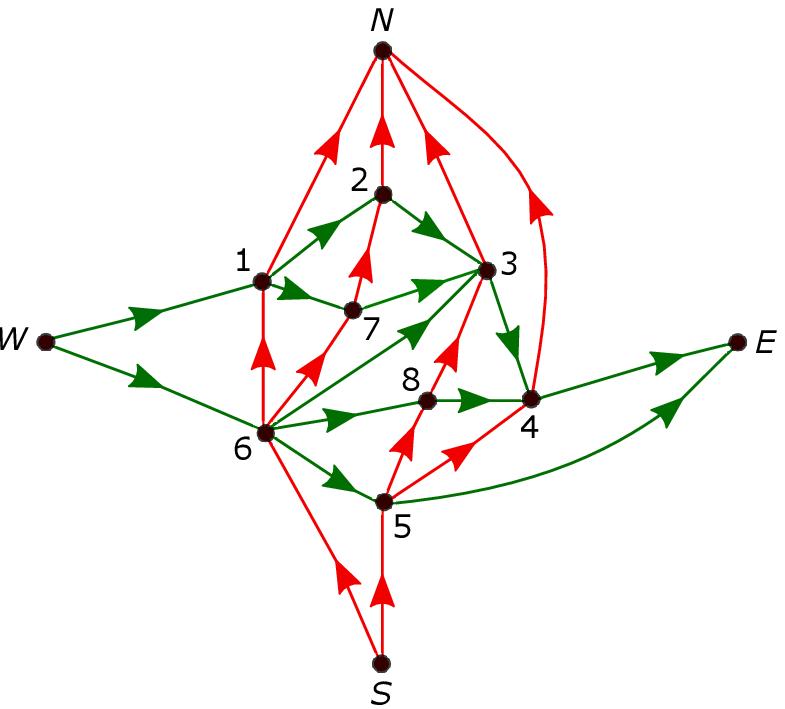}
\caption{REL obtained after flipping edge $(6, 3)$}
\label{drawing-52}
\end{subfigure}
\begin{subfigure}[b]{0.40\textwidth}
\includegraphics[width=.90\textwidth]{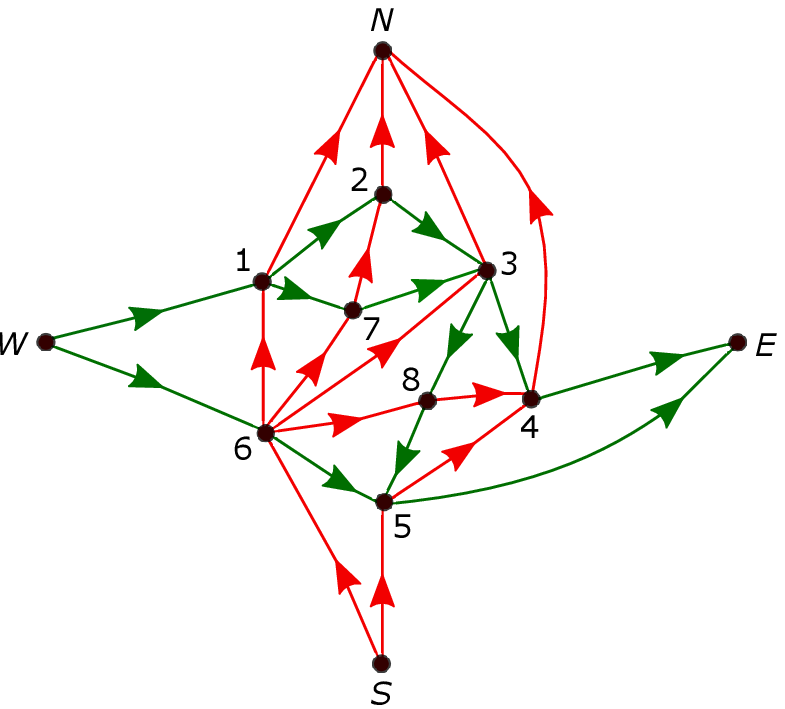}
\caption{REL obtained after flipping vertex $8$}
\label{drawing-53}
\end{subfigure}
\caption{ }
\label{drawing523}
\end{figure}

\begin{definition}{\rm
A \textit{flippable edge} \cite{eppstein2009area} is an edge $e$ that is not incident to a degree four vertex in a REL of a graph $G$, and the four-cycle surrounding $e$ is alternately labelled in the REL. Moreover, if an edge is flippable, its label can be modified. A four-degree vertex $v$ is  a \textit{flippable vertex} \cite{eppstein2009area} if the four-cycle around $v$ is alternately labelled in the REL. If a vertex is flippable the the labels of its incident edges can be modified.
 In Figures \ref{drawing-52} and \ref{drawing-53}, two different RELs are obtained from the REL shown in Figure \ref{drawing-5} by flipping the edge $(6, 3)$ and vertex $8$, respectively.
}
\end{definition}

\begin{definition}{\rm 
A \textit{floor-plan} \cite{rinsma1988existence} is a partition of a polygon by straight lines into component polygons. The polygon is called the boundary of a floor-plan and the component polygons are called \textit{modules}. The edges forming the perimeter of each module are termed \textit{walls}. Two modules in a floor-plan are \textit{adjacent} if they share a wall or a section of wall; it is not sufficient for them to touch at a point only.
An \textit{adjacency graph} \cite{rinsma1988existence} is a simple connected undirected planar graph providing a specific neighborhood of the given modules (Figure \ref{drawing-3} is a floor-plan corresponding to the adjacency graph in Figure \ref{drawing23}).
A \textit{rectangular floor-plan} (RFP)  \cite{rinsma1988existence} is a floor-plan in which the boundary and each module is  a rectangle. A RFP is also known as \textit{rectangular dual} (Figure \ref{drawing-3} illustrates a RFP).
An \textit{orthogonal floor-plan} (OFP)  \cite{rinsma1988existence} has a rectangular boundary with the walls of each module parallel to the sides of the outer boundary, i.e., an OFP may have some rectilinear modules, such as $L$-shaped, $T$-shaped etc. (Figure \ref{drawing-4} illustrates an OFP).}
\end{definition}

 In the past, most of the researchers have worked over the construction of floor-plans corresponding to PTGs only. The construction of floor-plans began with RFPs. The problem of finding a RFP has been studied in \cite{kozminski1988rectangular}, \cite{bhasker1987linear}, \cite{he1993finding}, \cite{alam2013computing}, etc. Ko{\'z}mi{\'n}ski and Kinnen \cite{kozminski1985rectangular} were the first to produce the necessary and sufficient conditions for the existence of RFPs corresponding to PTPGs. They presented an algorithm of O($n^2$) time complexity to obtain a RFP, if exists and proposed the following theorem:
\begin{theorem} {\rm A bi-connected PTPG has a RFP if and only if it has no more than four CIPs.}
\end{theorem}
Bhasker and Sahni \cite{bhasker1987linear, bhasker1986linear} improved the results given in \cite{kozminski1985rectangular} and developed a linear time algorithm to determine the existence of a RFP corresponding to a PTPG and to construct it. 
Later, Kant \textit{et al.} \cite{kant1997regular} applied the concept of REL to obtain a RFP. 
They presented a linear time algorithm to obtain a RFP from a REL, which is comparatively simpler than the previous known algorithms.
If the areas are also associated with the modules of a layout, then it has applications in cartography. Rectangular layouts which can be assigned any areas to its rectangles are called area-universal layouts. 
Eppstein \textit{et al.} \cite{eppstein2009area} established a simple necessary and sufficient condition for a rectangular layout to be area-universal and produced an algorithm to find an area-universal layout for a given PTG. They also introduced flippable items (\textit{flippable edge and flippable vertex}) in the RELs and obtained topologically distinct RELs from a REL by flipping an edge or a vertex. 
Later on, Steven \textit{et al.} \cite{steven2021morphing} defined a new operation \textit{rotation} in a REL. Using rotation, they were able to move between different RELs of a PTPG by swapping the colour and direction of the edges, which are lying inside an alternate coloured four cycle.

Recently, various ways of automatic generation of floor-plans have been explored during the previous few decades \cite{xiao2018Customization},
\cite{nitant2020automated}. 
Xiao \textit{et al.} \cite{xiao2018Customization} provided a graph approach to design generation (GADG) of RFPs based on existing legacy floor-plans using dual graphs of PTPGs. GADG uses a rectangular dual finding technique to automatically reproduce a new set of floor-plans while preserving the original connectivity information, which can be further improved and customized. 
Upasani \textit{et al.} \cite{nitant2020automated} considered the problem of generating dimensioned RFPs for user-specified adjacency graph and used linear optimization techniques to achieve a feasible solution. In addition, the input adjacency connections were arranged in a dimensionless rectangular configuration and each room's minimum width and aspect ratio range were used to create dimensional limitations. To obtain maximum adjacencies in the RFPs between modules, Vinod \textit{et al.} \cite{kumar2021transformations} presented a polynomial time algorithm and obtained different RFPs.

Further, the graphs for which RFPs do not exist or if the user desires for different shaped rooms, researchers have worked over the construction of OFPs and presented graph theoretical techniques for the construction of such floor-plans while satisfying adjacency constraints.
Sun \textit{et al.} \cite{sun1993floorplanning} found that there exist PTGs for which a RFP does not exist and for those PTGs, there may exist a floor-plan having rectangular and $L$-shaped modules. To obtain such floor-plans, they used dualization technique and gave an algorithm that checks if graph $G$ admits an OFP with $L$-shaped modules in O$(n^{3/2})$ time and presented its construction, if exists in O$(n^2)$ time. Later on, Yeap \textit{et al.} \cite{yeap1993floor} claimed that there exist PTGs, for which it is not possible to construct an OFP with rectangular and $L$-shaped modules only. Furthermore, they showed that it is necessary and sufficient to use $2$-bend modules ($T$-shaped and $Z$-shaped) including rectangular modules and $L$-shaped modules in order to find an OFP for any bi-connected PTG and presented a linear time algorithm for the construction. After that, He \cite{he1999floor} improved the previous results for OFPs and established a linear time algorithm for finding an OFP corresponding to the PTGs, using rectangle modules, $L$-modules and $T$-modules only.
Many researchers used weighted graphs as an input for generating floor-plans with required areas of modules \cite{watson1997vertex}, whereas the weight of a vertex represent the area of the respective module in the floor-plan.
Watson \textit{et al.} \cite{watson1997vertex} generated OFPs with specified areas of the modules while taking a weighted PTGs as an input. 
Liao \textit{et al.} \cite{liao2003compact} presented a linear time algorithm for constructing an OFP for any PTG which uses $I$-modules (rectangular), $L$-modules  and $T$-modules only, which was relatively simpler than the previous algorithms. In addition, the obtained floor-plan can be fitted in the area of a rectangle of size $(n-1).(2n + 1)/3$.  
Kurowski \cite{kurowski2003simple} presented a linear time algorithm to produce a floor-plan for any $n$-vertex PTG having area at most $n(n-1)$ and also bounded the number of $T$-modules to be at most $(n-2)/2$ in the floor-plan. 
Alam \textit{et al.} \cite{alam2013computing} considered weighted planar graphs and presented a linear time algorithm to obtain OFPs using at most $8$-sided rectilinear polygons as modules, where the area of each module is equal to a pre-specified weight of the respective vertex. They defined area-universality in OFPs and proved that the obtained OFPs are area-universal as well.
 
During recent times, the VLSI circuits and architectural floor-plans may not be restricted to rectangular boundaries only. In the last few decades, various approaches have been used to generate floor-plans with non-rectangular boundary. Baybars \textit{et al.} \cite{baybars1980enumerating} generalized the concept of geometric dual of a graph for obtaining the architectural arrangement from its underlying graph. They generated layouts with circulations and random non-rectangular boundaries corresponding to the adjacency graphs. Nummenmaa \cite{nummenmaa1992constructing} presented a linear-time algorithm to construct layouts with a rectilinear boundary for some specific PTGs. The approach is based on the canonical representation of planar graphs, where the vertices were taken as horizontal segments and edges were taken as vertical segments of the layout.
Miura \textit{et al.} \cite{kazuyuki2006inner} used graph drawing techniques to construct floor-plans with non-rectangular boundaries. For the input adjacency graph, they obtained inner rectangular drawing \footnote{A drawing of a plane graph is known as an inner rectangular drawing if every edge is represented as a horizontal or vertical line segment such that every inner face is a rectangle.}, which was represented as a floor-plan corresponding to the geometric dual graph of $G$. 
Shekhawat \textit{et al.} \cite{shekhawat2017rectilinear} provided graph theoretical tools to solve the problem of allocating rooms (with predefined adjacency and  required sizes) within user-specified contour shape. The adjacent graph with constraints was used as a distance  matrix for the input and dimensional floor-plans were generated.
Wu \textit{et al.} \cite{wu2018based} used a hierarchical approach for generating the building floor-plans with predefined high-level constraints such as size of the rooms, position of the rooms, adjacency between rooms and the outline of the building (outer boundary). They considered the layout's outer boundary as a polygon and further decomposed it into smaller rectangles and obtained the solution based on a mixed integer quadratic programming (MIQP) formulation. Moreover, after generating a floor-plan satisfying all the constraints, they also showed $3$-$D$ models of the floor-plans to visualize the results. 
Wu \textit{et al.} \cite{wu2019data-driven} provided a data-driven technique for automatically and efficiently generating floor-plans of residential buildings. For the input, they considered only the boundary of the layout. To generate floor-plans they located the rooms position first, then using encoder-decoder network, walls of the rooms were determined while preserving the user-specified outer boundary. Initially, one solution was generated, and then to generate multiple floor-plans with the same outer boundary, they generated a sample based on the probability distribution of rooms. 
Wang \textit{et al.} 
\cite{wang2020generating} demonstrated a generic approach for the automated generation of floor-plans with non-rectangular boundary corresponding to the user-specified design requirements. They used a formal mechanism to generate the graphs first according to the user's specification and presented a set of algorithms to place rooms with rectangular or non-rectangular boundaries satisfying the set of constraints. Then, adding dummy rooms (if required) and merging the rooms, non-rectangular outer boundaries were generated in order to find various floor-plans.
Hu \textit{et al.} \cite{hu2020graph2plan} introduced deep neural network graph$2$plan approach for automated floor-plan generation from layout graphs. To generate such layouts, they allowed users to specify the outer boundary, room's counts and room's connectivity as inputs. Using generative modelling, they retrieved a set of floor-plans correspond to the layout graph that fulfills both the layout and boundary constraints.
Rahbar \textit{et al.} \cite{rahbar2021architectural} demonstrated a new hybrid technique for generating automated $2$D architectural layouts with specified constraints such as the layout graph and the layout's outer boundary (footprint). Furthermore, the adjacency matrix were taken as an input for a layout graph. They created bubble diagrams corresponding to the layout graph using agent-based modelling, and then used a rule-based approach to convert the bubble diagram into a heat map. The interior of the building footprint was then divided into small grid cells, with the size of the grid cells determined by the distance between the bubbles and the layouts were generated. Instead of using conventional graph-theoretic and mathematical approaches, they applied agent-based modelling and deep learning techniques to generate such layouts.

It can be observed from the existing literature that there does not exist conditions for the existence of NRFPs and there do not exist efficient algorithms for constructing non-trivial NRFPs for a given adjacency graph. In this work, we derive a necessary and sufficient condition for the existence of non-trivial NRFPs with one concave corner and develop an algorithm for its construction. In the future, we will consider NRFPs with two and more concave corners.
\subsection{Preliminaries}\label{1.1}

Now we introduce a few new terminologies to better understand the work.

\begin{definition}{\rm
Triplet $(a, b, c)$ denotes three vertices $a, b, c$ of a graph $G$ such that $(a, b)$ and $(b, c)$ are two edges in $G$. In general, $(a_1, a_2,\dots, a_n)$ represents $n$ vertices $a_1, a_2,\dots, a_n$ such that $(a_i, a_{i+1})$ is an edge in $G$, $ 1 \le i \le {n-1}.$}
\end{definition}

\begin{figure}[H]
 \centering
 \begin{subfigure}[b]{0.32\textwidth}
  \includegraphics[width=.7\textwidth]{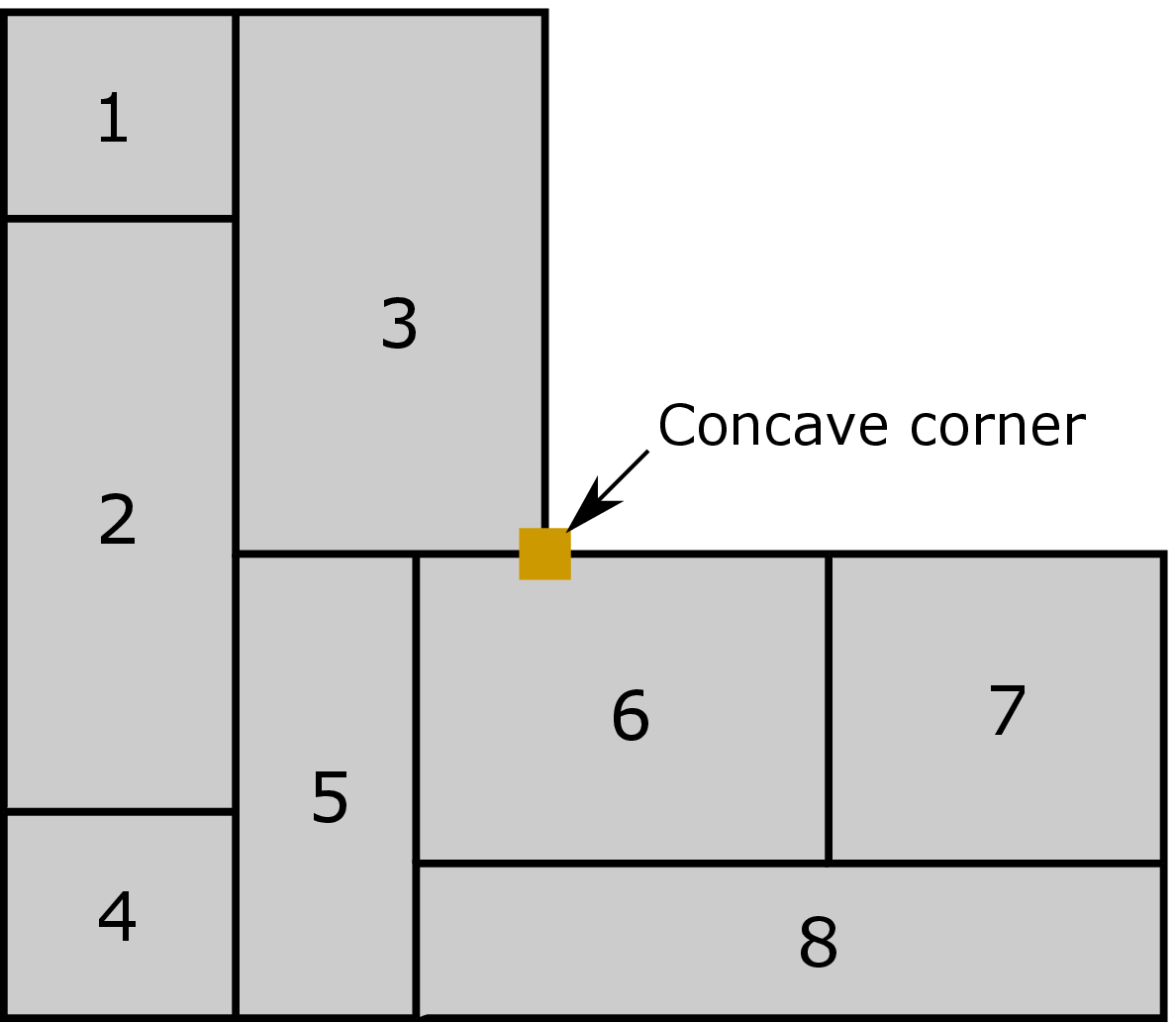}
  \caption{L-shaped}
  \label{drawing-1}
 \end{subfigure}%
 \begin{subfigure}[b]{0.34\textwidth}
  \includegraphics[width=.75\textwidth]{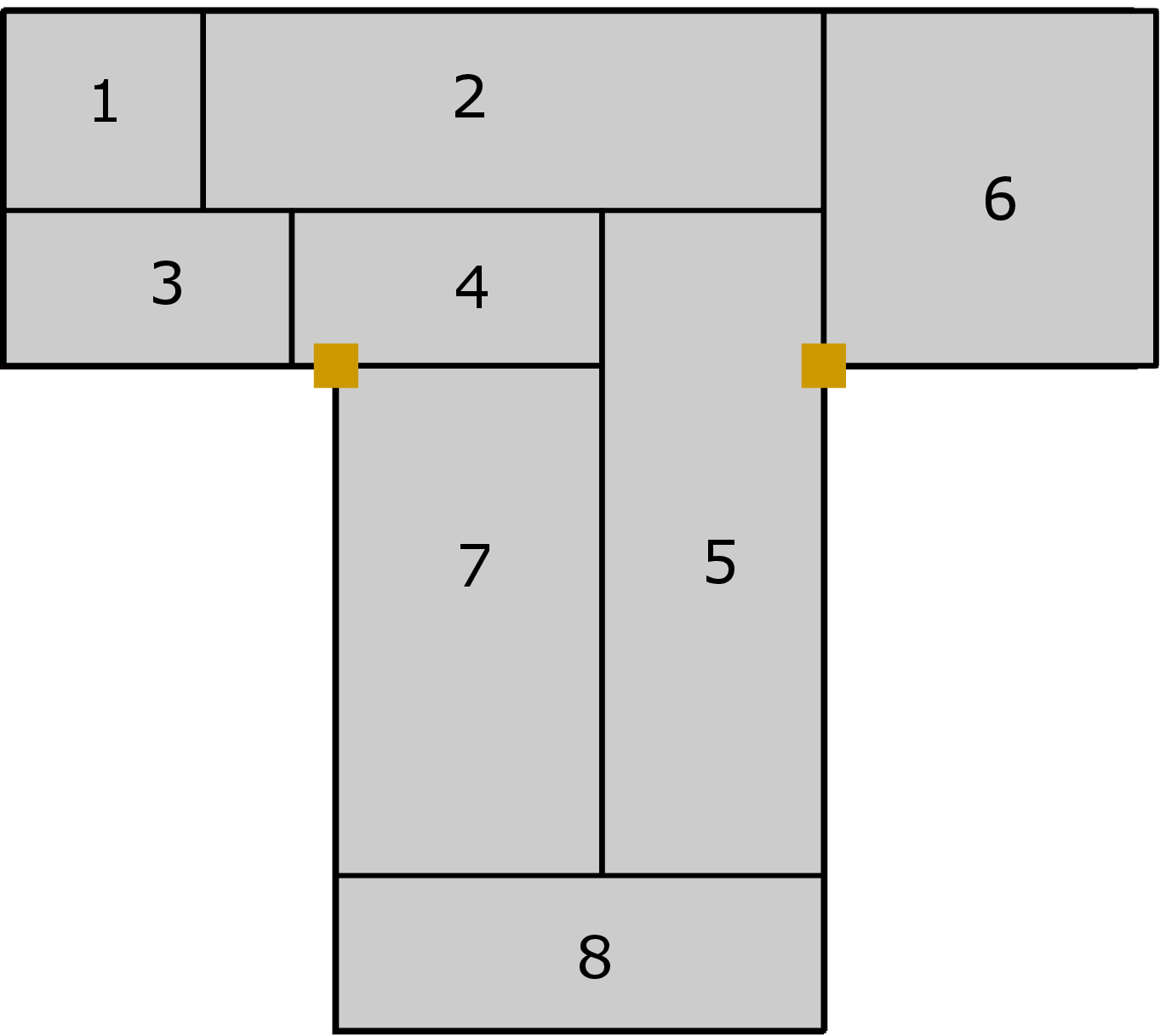}
  \caption{T-shaped}
  \label{drawing2}
 \end{subfigure}
 \begin{subfigure}[b]{0.32\textwidth}
  \includegraphics[width=.75\textwidth]{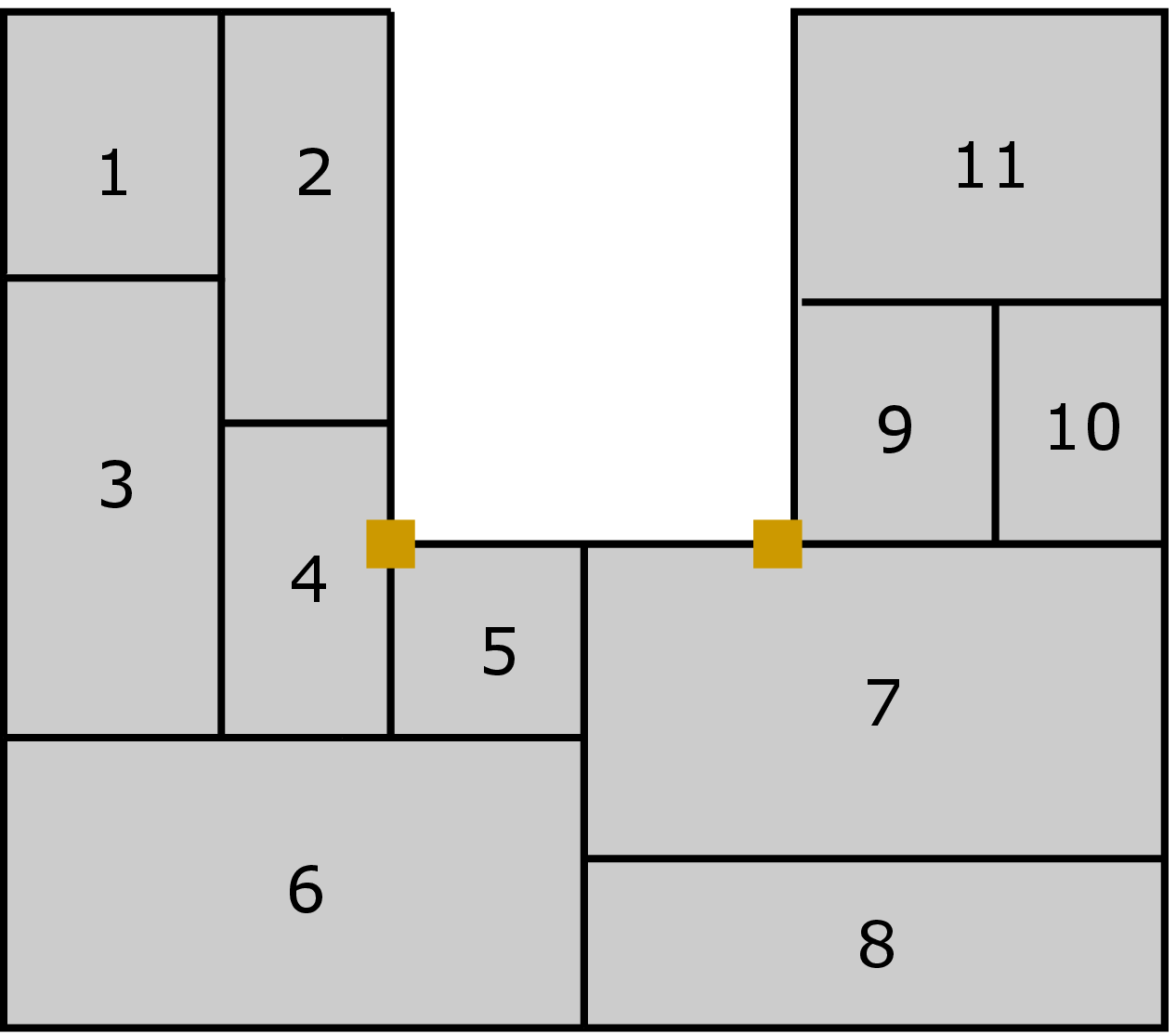}
  \caption{U-shaped}
  \label{drawing3}
 \end{subfigure}
 \begin{subfigure}[b]{0.32\textwidth}
  \includegraphics[width=.79\textwidth]{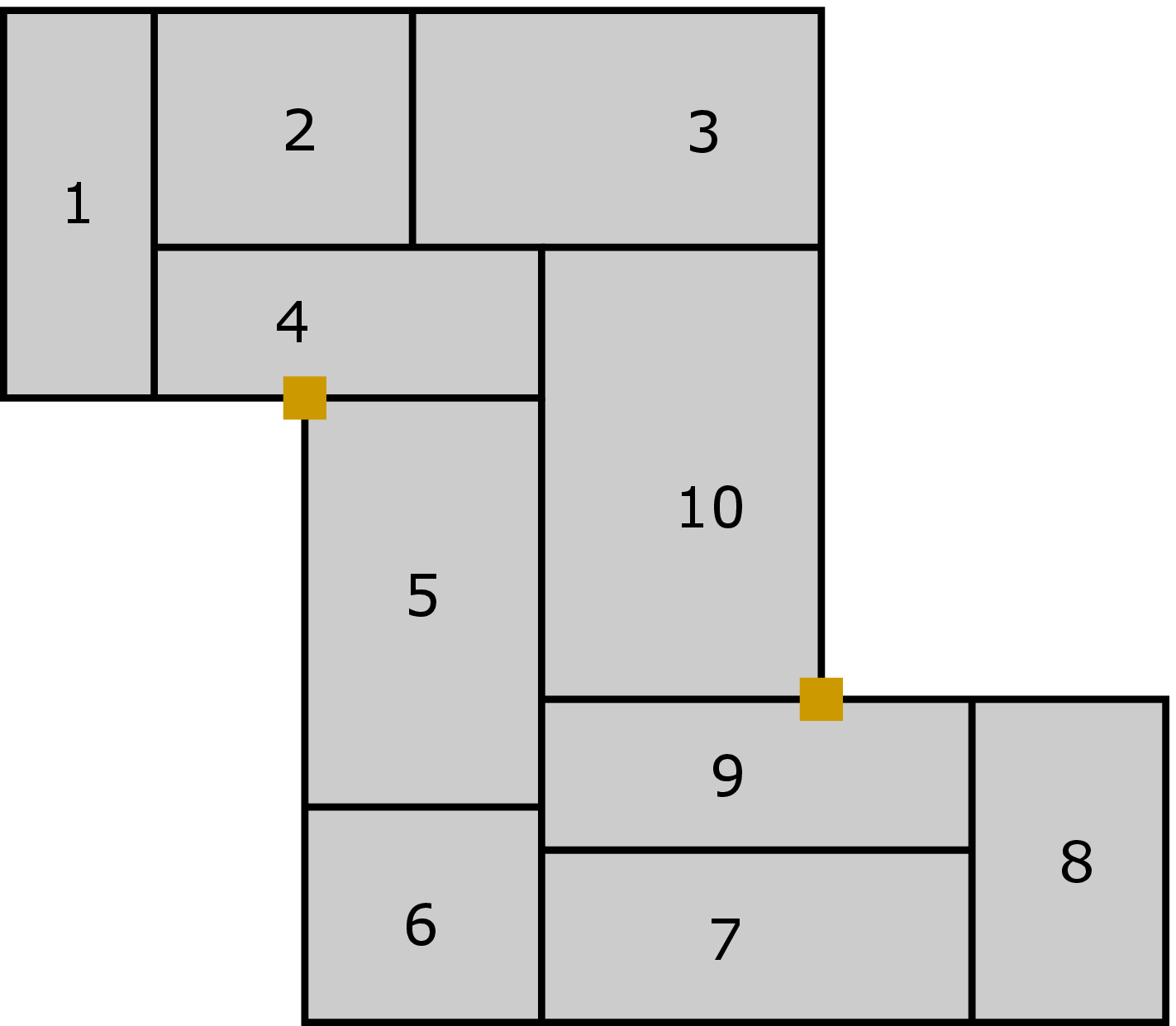}
  \caption{Z-shaped}
  \label{drawing4}
 \end{subfigure}
 \begin{subfigure}[b]{0.32\textwidth}
  \includegraphics[width=.82\textwidth]{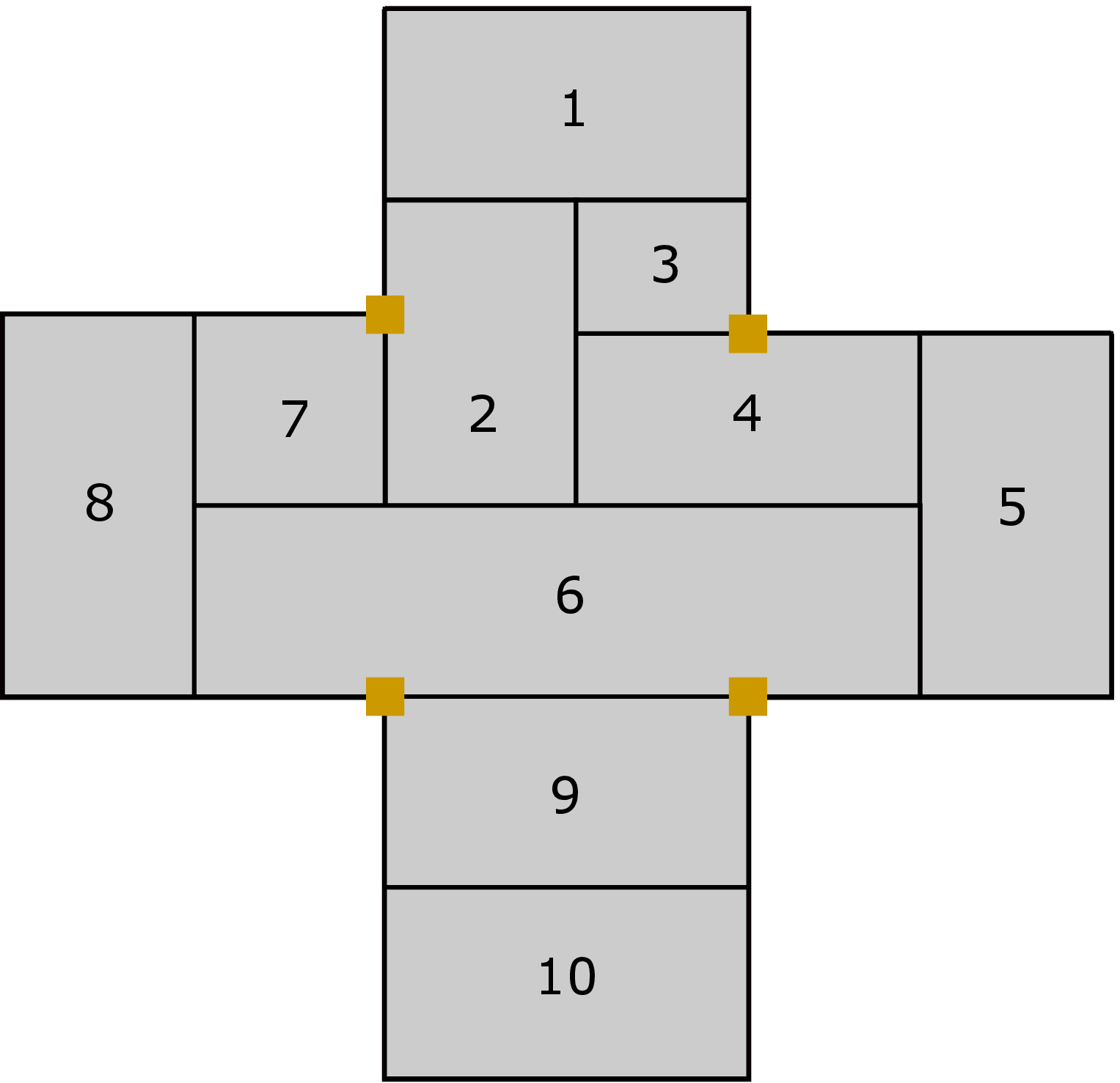}
  \caption{Plus-shaped}
  \label{drawing5}
 \end{subfigure}
 \begin{subfigure}[b]{0.34\textwidth}
  \includegraphics[width=.7\textwidth]{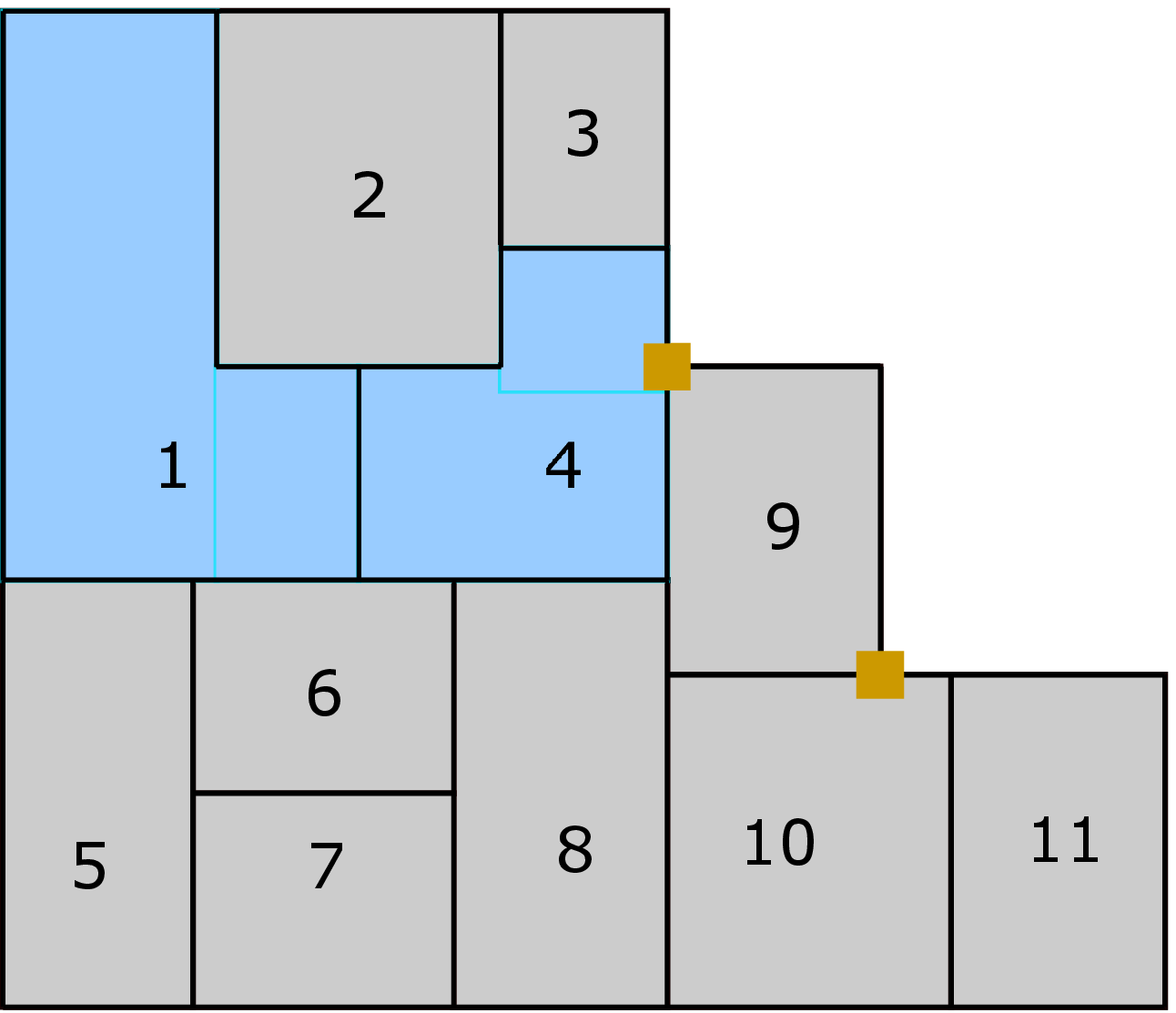}
  \caption{Stair-shaped}
  \label{drawing6}
 \end{subfigure}
 \caption{Classification of NRFPs}
 \label{4}
 \end{figure}

\begin{definition}{\rm 
A \textit{non-rectangular floor-plan} (NRFP) is a floor-plan in which the outer boundary is taken rectilinear with at least one \textit{concave corner} (refer to Figure \ref{4} where different shapes of NRFPs are shown).}
\end{definition}

\begin{definition}{\rm 
The outer boundary of a NRFP contains $2$-joints and $3$-joints \cite{rinsma1988existence} only. Corresponding to any $3$-joint (at the boundary), if the outer angle is $90^{\circ}$, then it is called {\it concave corner} of the NRFP (a concave corner is shown in Figure \ref{drawing-1}).
The NRFP in Figure \ref{drawing-1} has only one concave corner whereas NRFPs in Figures \ref{drawing2}, \ref{drawing3}, \ref{drawing4}, \ref{drawing6} have two concave corners. }
\end{definition}

In a NRFP, exactly two modules have an endpoint at any concave corner, whereas at any convex corner, only one module has an end point.

\begin{lemma}
If the concave corner is formed at the intersection of two boundary walls $W_1$ and $W_2$  of an NRFP, then the module sharing a common wall segment with $W_1$ can not have a common wall segment with $W_2$ and conversely (all the modules are assumed to be rectangles only).
\end{lemma}

\begin{definition}{\rm 
A floor-plan with $k\ge0$ concave corners on its outer boundary is called $k$-concave floor-plan ($k$-CFP). 
 A $k$-CFP is a {\it trivial} NRFP if it can be transformed into a $s$-CFP  with $s<k$  by stretching or shrinking its outer walls while preserving the adjacencies between modules, otherwise it is a non-trivial NRFP.
 
The NRFP in Figure \ref{drawing-1} illustrates a $1$-CFP and all the NRFPs in Figure \ref{4}  are non-trivial while the NRFP in Figure \ref{drawing--7}(b) is trivial.}
\end{definition}

{\it Remark.}
For any NRFP, the number of concave corners can always be increased and from here onward, a NRFP refers to a $non$-$trivial$ NRFP. 

\subsection{Overview}

The rest of the paper is organized in the following manner. In Section \ref{sec2}, we define $L$-shaped floor-plan ($\mathbb{L}$) and categorize it into two forms, trivial and non-trivial. 
Since not all PTPGs admit a non-trivial $\mathbb{L}$, in Section \ref{nece} we address the existence criteria of a non-trivial $\mathbb{L}$ corresponding to a given PTPG $G$. 
Necessary conditions for the existence of a non-trivial $\mathbb{L}$ are given in Theorem \ref{th3}. In Section \ref{algo}, we derive an $O(n^2)$ algorithm to obtain a non-trivial $\mathbb{L}$ corresponding to a given PTPG $G$, if it exists. In Section \ref{set}, we describe a method for selecting the five set of paths for the outer boundary of an $\mathbb{L}$ with respect to a triplet $(a, b, c)$. Since, not all set of paths produce a non-trivial $\mathbb{L}$, in Section \ref{sec 2.3} we identify the characteristics of the set of paths (refer to Theorem \ref{th4}) correspond to which a non-trivial $\mathbb{L}$ will always exist with respect to a triplet $(a, b, c)$. In Section \ref{2.4}, we give sufficient condition for the existence of a non-trivial $\mathbb{L}$ with respect to a triplet $(a, b, c)$. Non-triviality of an $\mathbb{L}$ with respect to a triplet $(a, b, c)$ depends on the graph as well as on the set of paths chosen for the outer boundary of $\mathbb{L}$. Therefore, we provide a method to select the set of paths satisfying all the characteristics (refer to Theorem \ref{th5}) to obtain a non-trivial $\mathbb{L}$ with respect to a triplet $(a, b, c)$. This method will be used in line $6$ of Algorithm \ref{EGDR1} to select five paths $P_1, P_2, P_3, P_4$ and $P_5$. After selecting set of paths, to obtain a REL of the modified graph $G'$ (obtained by adding a new vertex $\textit{NE}$ which is adjacent to all the vertices of path $P_1$ in $G$), in Section \ref{2.41}, we provide a method to compute four paths $P'_1, P'_2, P'_3, P'_4$ for the four-completion of $G'$.
In Section \ref{illus1}, we illustrate Algorithm \ref{EGDR1}. 
In Section \ref{2.5}, we derive an algorithm (refer to Algorithm \ref{EGDR}) to flip one of the edges among $(a, b)$ and $(b, c)$ if both are same labelled in the REL and obtain a RFP corresponding to the REL with edges $(a, b)$ and $(b, c)$  belongings to different sets (Algorithm \ref{EGDR} is a part of Algorithm \ref{EGDR1}). Then, by removing module $\textit{NE}$ from the RFP, we find a non-trivial $\mathbb{L}$. An illustration for Algorithm \ref{EGDR1} and Algorithm \ref{EGDR} is presented in Section \ref{illus2}.
In Section \ref{correctness0} and \ref{correctness}, we discuss the correctness and complexity of Algorithm \ref{EGDR1} and Algorithm \ref{EGDR}. In Section \ref{sec3}, we summarize our contributions and consider
a few open problems.

\subsection{Important Notations}

 In Table \ref{TABLE}, we define some notations which will be used in this work.
 \vspace{.5cm}
 
\begin{table}[H]
    \centering
    \begin{tabular}{ |p{1.6cm}|p{12cm}|  }
\hline 
 \textbf{Symbols}&\textbf{Description}   \\
 \hline
$a, b, c$   &  Vertices of a PTPG   \\
$A, B, C$ &  Modules in a floor-plan correspond to the vertices $a$, $b$ and $c$, respectively \\

PTG & Planar triangulated graph\\

PTPG & Properly triangulated planar graph\\

CIP & Corner implying path\\
 
RFP & Rectangular floor-plan\\
 
OFP & Orthogonal floor-plan \\

$\mathbb{L}$ & $L$-shaped floor-plan\\
 
NRFP  & Non-rectangular floor-plan \\
REL &   Regular edge labelling\\

$C_{cc}$ &  Concave corner \\

$C_{cv}$ &  Convex corner\\

\hline

\end{tabular}
\caption{Notations}
    \label{TABLE}
\end{table}

\section{$L$-shaped floor-plans}
\label{sec2}

A $1$-CFP, or a floor-plan with one ${C_{cc}}$ at the outer boundary, is known as an $L$-shaped floor-plan, indicated as $\mathbb{L}$. The boundary of an $\mathbb{L}$ is made up of six walls $W_1, W_2, W_3, W_4, W_5$ and $W_6$ (refer to Figure \ref{p3}). 
Consider $W_1$ and $W_2$ as the boundary walls of $\mathbb{L}$ meeting at ${C_{cc}}$. Other pair of walls $(W_2, W_3), (W_3, W_4), (W_4, W_5), (W_5, W_6)$ and $(W_6, W_1)$ meet at convex corners. In such a way an $\mathbb{L}$ has five convex corners and one convex corner. A trivial $\mathbb{L}$ can easily be transformed into a floor-plan with no concave corners, i.e., into a rectangular floor-plan. As a result, we exclusively analyze non-trivial $\mathbb{L}$s in this study. 

\begin{definition}{\rm
In any $\mathbb{L}$, two modules are said to be horizontally adjacent if they share a vertical wall segment, and two modules are said to be vertically adjacent if they share a horizontal wall segment.
}
\end{definition}

{\it Remark.}
In $\mathbb{L}$, all the modules are considered rectangles and the concave corner is assumed to be fixed in the North-East direction.
 
\begin{theorem} 
\label{th2}
{\rm An $\mathbb{L}$ is non-trivial if and only if there exist three exterior modules $A$, $B$ and $C$ sharing their boundary wall segments with $W_1$ or $W_2$ such that $A$ and $B$ are vertically (horizontally) adjacent while $B$ and $C$ are horizontally (vertically) adjacent.}
\end{theorem}

\proof 
First, suppose that $\mathbb{L}$ is non-trivial. Then there exist at least three exterior modules sharing their boundary wall segments with $W_1$ or $W_2$, since an $\mathbb{L}$ with only two exterior modules, sharing their boundary wall segments with $W_1$ or $W_2$, is trivial as it can always be transformed to a rectangular floor-plan. 
Let $A_1, A_2,\dots, A_n$; $n \geq 3$ (all modules on the boundary are taken in clockwise order) be $n$ exterior modules, sharing their boundary wall segments with $W_1$ or $W_2$, where $A_i, A_j$ are adjacent modules for $1 \le i \le {n-1}$ and $j = i+1$.
In a floor-plan, two modules can either be horizontally adjacent or vertically adjacent. Hence, any two adjacent modules which share their boundary wall segments with $W_1$ will be vertically adjacent while any two adjacent modules which share their boundary wall segments with $W_2$ will be horizontally adjacent. Based on the adjacency of the modules $A_i$ and $A_j$, the following scenarios are possible:
\begin{enumerate}[i.]

\item
When each pair of modules $A_i$ and $A_j$ are horizontally adjacent.

In this case, only one module (say $A_1$) shares its boundary wall segment with $W_1$, while rest of the modules will be sharing their boundary wall segments with $W_2$. Therefore, $W_2$ can be stretched to form a trivial $\mathbb{L}$ (RFP), which is a contradiction. Hence, this case can not occur.

\item
When each pair of modules $A_i$ and $A_j$ are vertically adjacent (similar to case (i)).

\item 
When some pairs are horizontally adjacent and others are vertically adjacent.

Assume that there are $r$ pairs of modules, which are vertically adjacent then $n-1-r$ pairs of modules will be horizontally adjacent. Since there is a concave corner between $W_1$ and $W_2$, the adjacency between modules will alter exactly once. It implies that modules $A_i$ and $A_j$, $1 \le i \le {r}$,  $2 \le j \le {r+1}$; $j = i+1$ are vertically adjacent, modules $A_i$ and $A_j$ for ${r+1} \le i \le {n-1}$,  ${r+2} \le j \le {n}$; $j = i+1$ are horizontally adjacent. Therefore $A_r$, $A_{r+1}$ and $A_{r+2}$ are three modules such that $A_r$, $A_{r+1}$ are vertically adjacent and $A_{r+1}$, $A_{r+2}$ are horizontally adjacent. If we replace $A_r = A$, $A_{r+1} = B$ and $A_{r+2} = C$, we obtain the desired condition.

\end{enumerate}

Conversely, assume that there exist three exterior modules $A$, $B$ and $C$ sharing their boundary wall segments with $W_1$ or $W_2$ such that $A$ and $B$ are vertically (horizontally) adjacent while $B$ and $C$ are horizontally (vertically) adjacent. In such a case, if we stretch any of the wall $W_1$ or $W_2$, then extra adjacencies will be introduced, i.e. it can not be reduced to a RFP. Hence, $\mathbb{L}$ is non-trivial. 
$\square$

\begin{figure}[]
 \centering
  \begin{subfigure}[b]{0.18\textwidth}
  \includegraphics[width=.80\textwidth]{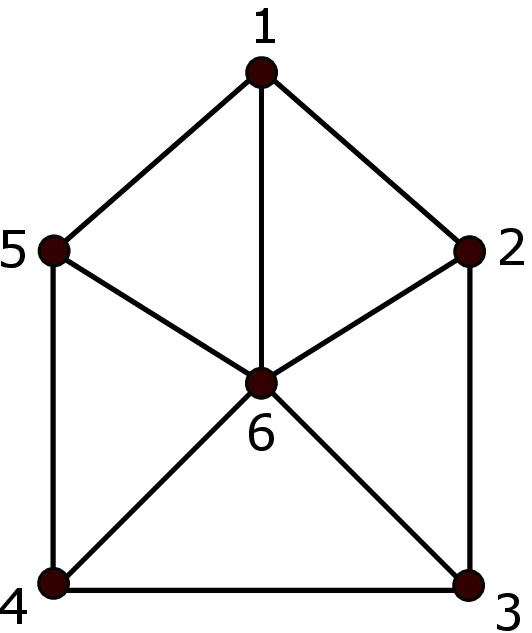}
  \caption{}
  \end{subfigure}
 \begin{subfigure}[b]{0.30\textwidth}
  \includegraphics[width=.85\textwidth]{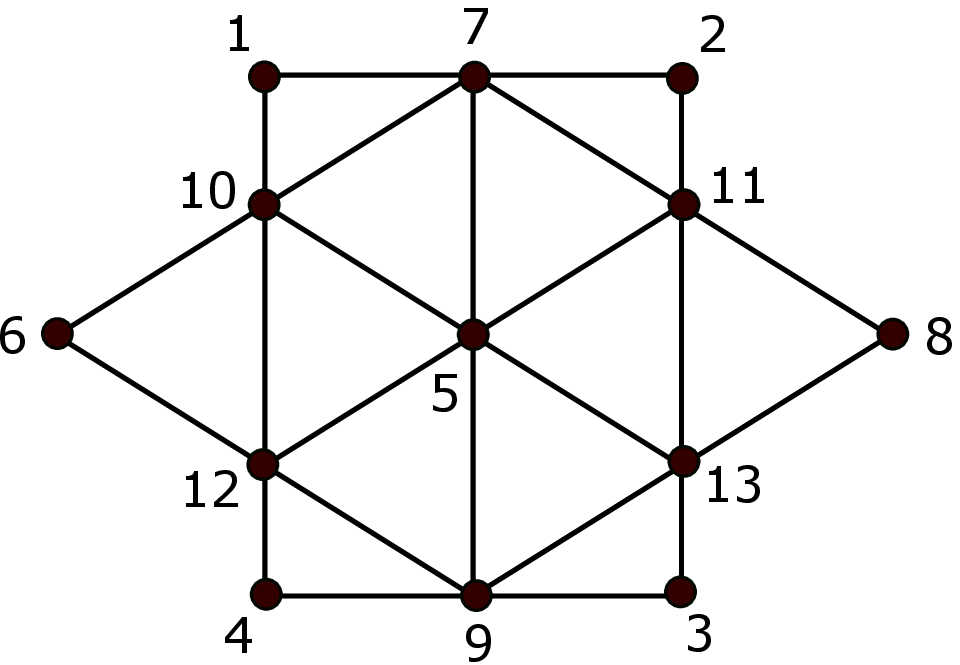}
  \caption{}
  \end{subfigure}
 \caption{Graphs for which non-trivial $\mathbb{L}$ does not exist }
 \label{p12}
 \end{figure}

\begin{figure}[]
 \centering
 \begin{subfigure}[b]{0.32\textwidth}
  \includegraphics[width=.90\textwidth]{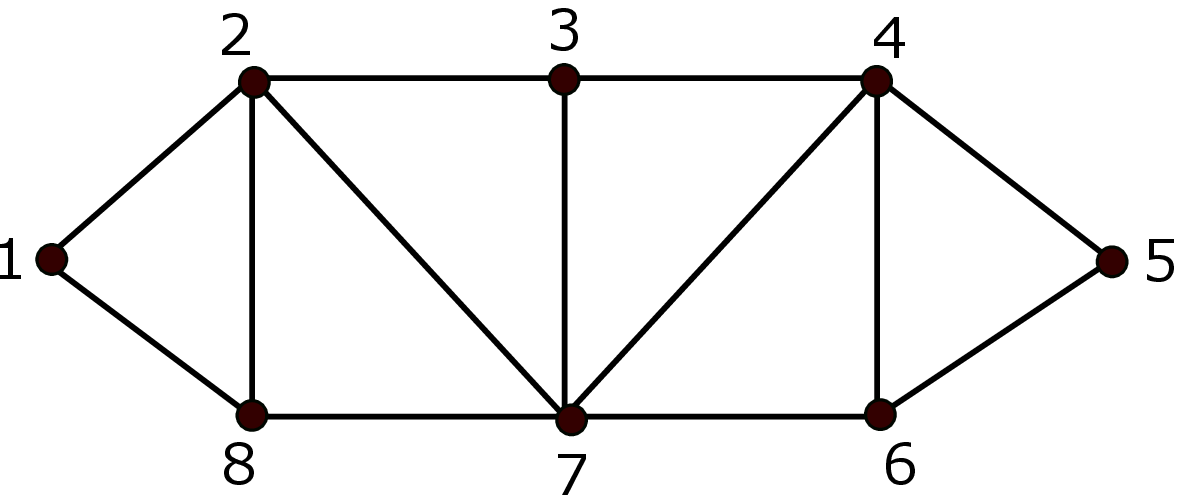}
  \caption{}
  \label{P2}
 \end{subfigure}
 \begin{subfigure}[b]{0.34\textwidth}
  \includegraphics[width=.74\textwidth]{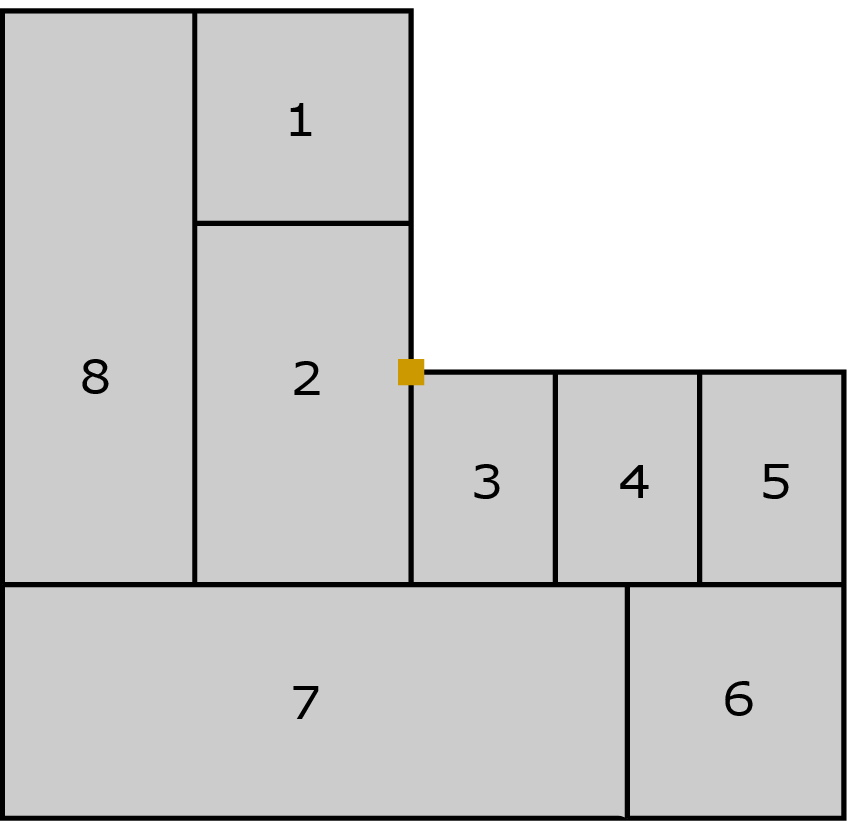}
  \caption{}
  \label{p1}
 \end{subfigure}
 \begin{subfigure}[b]{0.32\textwidth}
  \includegraphics[width=.74\textwidth]{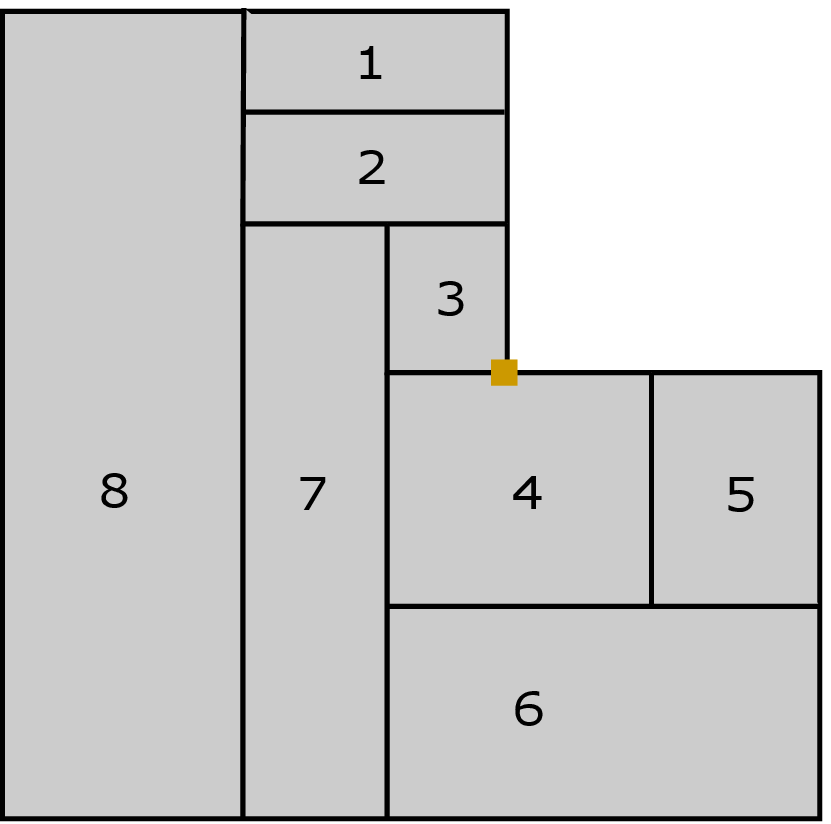}
  \caption{}
  \label{p}
 \end{subfigure}
 \caption{A PTPG $G$ (a), with its two different non-trivial $\mathbb{L}$s (b) and (c).}
 \end{figure}

As we have seen, it is possible to determine whether a given $\mathbb{L}$ is non-trivial, using Theorem \ref{th2}. However, it is not trivial to identify the existence of a non-trivial $\mathbb{L}$ corresponds to a given PTPG. To have a non-trivial $\mathbb{L}$ corresponding to a PTPG, the PTPG must meet certain conditions, known as necessary conditions (refer to Theorem \ref{th3}). 
In Figure \ref{p12}, two PTPGs are drawn, corresponding to which non-trivial $\mathbb{L}$s do not exist since PTPG in Figure \ref{p12}a violets condition (ii) of Theorem \ref{th3} and PTPG in Figure \ref{p12}b violets the condition (i) of Theorem \ref{th3}. At the same time, there might exist PTPGs for which multiple non-trivial $\mathbb{L}$s exist. Figures \ref{p1} and \ref{p} illustrate two different  non-trivial $\mathbb{L}$s for a PTPG given in Figure \ref{P2}. 

In the next Section, we provide graph-theoretic characterization for the existence of a non-trivial $\mathbb{L}$ corresponds to a given PTPG.

\subsection{Existence of a non-trivial $\mathbb{L}$ for a given PTPG \textit{G}}
\label{nece}
\begin{definition}{\rm
An $\mathbb{L}$ is said to be \textit{non-trivial with respect to a triplet $(a, b, c)$}, if modules $A$ and $B$ are vertically adjacent and modules $B$ and $C$ are horizontally adjacent in the corresponding $\mathbb{L}$.

For a PTPG \textit{G} given in Figure \ref{P2}, there are many possibilities for choosing a triplet which satisfy the condition (ii) of Theorem \ref{th3}.
Figures \ref{p1} and \ref{p} demonstrate two distinct non-trivial $\mathbb{L}$s with respect to triplets $(1, 2, 3)$ and $(3, 4, 5)$, respectively for a PTPG given in Figure \ref{P2}. 
 As a result, any non-trivial $\mathbb{L}$ corresponding to a PTPG $G$ is always attributable to a triplet (satisfying condition (ii) of Theorem \ref{th3}) of $G$.
}
\end{definition}

\begin{theorem}
\label{th3}{\rm 
If a non-trivial $\mathbb{L}$ exists correspond to a PTPG $G$, then the following conditions hold:
\begin{enumerate}[i.]

    \item $G$ contains a maximum of five CIPs,
    
    \item $G$ contains at least one triplet of exterior vertices $(a, b, c)$ with $a$ not adjacent to $c$. Furthermore, $a$ and $c$ have no common neighbour except $b$.
    
\end{enumerate}}
\end{theorem}

\proof 
In a PTPG $G$, if there is a CIP then in the corresponding floor-plan, the modules correspond to the vertices of a CIP form at least one convex corner at the boundary. If $G$ contains more than five CIPs (say $k$), the floor-plan must have at least $k$ convex corners. Therefore, if an non-trivial $\mathbb{L}$ exists, $G$ can have at most five CIPs (due to the fact that $\mathbb{L}$ has five convex corners).

\begin{figure}[]
\centering
\includegraphics[width=8.7cm,height=3.3cm]{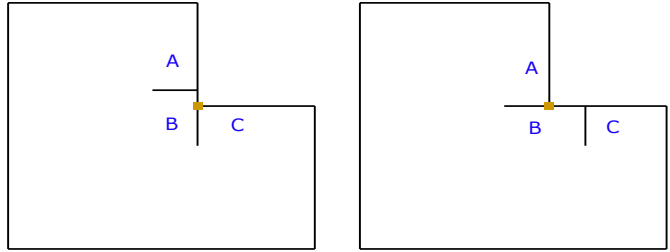}
\caption{Possible ways for different adjacencies of modules $A, B$ (vertically) and $B, C$ (horizontally) in a non-trivial $\mathbb{L}$ with respect to a triplet $(a, b, c)$}
\label{eg16}
\end{figure}

Now, assume that there is no such triplet $(a, b, c)$, possessing condition (ii) in PTPG \textit{G}. Then for all possible triplets $(a, b, c)$ on the outer boundary, either $a$ is adjacent to $c$, or $a$ and $c$ have a common neighbour other than $b$ in \textit{G}. If $a$ is adjacent to $c$ then vertices $a$, $b$ and $c$ form a triangle and the modules correspond to the vertices $a$, $b$ and $c$ ($A$, $B$ and $C$) can not be placed to form a concave corner in $\mathbb{L}$. Since all the modules are rectangles,  module $B$ will not be an outer module anymore, it implies that $\mathbb{L}$ does not exist correspond to the triplet $(a, b, c)$. Now, if $a$ and $c$ have a common neighbour other than $b$, say $b'$, then in the corresponding floor-plan at least one of the modules among $A$, $C$ or $B'$ ($B'$ represents a module in the floor-plan corresponding to a vertex $b'$) will not be a rectangle while assuming modules $A$, $B$ are vertically adjacent and modules $B$, $C$ are horizontally adjacent (see Figure \ref{eg16}). This contradicts our pre-assumption as we consider all the modules to be rectangles. 

Hence, both the conditions (i) and (ii) are necessary to exist a non-trivial $\mathbb{L}$ correspond to a PTPG $G$.

\subsection {Construction of a non-trivial $\mathbb{L}$ for a given PTPG $G$}
\label{algo}
\textbf{Outline of the Algorithm.}

To generate a non-trivial $\mathbb{L}$ corresponding to a PTPG \textit{G}, initially we search for a triplet $(a, b, c)$ which holds condition (ii) of Theorem \ref{th3}. After identifying a triplet $(a, b, c)$, we select five paths $P_1, P_2, P_3, P_4,$ and $P_5$ (refer to Section \ref{set} and Theorem \ref{th5}) made up of outer vertices of $G$, for the outer boundary of $\mathbb{L}$, satisfying some characteristics (see Theorem \ref{th4}). After that, a  new graph $G'$ is constructed by inserting a new vertex $\textit{NE}$ in $G$ which is adjacent to all the vertices of path $P_1$, and an REL is produced (using the method described in \cite{kant1997regular}) for $G'$. If edges $(a, b)$ and $(b, c)$ in the REL belong to different sets $T_1$ and  $T_2$, respectively, we produce a RFP with an extra module $\textit{NE}$ from the REL (using the method provided in \cite{bhasker1986linear}), otherwise we use Algorithm \ref{EGDR} to obtain such an REL (edges $(a, b)$ and $(b, c)$ belonging to different sets). Moreover, an edge $e$ in the REL belonging to set $T_1$ or $T_2$ represents vertical or horizontal adjacency between the modules (which are corresponding to the end vertices of edge $e$) in the RFP, respectively. By eliminating module $\textit{NE}$ from the RFP, we obtain a non-trivial $\mathbb{L}$ corresponding to the given PTPG $G$.

Many steps involved in Algorithm \ref{EGDR1} will be addressed in coming sections. An illustration for Algorithm \ref{EGDR1} is given in Section \ref{illus1}.

\begin{algorithm}[H]
\DontPrintSemicolon
  \KwInput{A PTPG \textit{G}}
  \KwOutput{A non-trivial $\mathbb{L}$ corresponding to \textit{G}}
  \caption{Construction of a non-trivial $\mathbb{L}$ for a given PTPG $G$}
  \label{EGDR1}
 Search for a triplet $(a,b,c)$ (holding condition (ii) in Theorem \ref{th3}) for a concave corner
 
   \If
       {such triplet exists} 
       {Go to step $6$ } 
 \Else 
 { A non-trivial $\mathbb{L}$ does not exist}
           Select five paths $P_1, P_2, P_3, P_4, P_5$ for the outer boundary of $\mathbb{L}$ (see Section \ref{set} and Theorem \ref{th5})
           
           Insert a new vertex \textit{NE} adjacent to all the vertices of $P_1$ in the graph \textit{G}
           
        Compute four paths $P_1', P_2', P_3', P_4'$ for four-completion of modified graph \textit{G'} (see Section \ref{2.41})
        
     Obtain a REL for $\textit{G'}$ (refer to \cite{kant1997regular})
 
     Check the labels of the edges $(a,b)$ and $(b,c)$ in the REL

     \If 
      {both the labels are different} 
     { Go to step $15$}
      \Else
      {Call Algorithm \ref{EGDR} }

  Obtain a RFP from the REL with edges $(a, b)$ and $(b, c)$ differently labelled (refer to \cite{bhasker1986linear})
  
  Remove the module $\textit{NE}$
    
  Get the non-trivial $\mathbb{L}$ for graph \textit{G}
  
 Exit
  \end{algorithm}

\subsection{Method for selecting the set of paths for the boundary of an $\mathbb{L}$} 
\label{set}
 
From \cite{kant1997regular}, we know that for obtaining a RFP using REL approach, it is required to choose four edge-disjoint paths $P_1, P_2, P_3$ and $P_4$, made up of exterior vertices of $G$.  After selecting these four paths, four new vertices $\textit{N}, \textit{E}, \textit{S}, \textit{W}$ are added to $G$, which are adjacent to all the vertices of the respective paths. The corresponding modules of these exterior vertices form boundary walls of the RFP. This is known as the four-completion process. Then a REL is obtained corresponding to the graph produced by the four-completion process. As a result, we extend this notion to generate L-shaped boundary of a floor-plan. Since the number of convex corners in an $\mathbb{L}$ are five, we must identify five edge-disjoint paths, say $P_1, P_2, P_3, P_4$ and $P_5$, made up of the outer vertices of a PTPG.

\begin{figure}[]
\includegraphics[width=7cm,height=6cm]{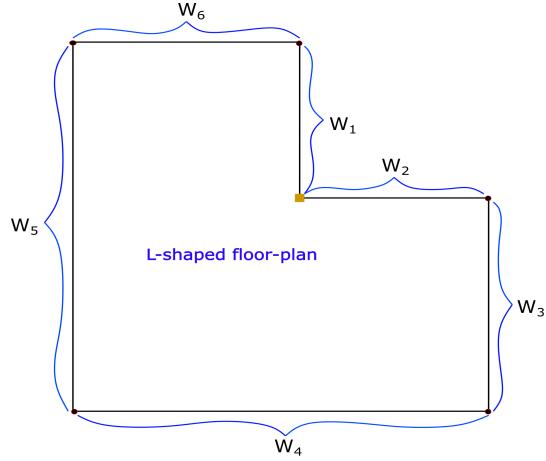}
\centering
\caption{ Boundary walls of an $\mathbb{L}$}
\label{p3}
\end{figure}

In Figure \ref{p3}, the boundary of $\mathbb{L}$ is divided into six boundary walls, say $W_1, W_2, W_3, W_4, W_5$ and $W_6$, where the modules correspond to the vertices of $P_1$ form the boundary walls $W_1$ and $ W_2$, and the modules correspond to the vertices of $P_i$ form the boundary wall $W_j$; $j=i+1$ for $i= 2, 3, 4, 5$.

\begin{figure}[H]
 \begin{subfigure}[b]{0.23\textwidth}
  \includegraphics[width=.75\textwidth]{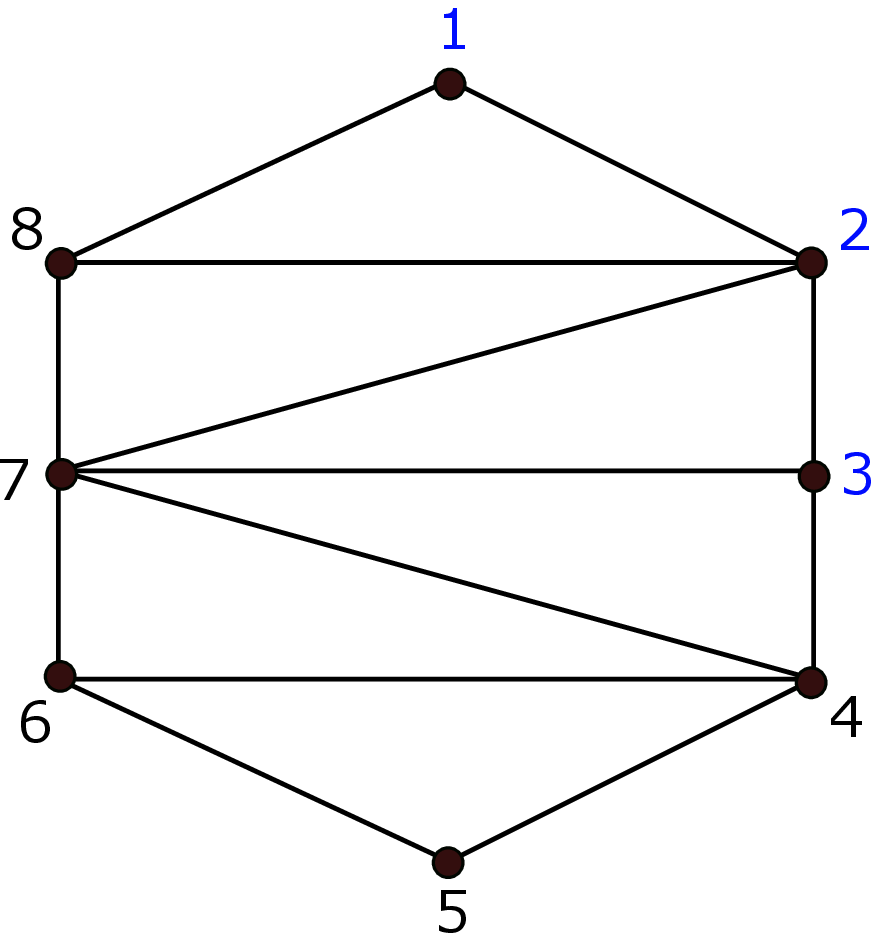}
  \caption{}
  \label{P4}
 \end{subfigure}
 \begin{subfigure}[b]{0.24\textwidth}
  \includegraphics[width=.85\textwidth]{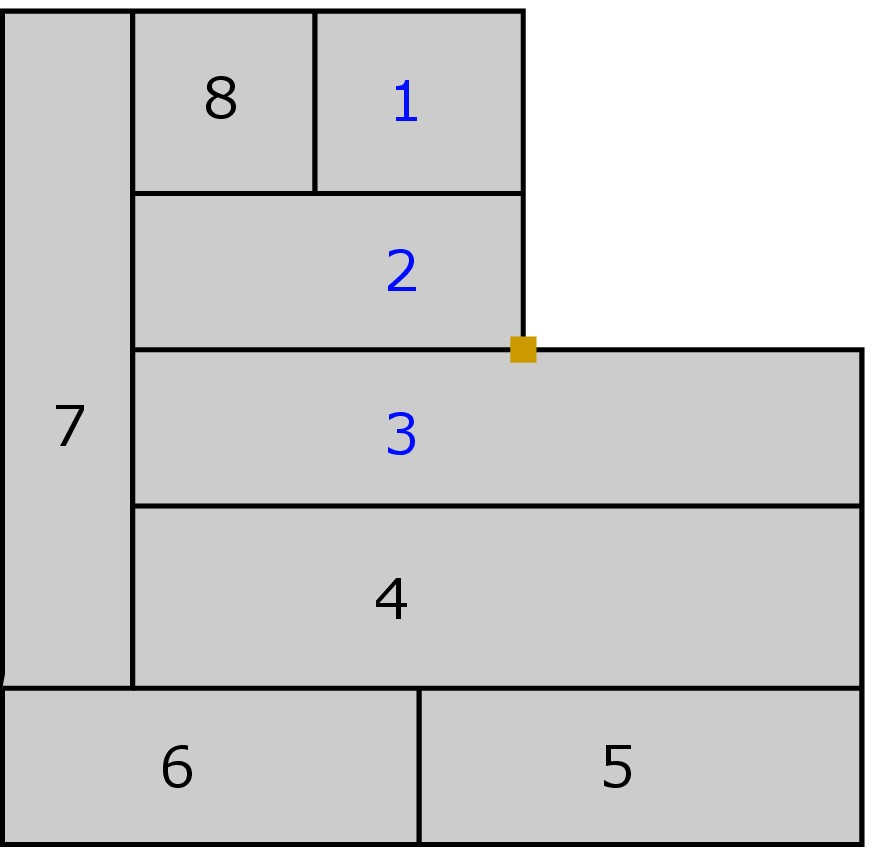}
  \caption{}
  \label{p5}
 \end{subfigure}
 \begin{subfigure}[b]{0.24\textwidth}
  \includegraphics[width=.85\textwidth]{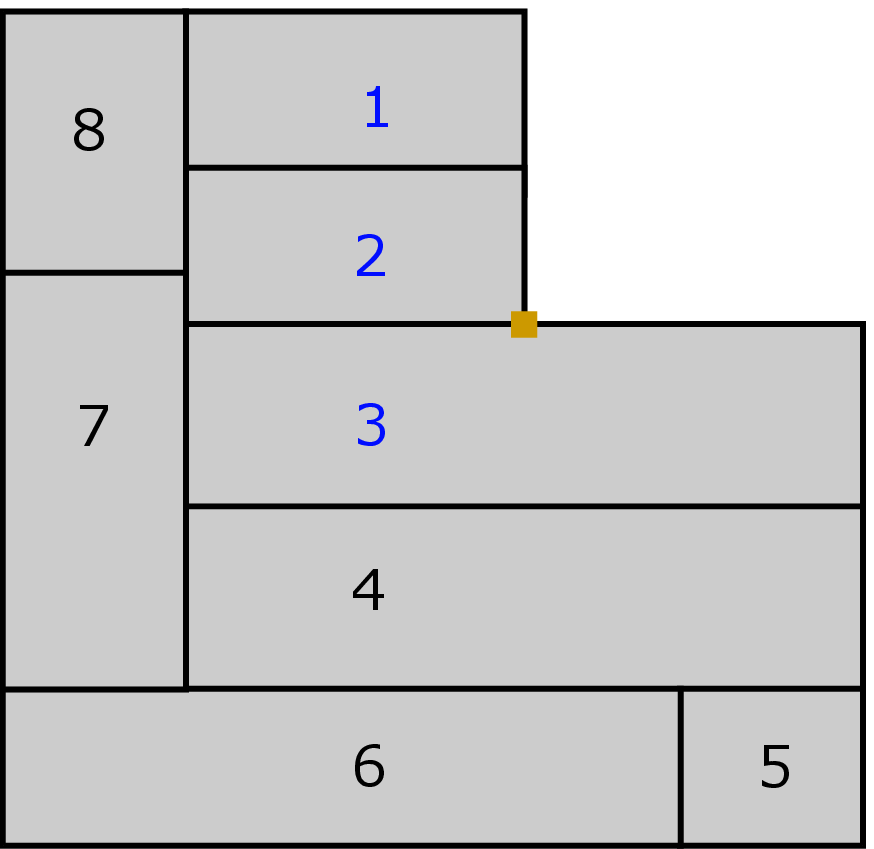}
  \caption{}
  \label{5p}
 \end{subfigure}
 \begin{subfigure}[b]{0.24\textwidth}
  \includegraphics[width=.85\textwidth]{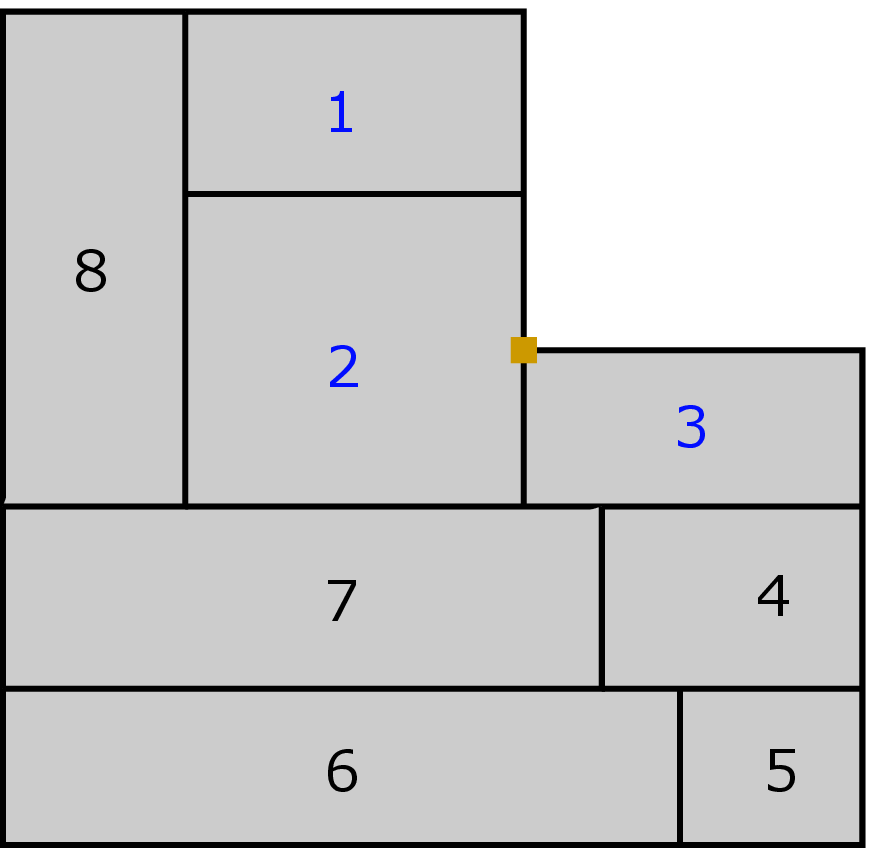}
  \caption{}
  \label{p6}
 \end{subfigure}
 \caption{A PTPG $G$ (a), with its two trivial $\mathbb{L}$s (b) and (c), (d) A non-trivial $\mathbb{L}$ with the same set of paths, used in (c) }
 \end{figure}

In a PTPG $G$, to find a set of five paths, $\{P_1, P_2, P_3, P_4, P_5\}$, where all the vertices of these paths are taken in clockwise order, we first check the number of CIPs in $G$. If the number of CIPs are five, then five paths can be obtained by putting all the consecutive exterior vertices in one path which do not form a cycle. Now, if the number of CIPs are less than five, then we need to split some of the paths to obtain five paths. We can choose any path to split such that the triplet vertices of $G$ would not be separated. Moreover, we assume that the triplet vertices $a,b,c$ always belong to $P_1$.
In Figure \ref{P4}, a PTPG \textit{G} is drawn and its two trivial $\mathbb{L}$s are shown in Figures \ref{p5} and \ref{5p}. These floor-plans are obtained by taking different set of five paths, $\{(1,2,3), (3,4,5), (5,6), (6,7), (7,8,1)\}$ and 
$\{(1,2,3), (3,4,5), (5,6), (6,7,8), 
(8,1)\}$, respectively.

For a PTPG $G$, there may not exist a non-trivial $\mathbb{L}$ correspond to the chosen set of five paths with respect to triplet $(a, b, c)$. At the same time, there may exist another set of paths corresponding to which a non-trivial $\mathbb{L}$ exists with respect to the same triplet $(a, b, c)$.
In Figure \ref{p6}, a non-trivial $\mathbb{L}$ is obtained by selecting the same set of paths (used in Figure \ref{5p}). However, corresponding to the set of paths which are used in Figure \ref{p5}, it is not possible to obtain a non-trivial $\mathbb{L}$ (refer to Theorem \ref{th4}). It indicates that we have to select five paths with some specific properties in order to obtain a non-trivial $\mathbb{L}$ with respect to triplet $(a, b, c)$ for a PTPG $G$. The characteristics of the set of paths to exist a non-trivial $\mathbb{L}$ are presented in Theorem \ref{th4}.

\subsubsection{Characteristic of the set of paths}
\label{sec 2.3}

\begin{theorem}
\label{th4}
{\rm In a PTPG $G$, let the set of five paths be \{$P_1, P_2, P_3, P_4, P_5$\} chosen in the clockwise direction, and the triplet vertices be $(a, b, c)$ belonging to $P_1$.
Then there exists a non-trivial $\mathbb{L}$ corresponding to the chosen set of paths in $G$ if and only if  the following conditions are satisfied:

\begin{enumerate}[i.]

\item there does not exist any vertex which is common in any four consecutive paths.

\item there is no shortcut between any vertex of $P_5$ and vertex $c$ or between any vertex of $P_5$ and any vertex of $P_1$ coming after $c$ in clockwise order.

\item there is no shortcut between any vertex of $P_2$ and vertex $a$ or between any vertex of $P_2$ and any vertex of $P_1$ coming after $a$ in counter-clockwise order. 
 
\item there is no common neighbor of vertices $a, a_1, a_2\dots, a_n$ and $c, c_1, c_2,\dots, c_m$, where $a_1, a_2\dots, a_n$ are the vertices of $P_1$ coming after $a$ in counter-clockwise order and $c_1, c_2,\dots, c_m$ are the vertices of $P_1$ coming after $c$ in clockwise order.

\end{enumerate}
}
\end{theorem}

\proof

We will prove this theorem by contradiction. Firstly, assume that for a PTPG $G$, a non-trivial $\mathbb{L}$ exists with respect to the triplet $(a, b, c)$ and the set of five paths is \{$P_1, P_2, P_3, P_4, P_5$\}, and at least one of the above conditions is not true.

\begin{enumerate}[i.]

\item Assume that there is a vertex in the set of paths which is common in four consecutive paths then, to exist an $\mathbb{L}$ either this vertex will belong to paths \{$P_1, P_2, P_3, P_4$\} or to paths \{$P_3, P_4, P_5, P_1$\}. In both cases, the corresponding module will be sharing a line segment with the complete wall $W_1$ or $W_2$ and the resultant floor-plan will be trivial, which is not possible. Hence, there can not exists any vertex which is common in any four consecutive paths.

\item Now, assume that in $G$, there is a shortcut between a vertex of $P_5$ (say $s$) and vertex $c$ of $P_1$. 
 In any $\mathbb{L}$, exactly two modules have end points at the concave corner. More specifically, if $\mathbb{L}$ is non-trivial with respect to the triplet $(a, b, c)$, then in $\mathbb{L}$ two modules of $\{A, B, C\}$ (either $A$ and $B$ or $B$ and $C$) have end points at the concave corner. For both cases, modules $S$ and $C$ are adjacent (since $(s, c)$ is a shortcut), which is possible only when $B$ and $C$ are vertically adjacent, which is not true as $\mathbb{L}$ is non-trivial with respect to the triplet $(a, b, c)$, i.e. modules $A$ and $B$ are vertically adjacent and module $B$ and $C$ are horizontally adjacent. 
  
Hence, there can not exist a shortcut $(s, c)$ in $G$; otherwise, the resultant $\mathbb{L}$ will be trivial with respect to the triplet $(a, b, c)$. Similarly, the result can be proved, if there is a shortcut between a vertex $s$ of $P_5$ and any vertex $c_r$, $1\leq r \leq m$.
\item It can be proved in a similar way as case (ii). 

\item If $(a, b, c)$ is a triplet then clearly $a$ and $c$ can not have any common neighbour. Other than this, if $a$ has a common neighbour with any of the vertices $c_1, c_2,\dots, c_m$ then in the corresponding $\mathbb{L}$ either both pairs of modules $A, B$ and $B, C$ will be horizontally adjacent or vertically adjacent.
It implies the floor-plan will always be trivial, which is a contradiction. Similarly the result can be proved for other vertices having common neighbours.
 \end{enumerate}
 
To prove the converse part we use the correctness of Algorithm \ref{EGDR1} and Algorithm \ref{EGDR}, from which we can conclude that if a set of paths $\{P_1, P_2, P_3, P_4, P_5\}$ is satisfying the characteristics given in Theorem \ref{th4}, then a non-trivial $\mathbb{L}$ can always be obtained.
$\square$ 

\begin{figure}[]
 \begin{subfigure}[b]{0.495\textwidth}
  \includegraphics[width=.95\textwidth]{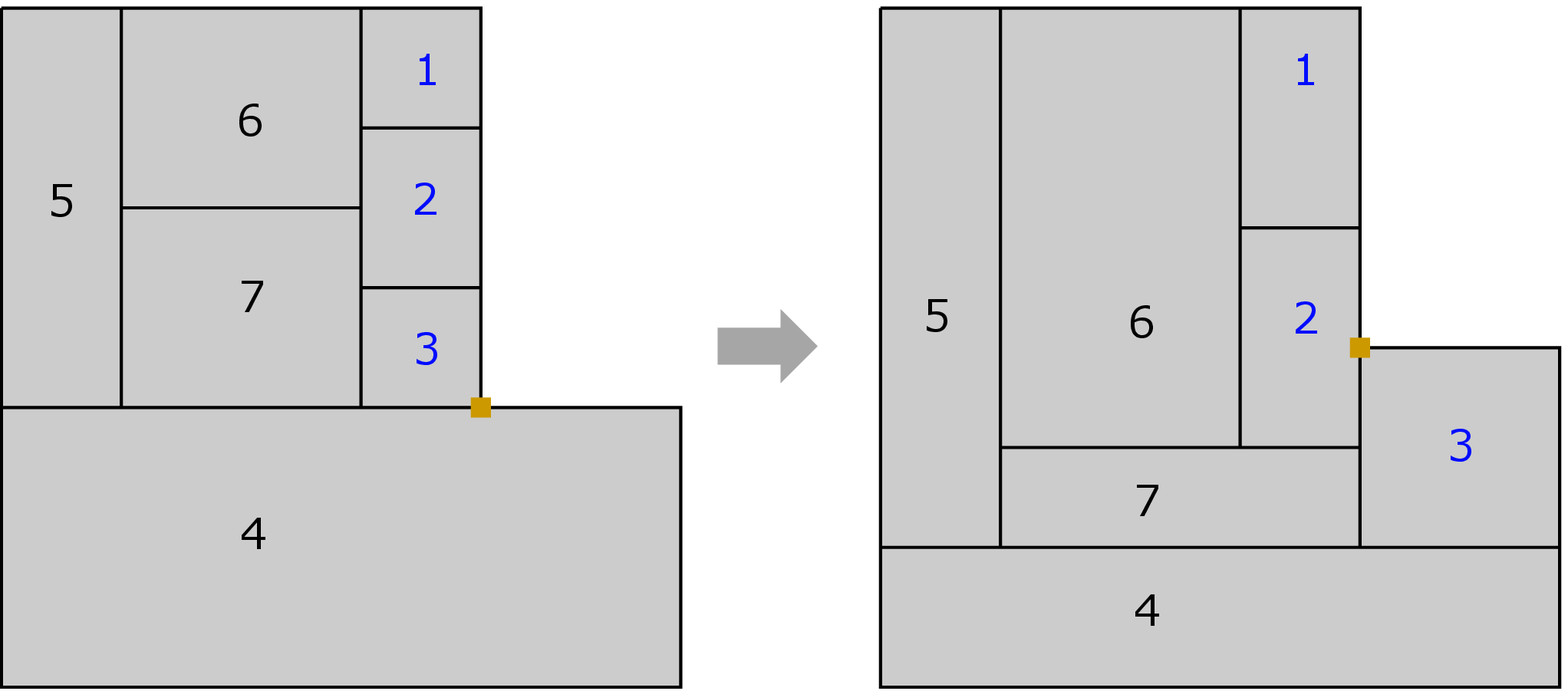}
  \caption{   }
  \label{}
 \end{subfigure}
 \begin{subfigure}[b]{0.495\textwidth}
  \includegraphics[width=.95\textwidth]{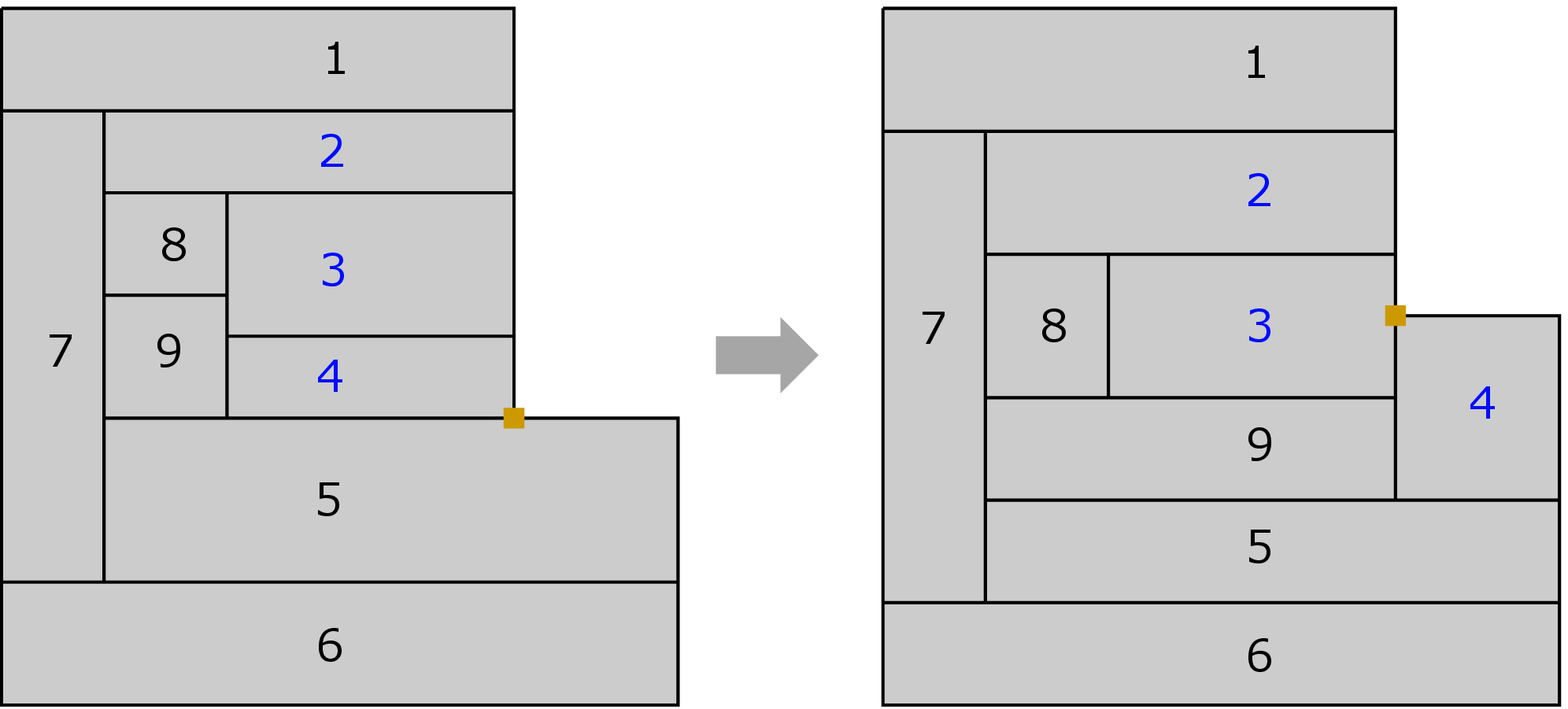}
  \caption{}
  \label{}
 \end{subfigure}
 \caption{Non-trivial $\mathbb{L}$s obtained from trivial $\mathbb{L}$s by modifying the set of paths}
 \label{P7}
 \end{figure}

It is evident from Figure \ref{P7}a (on the left) that a non-trivial $\mathbb{L}$ can not exist with respect to triplet $(1, 2, 3)$ without removing the module $4$ from being common in four consecutive paths $P_1, P_2, P_3$ and $P_4$. Whereas, in Figure \ref{P7}b (on the left), modules $1$ and $5$ both are adjacent to module $7$ (i.e., in the dual graph, the vertices $1$ and $5$ have a common neighbor $7$ which violets the condition (iv) of Theorem \ref{th4}). As a result, a non-trivial $\mathbb{L}$ can not exist without modifying the set of paths with respect to triplet $(2, 3, 4)$. 
In the same way, it can be seen in Figure \ref{p5} why a non-trivial $\mathbb{L}$ is not possible corresponding to the same set of paths (condition (ii) of Theorem \ref{th4} is not holding since vertex $3$ has a shortcut with vertex $7$).

However, by selecting another set of paths, non-trivial $\mathbb{L}$s can be obtained correspond to the $\mathbb{L}$s in Figures \ref{P7}a (on the left) and \ref{P7}b (on the left), shown in Figures \ref{P7}a (right) and \ref{P7}b (right), respectively.

 
\subsection{Sufficient conditions for the existence of a non-trivial $\mathbb{L}$ for a PTPG $G$ }
\label{2.4}
 
The non-triviality of an $\mathbb{L}$ is determined by the associated graph as well as by the set of paths chosen for the outer boundary. The necessary conditions in Theorem \ref{th3} only examine graph-related properties. Therefore, they are not sufficient for every graph to have a non-trivial $\mathbb{L}$.
  If the chosen set of paths does not meet the conditions specified in Theorem \ref{th4}, in the case of four or less CIPs, then the set of paths can always be modified, or a new set of paths can be chosen (refer to Theorem \ref{th5}) to produce a non-trivial $\mathbb{L}$. In the case of five CIPs, however, it is not possible (refer to Theorem \ref{th6}). More precisely, the necessary conditions given in Theorem \ref{th3} are sufficient for the PTPGs with at most four CIPs.

\begin{theorem} 
\label{th5}

{\rm For a PTPG \textit{G} with at most four CIPs and satisfying the necessary conditions given in Theorem \ref{th3}, there exist a set of five paths (made up of exterior vertices), corresponding to which a non-trivial $\mathbb{L}$ can be obtained.}
\end{theorem}

\proof 
To prove the theorem, firstly we will obtain five set of paths correspond to any PTPG with four or less CIPs. After that we will prove that for the obtained set of paths, a non-trivial $\mathbb{L}$ exists. Now, if the number of CIPs are $k$, $k \leq 4$, then $k$ initial paths can be computed, say $P'_1$, $P'_2$,$\dots$, $P'_k$ (refer to Section \ref{set}).
Since, the PTPG has at most four CIPs, it is required to split one or more paths to have five paths.

\begin{figure}
    
\includegraphics[width=17cm,height=21cm]{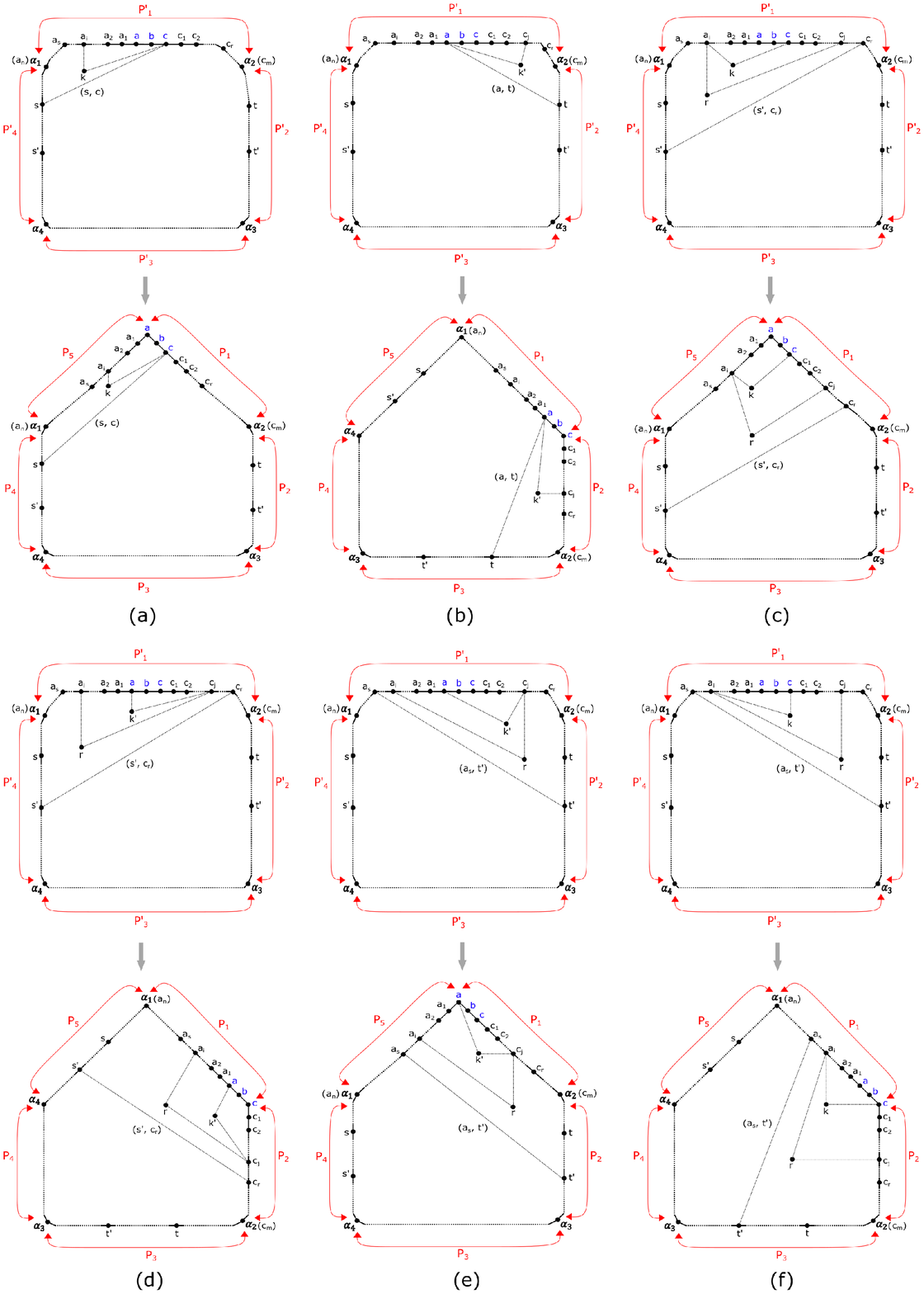}
 \caption{PTPGs with outer boundary divided into five paths by splitting a path in the PTPGs with outer boundary divided into four paths}
 \label{th4eg1}
 \end{figure}

If the number of CIPs are four then initial paths are $P'_1$, $P'_2$, $P'_3$ and $P'_4$ and $(a, b, c)$ is a triplet. Also, assume that $\alpha_1$, $\alpha_2$, $\alpha_3$, $\alpha_4$ are the common end vertices between paths $\{P'_4, P'_1\}$, $\{P'_1, P'_2\}$, $\{P'_2, P'_3\}$ and $\{P'_3, P'_4\}$, respectively (refer to Figure \ref{th4eg1}). The triplet vertices $(a, b, c)$ will be taken in path $P'_1$. Apart from $a$, $b$ and $c$, there can be other vertices in $P'_1$. To determine the set of paths \{$P_1, P_2, P_3, P_4, P_5 $\}, we need to split only one path. So, we choose path $P'_1$ to be splitted while keeping either $a$ or $c$ as an end point of both the new paths.

If path $P'_1$ is splitted from $a$ (refer to Figures \ref{th4eg1}a, \ref{th4eg1}c, \ref{th4eg1}e), then the two new paths will be called $P_5$ and $P_1$ such that $a$ is the common end vertex for both of the paths. The vertices from $\alpha_1$ to $a$ will be taken in $P_5$ and vertices from $a$ to $\alpha_2$ will be taken in path $P_1$ (since $\alpha_1$ and $\alpha_2$ are end points of $P'_1$). Also, we call rest of the paths as $P_2$, $P_3$ and $P_4$ in place of $P'_2$, $P'_3$ and $P'_4$, respectively.
 
If path $P'_1$ is splitted from $c$ (refer to Figures \ref{th4eg1}b, \ref{th4eg1}d, \ref{th4eg1}f), then the two new paths will be called $P_1$ and $P_2$ such that $c$ is the common end vertex for both of the paths. The vertices from $\alpha_1$ to $c$ will be taken in $P_1$ and vertices from $c$ to $\alpha_2$ will be taken in path $P_2$. Also, we call rest of the paths as $P_3$, $P_4$ and $P_5$ in place of $P'_2$, $P'_3$ and $P'_4$, respectively.

The following cases may occur while splitting the path $P'_1$, where $s$ and $s'$ are the vertices from path $P'_4$, $t$ and $t'$ are the vertices from path $P'_2$.
$a_1$, $a_2$,$\dots$,$a_i$,$\dots$,$a_s$,$\dots$,$a_n = \alpha_1$ are consecutive vertices from $P'_1$ in the anticlockwise order coming after $a$. $c_1$, $c_2$,$\dots$,$c_j$,$\dots$,$c_r$,$\dots$,$c_m = \alpha_2$ are consecutive vertices from $P'_1$ in the clockwise order coming after $c$.

\begin{enumerate}[i.]

\item There is a shortcut $(s,c)$ but there does not exist any shortcut $(s', c_r)$, $ 1\leq r \leq m$ or there is a shortcut $(a, t)$ but there does not exist any shortcut $(a_s, t')$, $ 1\leq s \leq n$. 
\begin{enumerate}[(a)]

\item If any vertex $a_i$, $ 1\leq i \leq n$ and $c$ have a common neighbor (say $k$), then we split $P'_1$ from $a$ (see Figure \ref{th4eg1}a). The set of five paths will be $\{P_1 = \{a, b, c, c_1, c_2, \dots c_j,\dots,c_r,\dots, c_m = \alpha_2 \}, P_2 = \{\alpha_2,\dots,\alpha_3\}, P_3 = \{\alpha_3,\dots,\alpha_4\}, P_4 = \{\alpha_4,\dots, \alpha_1\}, P_5 = \{\alpha_1 = a_n,\dots,a_s,\dots,a_i,\dots, a_2, a_1, a\}\}$. The obtained set of paths is satisfying all the conditions given in Theorem \ref{th4}. Therefore, a non-trivial $\mathbb{L}$ will exist correspond to this set of paths. If any vertex $c_j$, $ 1\leq j \leq m$ and $a$ have a common neighbor (say $k'$), then we split $P'_1$ from $c$ (see Figure \ref{th4eg1}b). The set of five paths will be $\{P_1 = \{\alpha_1  = a_n,\dots, a_s,\dots,a_i,\dots,a_2, a_1, a, b, c\}, P_2 = \{c, c_1, c_2,\dots,c_j,\dots,c_r,\dots, c_m = \alpha_2\}, P_3 = \{\alpha_2,\dots,\alpha_3\}, P_4 = \{\alpha_3,\dots,\alpha_4\}, P_5 = \{\alpha_4,\dots, \alpha_1\}\}$. The obtained set of paths is satisfying all the conditions given in Theorem \ref{th4}, hence a non-trivial $\mathbb{L}$ will exist correspond to this set of paths.

\item If there is not a common neighbor of any vertex $a_i$ and $c$, $ 1 \leq i \leq n$ (in case of shortcut $(s, c)$), we split $P'_1$ from $a$. If there is not a common neighbor of any vertex $c_j$ and $a$, $ 1 \leq j \leq m$ (in case of shortcut $(a,t)$) then we split $P'_1$ from $c$. 
\end{enumerate}
 
\item If there is any shortcut $(s', c_r)$, $1 \leq r \leq m$  but there does not exist any shortcut $(s, c)$, or $(a_s, t')$, $1 \leq s \leq n$ is any shortcut but there does exist any shortcut $(a, t)$.
\begin{enumerate}

\item If any vertex $a_i$ ($1\leq i\leq n $) and $c$ have a common neighbor, we split $P'_1$ from $a$ (see Figure \ref{th4eg1}c) and if any vertex $c_j$ ($j\leq r$) and $a$ have a common neighbor, then we split $P'_1$ from $c$ (see Figure \ref{th4eg1}d). Both types of common neighbor can not appear together in any graph (since the graph would not be planar anymore), but vertices $a_i$ and $c_j$ can have a common neighbor with both of these types of common neighbors (in case of shortcut $(s', c_r)$).

In case of shortcut $(a_s, t')$, if any vertex $c_j$ and $a$ have a common neighbor, then we split $P'_1$ from $a$ (see Figure \ref{th4eg1}e) and if any vertex $a_i$ ($i\leq s$) and $c$ have a common neighbor, then we split $P'_1$ from $c$ (see Figure \ref{th4eg1}f). Both of these types of common neighbor can not appear together in any graph, but vertices $a_i$ and $c_j$ can have common neighbors with both of these types.

\item When neither any vertex  $c_j$, $ 1\leq j \leq m$ and $a$ have a common neighbor nor any vertex $a_i$, $ 1\leq i \leq n$ and $c$ have a common neighbor but vertices $a_i$ and $c_j$ may or may not have a common neighbor.
Then, we split either from $a$ or from $c$ (from both ways we can obtain a set of paths corresponding to which a non-trivial $\mathbb{L}$ exists but we prefer to split from that vertex which ensures least number of vertices in $P_1$).

\end{enumerate}

\item When $(s', c_r)$, $ 1\leq r \leq m$ and $(s, c)$ both shortcuts are present or $(a_s, t')$, $ 1\leq s \leq n$ and $(a, t)$ both shortcuts are present. 

This case is similar to case (ii) and hence the set of paths will be chosen in the same manner.

\item When there is not any shortcut $(s', c_r)$, $ 1\leq r \leq m$ and $(s, c)$ or $(a_s, t')$, $ 1\leq s \leq n$ and $(a, t)$ but common neighbors are there (as defined in case (ii)).

This case is also similar to case (ii) and hence the set of paths will be chosen in the same manner.

\item When there is not any of the shortcuts and not any common neighbors defined above, then we will split path $P'_1$ either from $a$ or from $c$, so that the number of vertices are least in $P_1$. From both ways of splitting, we will have a non-trivial $\mathbb{L}$ correspond to the obtained set of paths.

\end{enumerate}

When the number of CIPs in a PTPG are three, then we have initially three paths (say $P'_1$, $P'_2$, $P'_3$). By splitting these, we need two more paths in order to have five set of paths {$P_1$, $P_2$, $P_3$, $P_4$, $P_5$}. To split for the first time, we repeat the same process that was used in case of four CIPs. To split again, we can split any path among the four paths such that the triplet is in $P_1$. Similarly, we can obtain five set of paths in case of two CIPs or no CIPs.

Thus, in all the cases, a set of five paths $\{P_1, P_2, P_3, P_4, P_5\}$ is obtained which satisfies the conditions given in Theorem \ref{th4}, i.e. a non-trivial $\mathbb{L}$ can be obtained correspond to it. 
$\square$
\subsection {Method to compute four paths $P'_1, P'_2, P'_3, P'_4$ for the modified graph $G'$.}
\label{2.41} 
 For a given PTPG $G$ with four or less CIPs, we use the method discussed in Theorem \ref{th5} to obtain five paths $P_1, P_2, P_3, P_4, P_5$ in line $6$ of Algorithm \ref{EGDR1}. If a PTPG has five CIPs, then we select five paths using the method defined in Section \ref{set}. Now, after inserting a vertex $\textit{NE}$ adjacent to all the vertices of $P_1$, we identify four paths $P'_1, P'_2, P'_3, P'_4$ for four-completion of the modified graph $G'$. If $P_1$ contains outer vertices $a, b, c,\dots,c_{m-1}, c_m$ then $P_1'$ contains all the vertices of $P_5$, vertex $a$ and vertex $\textit{NE}$ (in the clockwise order). $P_2'$ will include $\textit{NE}$, $c_m$ and all the vertices of $P_2$, $P_3'$ will contain the vertices of $P_3$, and $P_4'$ will include the vertices of $P_4$. If $P_1$ contains outer vertices $a_n, a_{n-1},\dots,a, b, c$, then $P_1'$ contains all the vertices of $P_5$, $a_n$ and $\textit{NE}$ (in clockwise order). $P_2'$ will include $\textit{NE}$, $c$ and all the vertices of $P_2$, $P_3'$ will contain the vertices of $P_3$, and $P_4'$ will include the vertices of $P_4$. After identifying four paths $P'_1, P'_2, P'_3, P'_4$, we add four new vertices $N, E, S, W$ which are adjacent to all the vertices of the respective paths.

\begin{theorem}
\label{th6}

{\rm In a PTPG \textit{G} with five CIPs, if there is a set of paths corresponding to which a non-trivial $\mathbb{L}$ (with respect to triplet $(a, b, c)$) does not exist, then there does not exist a non-trivial $\mathbb{L}$ (with respect to fixed triplet $(a, b, c)$), corresponding to any other set of paths.
}
\end{theorem}

\proof

Suppose a PTPG \textit{G} has five CIPs and the triplet of vertices is $(a, b, c)$. A set of five paths \{$P_1, P_2, P_3, P_4, P_5$\} is given corresponding to which a non-trivial $\mathbb{L}$ does not exist. Then, there are five shortcuts corresponding to these CIPs. Assume that ($a_{1}, a_{2}$), ($b_{1}, b_{2}$), ($c_{1}, c_{2}$), ($d_{1}, d_{2}$), ($e_{1}, e_{2}$) are five shortcuts and the outer paths of these shortcuts are CIPs of $G$. Then, in each path $P_i$, $i = 1, 2, 3, 4, 5$,  two end vertices from different shortcuts will lie, otherwise there will be a cycle in between the vertices of a path which contradicts the assumption that \{$P_1, P_2, P_3, P_4, P_5$\} is the required set of paths.

\begin{figure}[]
\includegraphics[width=6.9cm,height=7.cm]{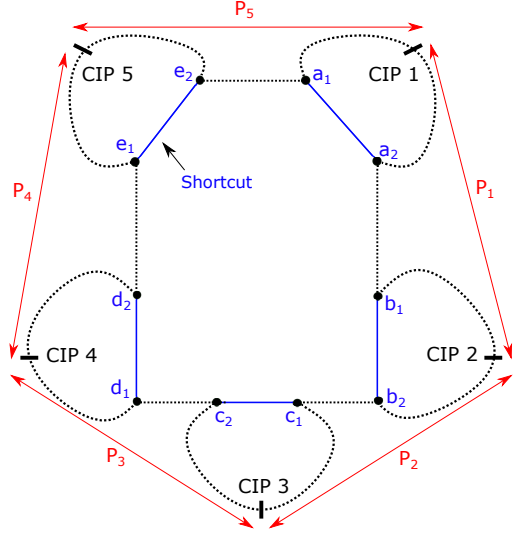}
\centering
\caption{A general PTPG with five CIPs}
\label{drawing-2}
\end{figure}

Without loss of generality, select vertices $a_{2}$ and $b_{1}$ in $P_1$, accordingly, $\{b_{2}, c_{1}\}, \{c_{2}, d_{1}\}, \{d_{2}, e_{1}\}$, and $\{e_{2}, a_{1}\}$ will be in the paths $P_2$, $P_3$, $P_4$ and $P_5$ respectively, and also take the triplet $(a, b, c)$ in $P_1$ (refer to Figure \ref{drawing-2}). Since corresponding to the given set of paths, a non-trivial $\mathbb{L}$ does not exist, it follows from Theorem \ref{th4} that the given set of paths does not satisfies at least one of the conditions given in Theorem \ref{th4}. In such a case, these set of paths can not be modified, because we have already five paths. Neither any path can be splitted nor any two paths can be merged further. Therefore, the given set of paths can not be modified in other set of paths, corresponding to which a non-trivial $\mathbb{L}$ can be obtained. Hence, there does not exist a non-trivial $\mathbb{L}$ (with respect to triplet $(a, b, c)$), correspond to any other set of paths for the PTPG \textit{G}.
$\square$  

\begin{figure}[H]
\includegraphics[width=5.5cm,height=4cm]{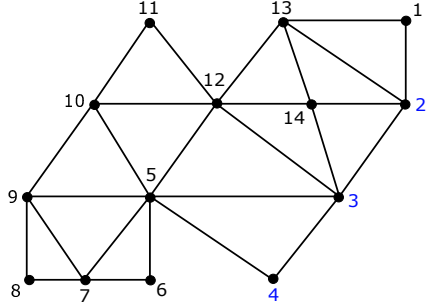}
\centering
\caption{A PTPG with five CIPs }
\label{p9}
\end{figure}

In Figure \ref{p9}, a PTPG is drawn which has five CIPs, namely, $\{13, 1, 2\}, \{3, 4, 5\}, \{5, 6, 7\}, \{7, 8, 9\}, \{10, 11, 12\}$. Triplet of vertices is $(2, 3, 4)$ and the set of paths is $\{ (1, 2, 3, 4), (4, 5, 6), (6, 7, 8), (8, 9, 10, 11), (11, 12, 13, 1)\}$. There is a vertex $12$ in $P_5$ which forms a shortcut with the vertex $3$ of $P_1$. It implies that a non-trivial $\mathbb{L}$ can not exist correspond to this triplet.

In case of four or less CIPs in $G$, we can always obtain a non-trivial $\mathbb{L}$ if it satisfies the necessary conditions because we can always select a set of five paths for which a non-trivial $\mathbb{L}$ can be obtained. 

\begin{figure}[H]
 \centering
 \begin{subfigure}[b]{0.28\textwidth}
  \includegraphics[width=.66\textwidth]{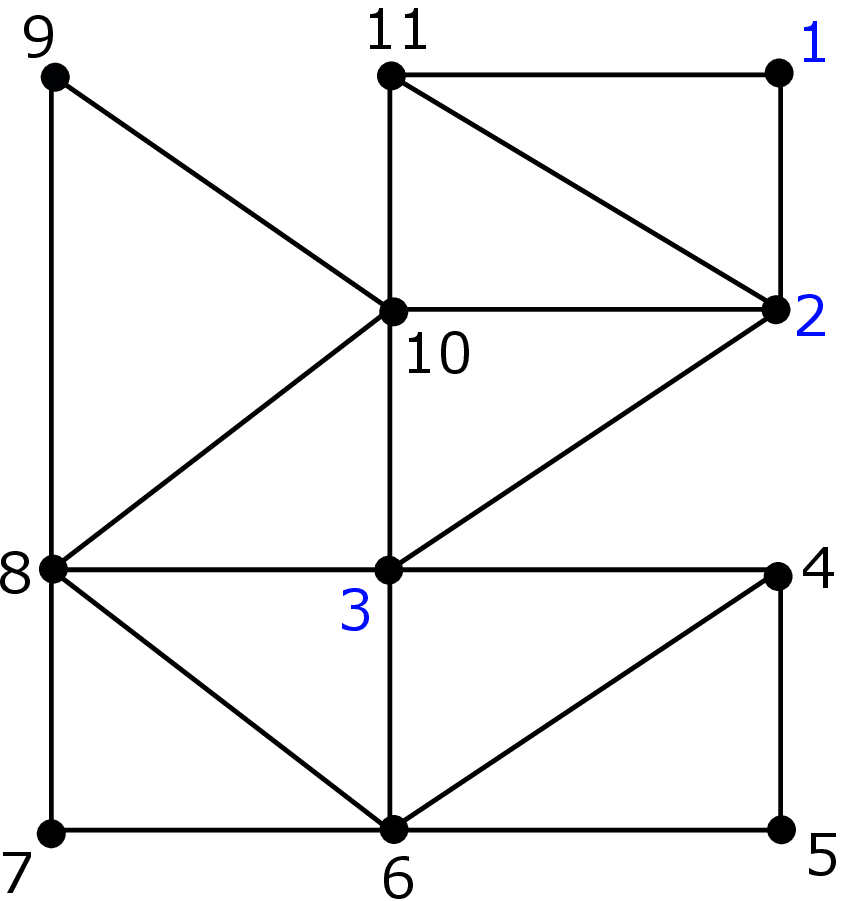}
  \caption{A PTPG with $4$ CIPs}
  \label{p10}
 \end{subfigure}
 \begin{subfigure}[b]{0.32\textwidth}
  \includegraphics[width=.78\textwidth]{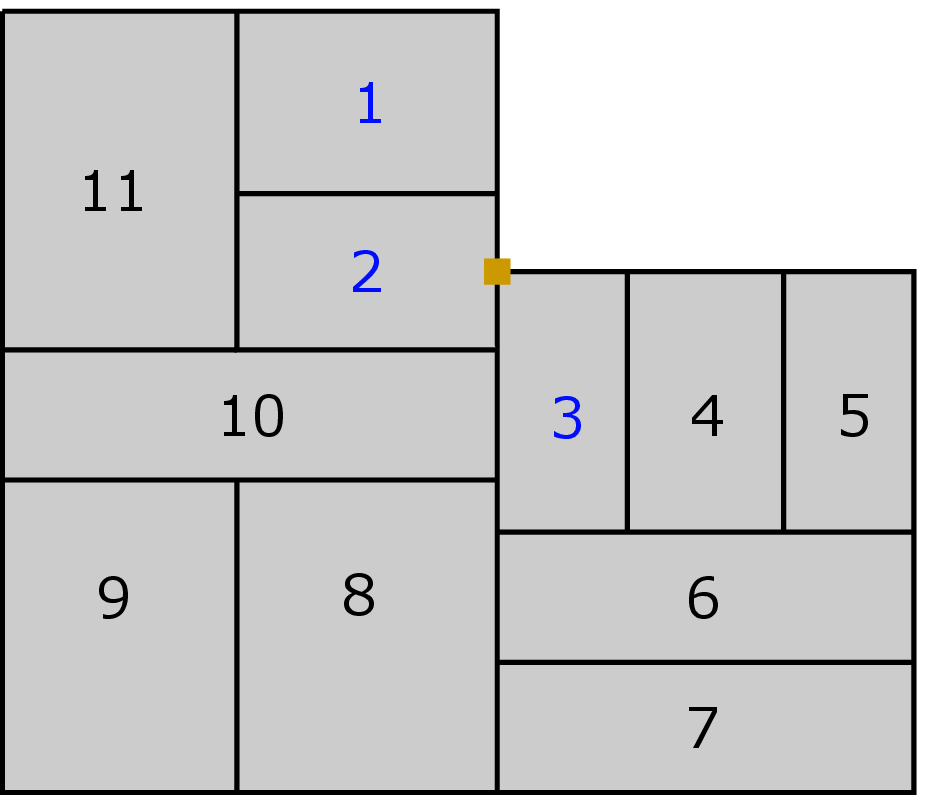}
  \caption{A non-trivial $\mathbb{L}$}
  \label{p11}
 \end{subfigure}
 \caption{}
 \end{figure}

In Figure \ref{p10}, a PTPG is shown, which has four CIPs. The initial set of paths is
\{$(1, 2, 3)$, $(3, 4, 5)$, $(5, 6, 7)$, $(7, 8, 9)$, $(9, 10, 11, 1)$\} and 
$(1, 2, 3)$ is chosen as a triplet. A shortcut $(10, 3)$ lies between a vertex of $P_5$ and vertex $3$. 
Now, according to Theorem \ref{th4}, a non-trivial $\mathbb{L}$ can not exist correspond to this set of paths. Hence, we choose another set of paths (using the proof of Theorem \ref{th5}) for which a non-trivial $\mathbb{L}$ exists. In Figure \ref{p11}, a non-trivial $\mathbb{L}$ is obtained corresponding to the set of paths $\{(1, 2, 3, 4, 5),
(5, 6, 7), (7, 8, 9), (9, 10, 11),
(11, 1)\}$.

\subsection{An illustration for Algorithm \ref{EGDR1}}
\label{illus1}

In Figure \ref{dig7}, a PTPG is shown.

\begin{enumerate}

\item Consider $(a,b,c)$ as a triplet.

\item There are two CIPs in PTPG $G$ (see Figure \ref{dig7}), $\{d, e, a, b\}$ and $\{b, c, d\}$. The set of five paths is $\{P_1=(a, b, c),  P_2=(c),  P_3=(c, d),  P_4=(d, e, a),  P_5=(a)\}$.

\item After adding a new vertex $\textit{NE}$ which is adjacent to all the vertices of path $P_1$, we obtain a modified graph $G'$ (see Figure \ref{dig8}).

\begin{figure}[]
\begin{subfigure}[b]{0.26\textwidth}
\includegraphics[width=.68\textwidth]{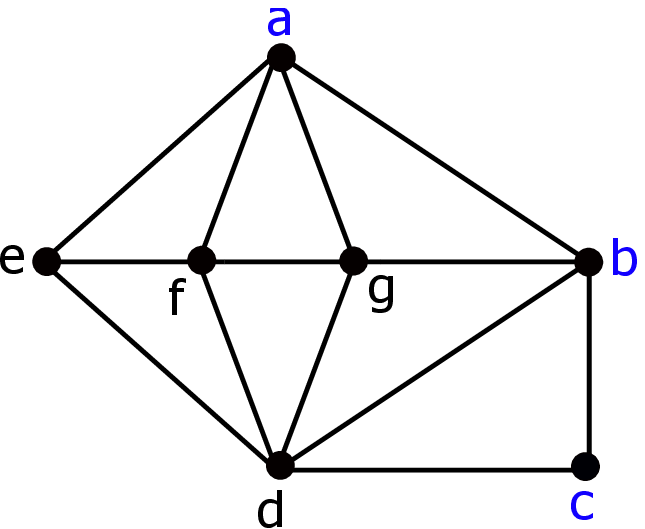}
\caption{}
\label{dig7}
\end{subfigure}
\begin{subfigure}[b]{0.32\textwidth}
\includegraphics[width=.73\textwidth]{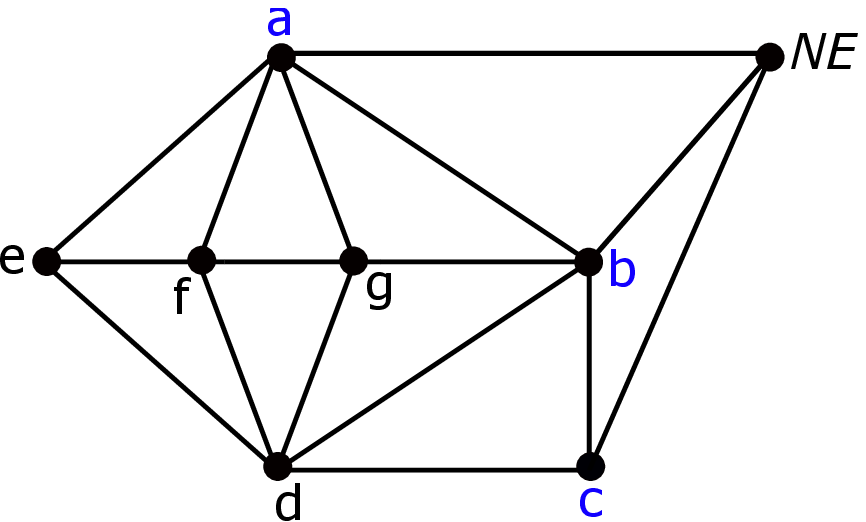}
\caption{
}
\label{dig8}
\end{subfigure} \begin{subfigure}[b]{0.41\textwidth}
\includegraphics[width=.88\textwidth]{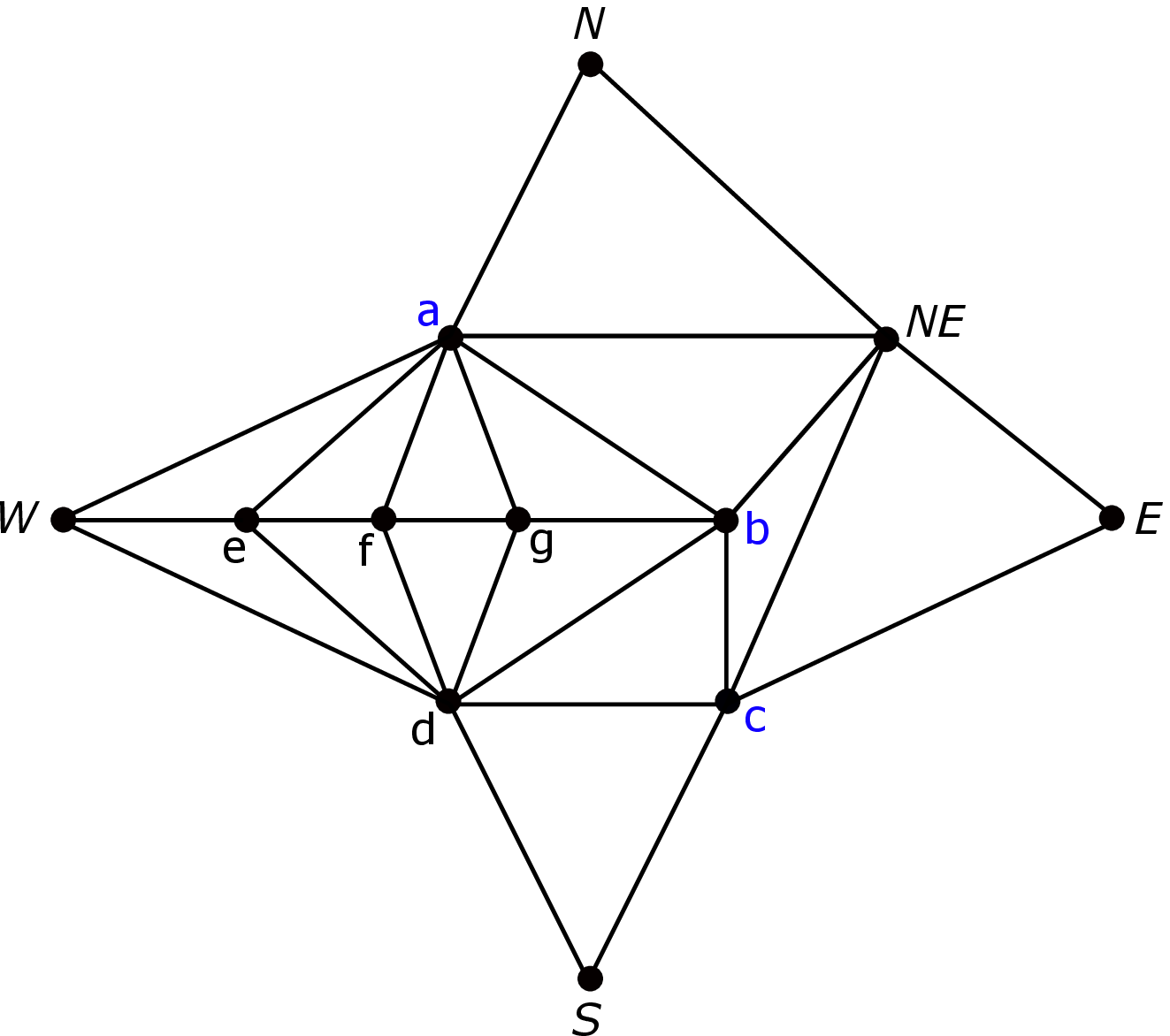}
\caption{}
\label{ex1}
\end{subfigure}
\caption{(a) A PTPG $G$, (b) Modified graph $G'$, (c) Graph obtained after adding \textit{N}, \textit{E}, \textit{S},
\textit{W}}
\end{figure}
 
\item To construct a RFP for the modified graph $G'$, four paths are computed $P'_1=(a, \textit{NE}), P'_2=(\textit{NE}, c), P'_3=(c, d), P'_4=(d, e, a)$ in four-completion process. After that four new vertices $N, E, W, S$ are added, which are adjacent to all the vertices of paths $P'_1, P'_2, P'_3, P'_4$, respectively (refer to Figure \ref{ex1}).

\item To obtain a REL, we perform edge contraction (see Figure \ref{rel}) and then edge expansion (see Figure \ref{finalexp}) \cite{kant1997regular}.

\begin{figure}[]
\includegraphics[width=17.6cm,height=18cm]{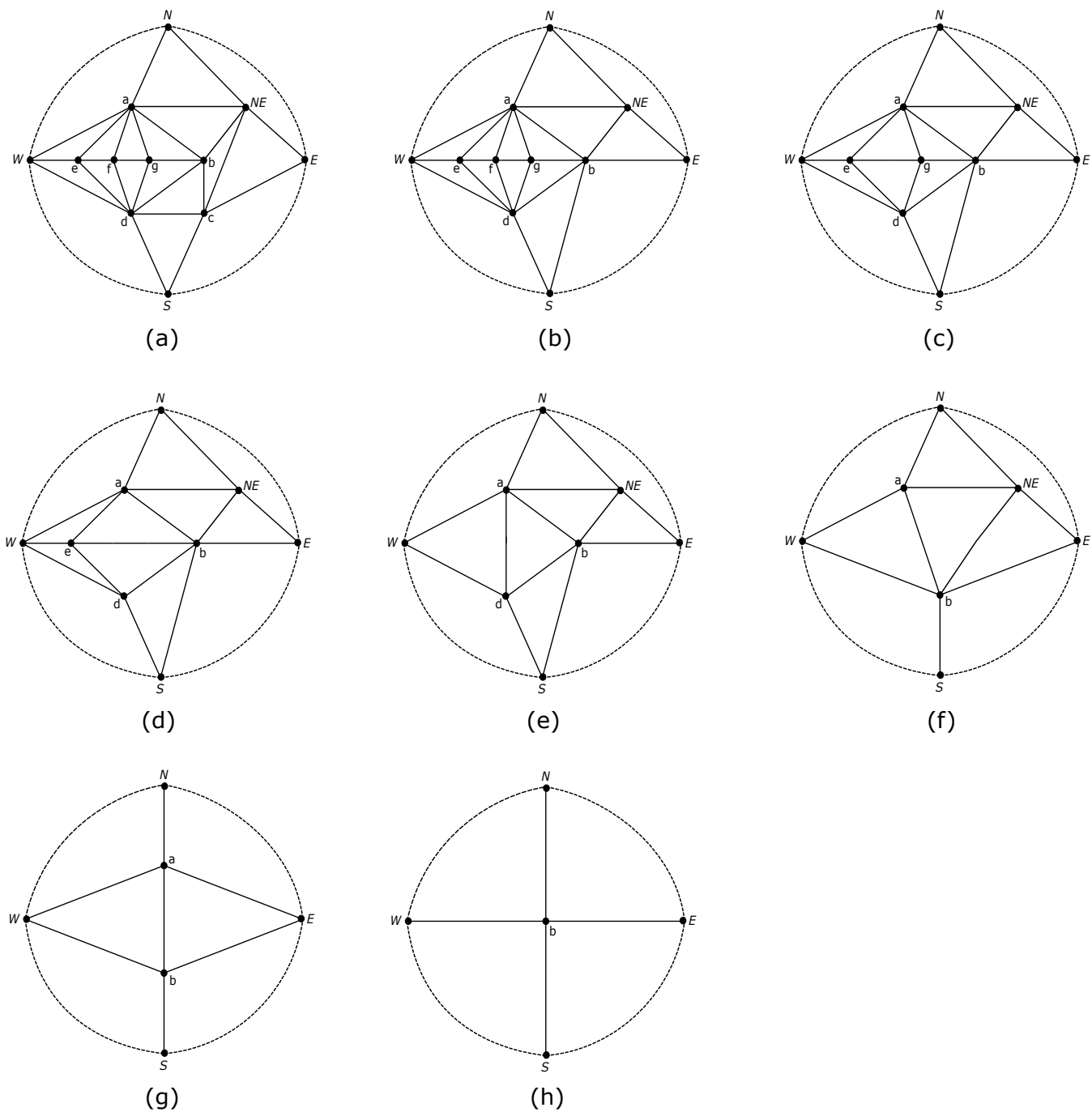}
\centering
\caption{ (a) Graph obtained after adding vertices $\textit{N}, \textit{E}, \textit{S}$ and $\textit{W}$, (b) Graph obtained after contracting edge $(b, c)$, (c) Graph obtained after contracting edge $(e, f)$, (d) Graph obtained after contracting edge $(e, g)$, (e) Graph obtained after contracting edge $(e, a)$, (f) Graph obtained after contracting edge $(b, d)$, (g) Graph obtained after contracting edge $(a,\textit{NE})$, (h) Graph obtained after contracting edge $(a, b)$}
\label{rel}

\end{figure}

\begin{figure}[]
\includegraphics[width=17.6cm,height=18cm]{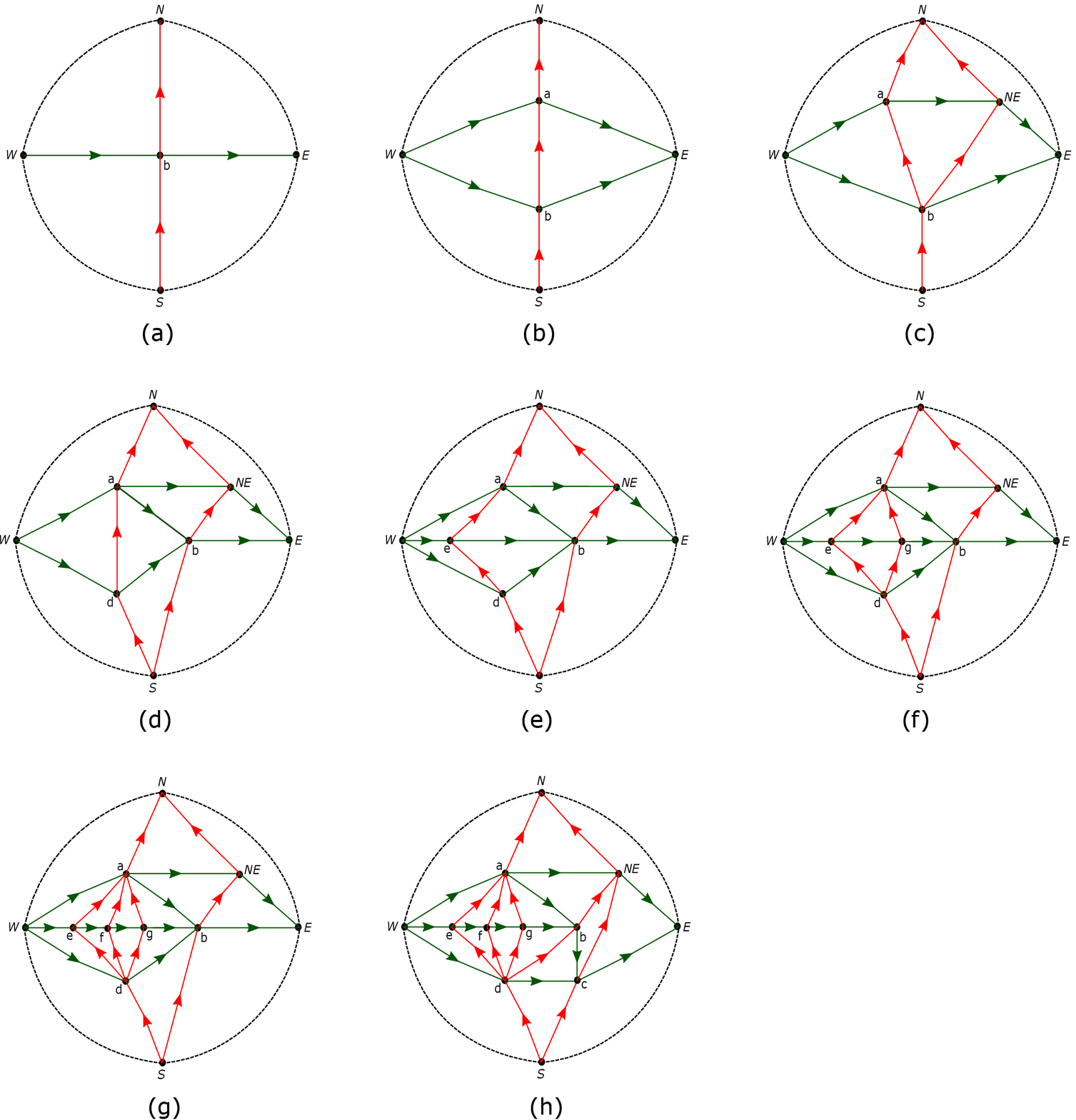}
\centering
\caption{(a) Trivial REL, (b) REL obtained after expanding edge $(a, b)$, (c) REL obtained after expanding edge $(a,\textit{NE})$, (d) REL obtained after expanding edge $(b, d)$, (e) REL obtained after expanding edge $(e, a)$, (f) REL obtained after expanding edge $(e, g)$, (g) REL obtained after expanding edge $(e, f)$, (h) REL obtained after expanding edge $(b, c)$}
\label{finalexp}
\end{figure}

\item The labels of the edges $(a, b)$ and $(b, c)$ are same (see Figure \ref{finalexp}h). We flip the edge $(a, b)$ and the modified REL with edges $(a, b)$ and $(b, c)$ belonging to different sets is shown in Figure \ref{nontriv}.
 
 \item From the modified REL, we have obtained a RFP correspond to the modified graph $G'$ (refer to Figure \ref{lshaped}a).

\item Remove the module \textit{NE} from the RFP.

\item A non-trivial $\mathbb{L}$ for the given PTPG $G$ is shown in Figure \ref{lshaped}b.
 
\begin{figure}[H]
\includegraphics[width=7cm,height=7cm]{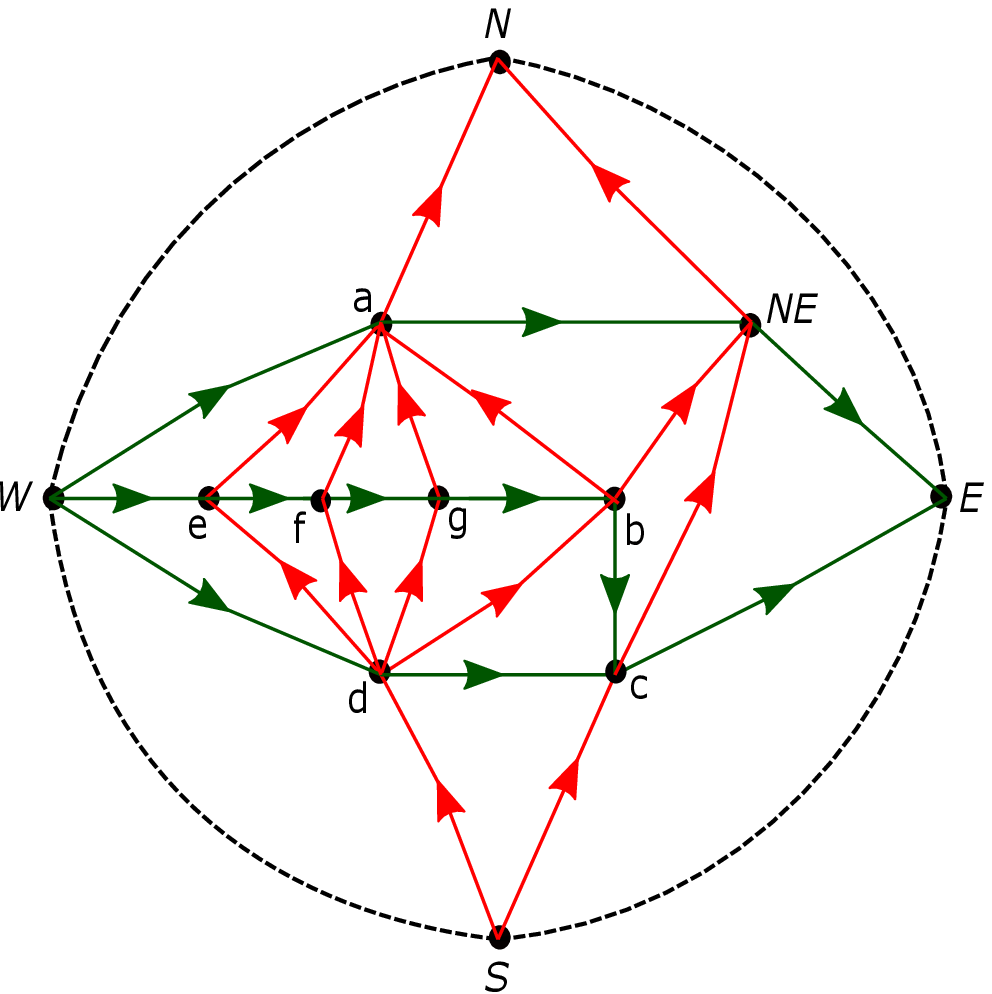}
\centering
\caption{REL obtained after flipping the edge $(a, b)$ }
\label{nontriv}
\end{figure}

\begin{figure}[]
 \centering
 \begin{subfigure}[b]{0.35\textwidth}
  \includegraphics[width=.83\textwidth]{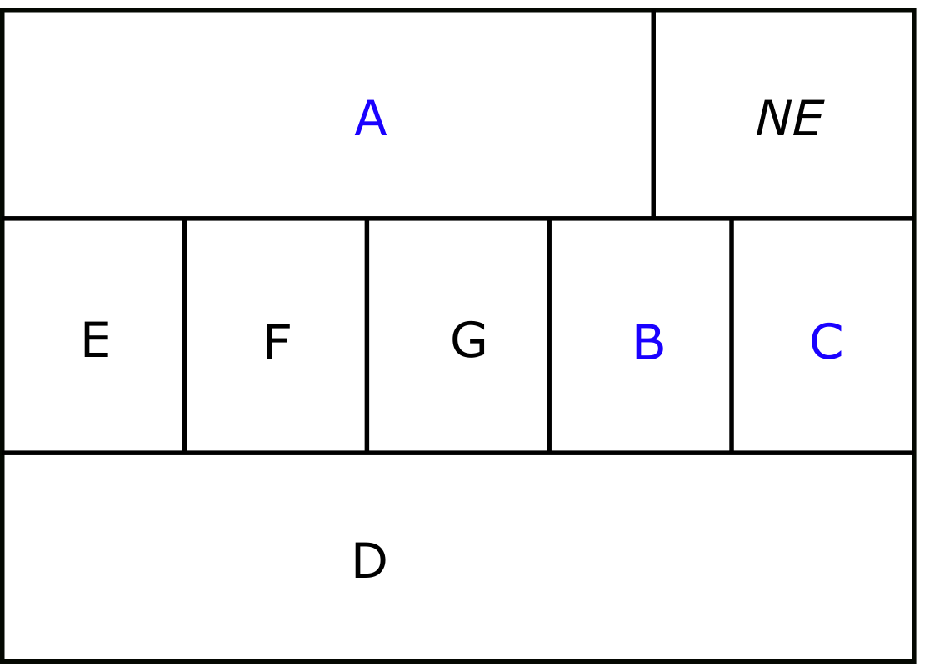}
  \caption{}
  \label{}
 \end{subfigure}
 \begin{subfigure}[b]{0.35\textwidth}
  \includegraphics[width=.83\textwidth]{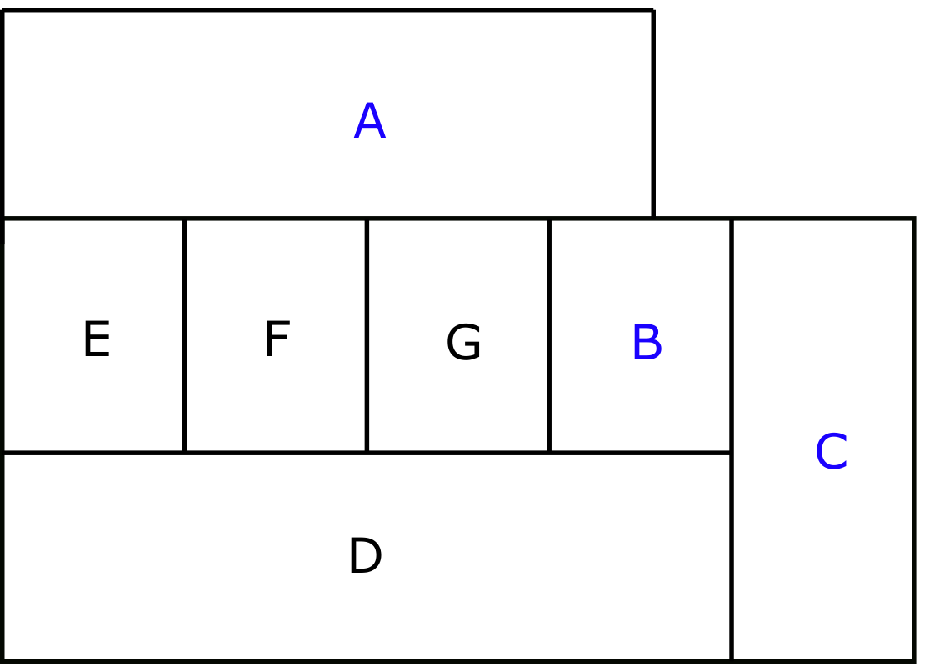}
  \caption{}
  \label{}
 \end{subfigure}
  \caption{(a) A RFP, (b) A non-trivial $\mathbb{L}$ for the PTPG $G$ obtained from (a) by removing module \textit{NE}}
  \label{lshaped}
 \end{figure}
\end{enumerate}

\subsection{Flipping Algorithm}
\label{2.5}

As it is known that for a given PTPG there can exist multiple RFPs, since there can have many possibilities to choose a set of four paths made up of outer vertices of the graph in the four-completion process \cite{kant1997regular}. Other than that corresponding to a particular set of paths, we can obtain multiple RFPs by flipping some of the edges or vertices in the REL.

\begin{figure}[H]
\includegraphics[width=13cm,height=10cm]{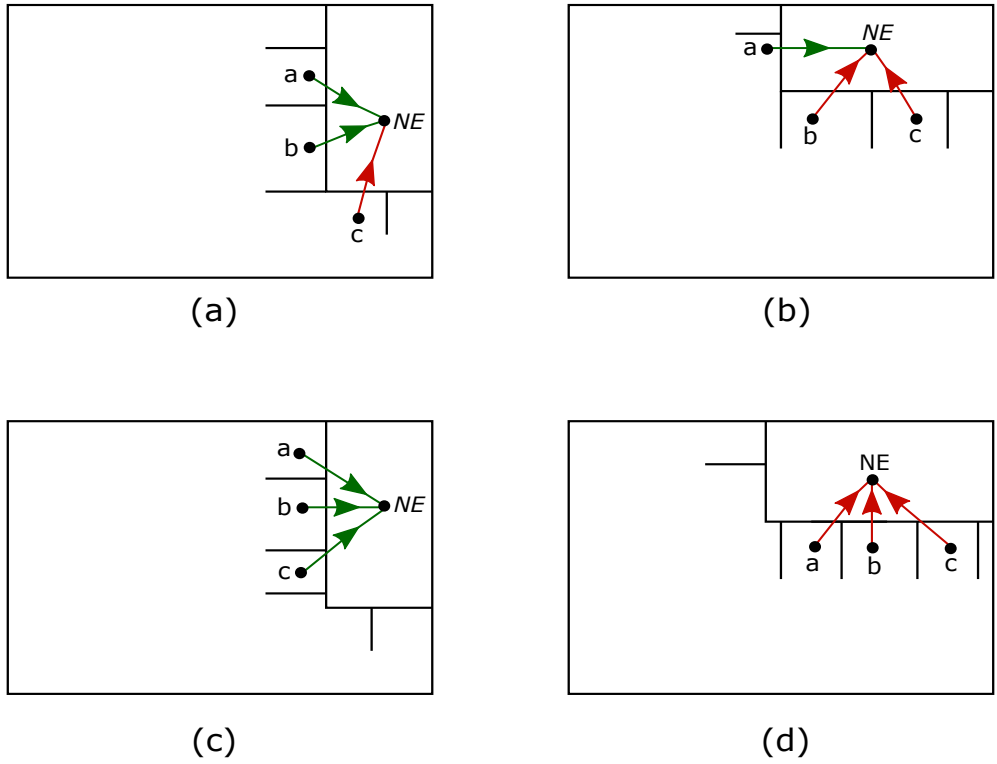}
\centering
\caption{Possible positions of the modules corresponding to triplet $(a, b, c)$}
\label{16}
\end{figure}

In our work, to obtain a non-trivial $\mathbb{L}$ for a PTPG $G$ with respect to triplet $(a, b, c)$, we obtain a REL for the modified graph $G'$ having an extra vertex $\textit{NE}$ and then get a RFP from the REL. By removing the corner module $\textit{NE}$, an $\mathbb{L}$ is obtained which may or may not be non-trivial. If it is trivial then in the REL, edges $(a,b)$ and $(b, c)$ belongs to the same set either $T_1$ or $T_2$. In such a case, we need to ensure that they must belong to different sets for a non-trivial $\mathbb{L}$. To transform a trivial $\mathbb{L}$ into a non-trivial $\mathbb{L}$, we use flipping algorithm with four cases (refer to Figure \ref{16}). In the first two cases (in Figures \ref{16}a and \ref{16}b), we will be using Algorithm \ref{EGDR} to flip edges $(b, c)$ and $(a, b)$ respectively.

{\it Remark.}
In case of Figure \ref{16}c and Figure \ref{16}d, when the modules corresponding to the triplet vertices are above concave corner or to the right of the concave corner, we need to flip edge $(c, \textit{NE})$ or $(a, \textit{NE})$ first, in order to flip edge $(b, c)$ or $(a, b)$, respectively.

\begin{figure}[]
 \centering
 \begin{subfigure}[b]{0.38\textwidth}
  \includegraphics[width=.78\textwidth]{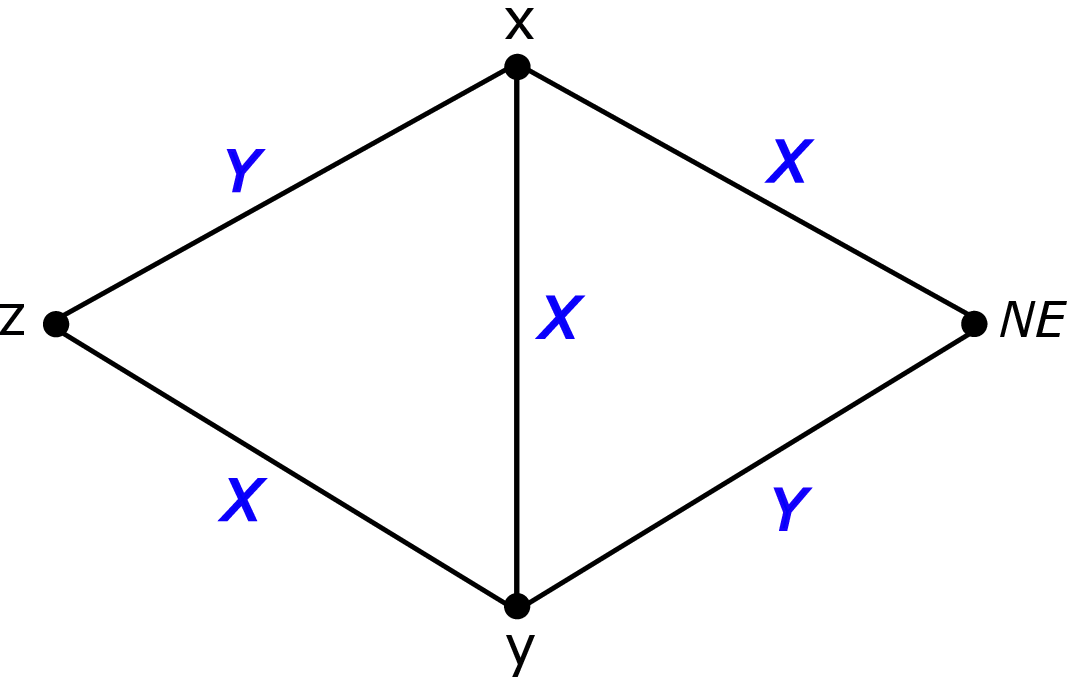}
  \caption{}
  \label{flipi1}
 \end{subfigure}
 \begin{subfigure}[b]{0.38\textwidth}
  \includegraphics[width=.78\textwidth]{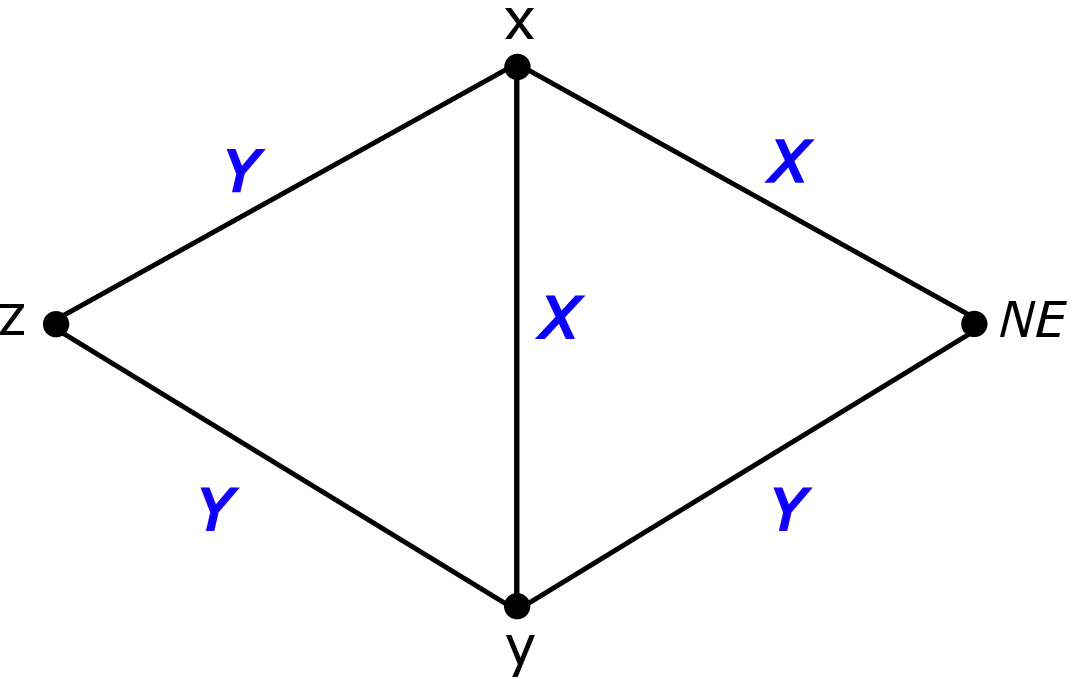}
  \caption{}
  \label{flipi2}
 \end{subfigure}
 \begin{subfigure}[b]{0.38\textwidth}
  \includegraphics[width=.78\textwidth]{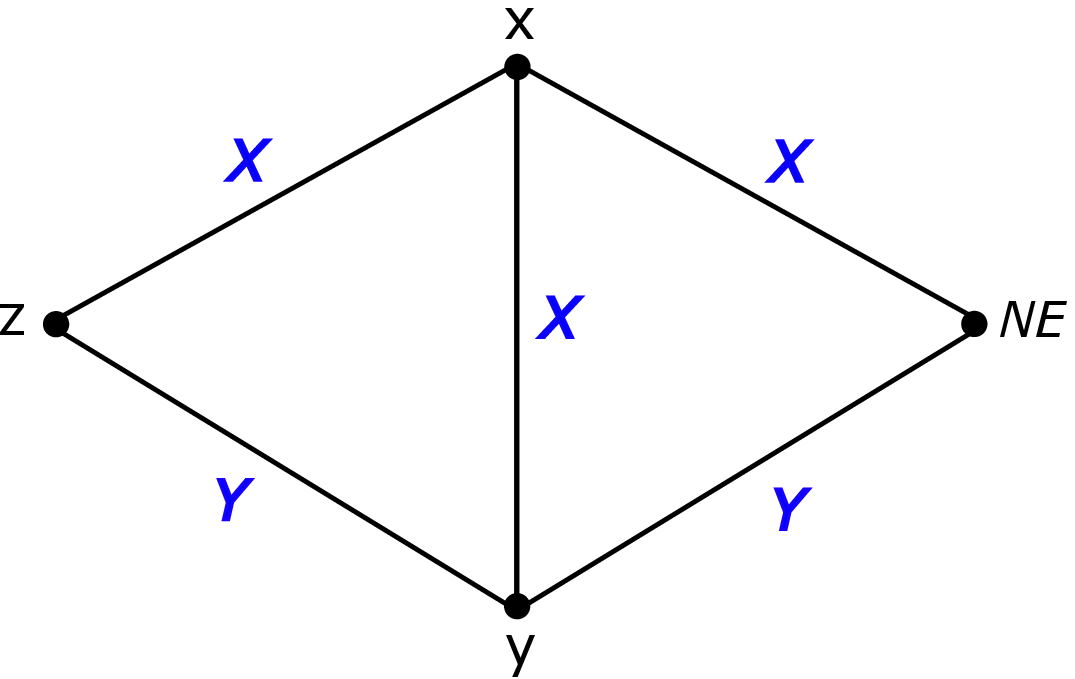}
  \caption{}
  \label{flipi3}
  \end{subfigure}
  \caption{Possible cases for the labels of edges $(x, z)$ and $(y, z)$, when edge $(x, y)$ is selected to flip}
  \label{FLIPI3}
 \end{figure} 
 
\begin{algorithm}[]
\DontPrintSemicolon
\KwInput{An REL with both the edges $(a, b)$ and $(b, c)$  belonging to same set ($T_1$ or $T_2$) and vertex \textit{NE}}
\KwOutput{An REL with edges $(a, b)$ and $(b, c)$ belonging to the different sets}
\caption{Flipping Algorithm }
\label{EGDR}
 A[ ]: An array list of the edges which are selected to be flipped
 
 C[ ]: An array list of the previous common neighbors of end vertices of the edges from $A$

 $A[i]$ = $(x_i, y_i)$ where $(x_i, y_i)$ is the $i$th selected edge in $A$ and $C[i]$ is the previous common neighbor of end vertices of the edge $(x_i, y_i)$
 
 Start $i$ = $0$
 
\If 
     {$(a, b)$ and $(b,c)$ are in $T_1$}  {$(x_0, y_0)$ = $(b, c)$ and $C[0] = \textit{NE}$}
\Else
      {$(x_0, y_0)$ = $(a, b)$ and $C[0] = \textit{NE}$}
      
Assume that edges $(x_i, y_i)$ and $(x_i, C[i])$ are in set $X$ and $(y_i, C[i])$ is in set $Y$, (where $(X, Y)$ can be $(T_1, T_2)$ or $(T_2, T_1)$) (refer to Figure \ref{FLIPI3})

Identify a new common neighbor of vertices $x_i$ and $y_i$ other than $C[i]$ (say $z[i]$).

\If {$(x_i, z[i])$ is in $Y$ and $(y[i], z[i])$ is in $X$}
 {edge $(x_i, y_i)$ is flippable, flip $A[i]$\
 
 \If {$i$ = $0$}
 {return exit}\

 \Else 
 {delete $A[i]$ \
 
 \If {deleted $A[i]$ is the only edge to be flipped to flip $A[i-1]$}
  {$i$ = $i-1$; go to step $12$}
  
 \Else {replace $(x_i, y_i)$ = $(y_i, z[i])$, $C[i]$ = $x_i$ and go to step $9$}
   }}
 
\ElseIf {$(x_i, z[i])$ and $(y_i, z[i])$ are in $Y$}
{$A[i+1]$ = $(y_i, z[i])$ and 
$C[i+1] = x_i$ \

\If {$A[i+1]$ = $A[j+1]$ for some $j < i$}
         {we identify four vertices forming an alternating four cycle in the REL.
         To identify these vertices, we first look for edges $A_k = (x_k, y_k), A_l = (x_l, y_l), A_m = (x_m, y_m), A_n = (x_n, y_n)$, which are in between the edges $A_{j+1}, A_{j+2},\dots,A_{i+1}$ in $A$
         and there requires to flip only one edge to flip each of them. Now, identify new common neighbors $z[k], z[l], z[m]$ and $z[n]$ (other than existing (previous) neighbors) of the end vertices of edges $A_k, A_l, A_m$ and $A_n$, respectively. The vertices $z[k], z[l], z[m]$ and $z[n]$ forms an alternating four cycle in the REL. For the REL, pick all the edges which lie inside the four-cycle $z[k], z[l], z[m]$, $z[n]$, and flip all these edges ($A[j+1]$ also lies inside the four cycle $z[k], z[l], z[m]$, $z[n]$), then delete the edges $A_{j+1}, A_{j+2},\dots,A_{i+1}$ from the array list.\
         
 \If {$A[j+1]$ is the only edge to be flipped to flip $A[j]$}
  {$i$ = $j$; go to step $12$ }
  \Else {replace $(x_{j+1}, y_{j+1}) = (y_{j}, z[j])$ and
  go to step $9$}
  }

     \Else 
     {go to step $9$}}

 \ElseIf 
     {$(x_i, z[i])$ is in $X$ and $(y_i, z[i])$ is in $Y$}
     {$A[i+1]$ = $(x_i, z[i])$ and $C[i+1]$ = $y_i$ \ 
     
     $i=i+1$; go to step $9$}

\end{algorithm}
 
\subsection{An illustration}
\label{illus2}

In Figure \ref{eg0}, a PTPG is shown, corresponding to which a non-trivial $\mathbb{L}$ is obtained in Figure \ref{nfp}b.
\begin{enumerate}
\item $(a,b,c)$ is chosen as a triplet.

\item The set of five paths is $\{(a, b, c), (c), (c, d), (d, e), (e, f, a)\}$.

\item After inserting a new vertex $\textit{NE}$ in PTPG $G$, a modified PTPG $G'$ is shown in Figure \ref{Eeg3}.
 
\item The four modified paths are
$(e, f, a, \textit{NE}), (\textit{NE}, c), (c, d), (d, e)$ and the graph with four new vertices $N, E, W, S$ is drawn in Figure \ref{eg3}.

 \item REL of graph $G'$ is obtained in Figure \ref{nontrivial}.

 \item In the REL in Figure \ref{nontrivial}, label of the edges $(a, b)$ and $(b, c)$ are same. We use Algorithm \ref{EGDR} to flip the edge $(b, c)$ (see Figure \ref{fe1}).
 
 \item From the new REL having differently labelled edges $(a, b)$ and $(b, c)$ (see Figure \ref{fe1}l), we obtain RFP (refer to Figure \ref{nfp}a).

\item Remove the module \textit{NE} from the RFP.

\item The output, a non-trivial $\mathbb{L}$ for the given PTPG $G$ is drawn in Figure \ref{nfp}b.

\begin{figure}[]
\begin{subfigure}[b]{0.30\textwidth}
\includegraphics[width=.78\textwidth]{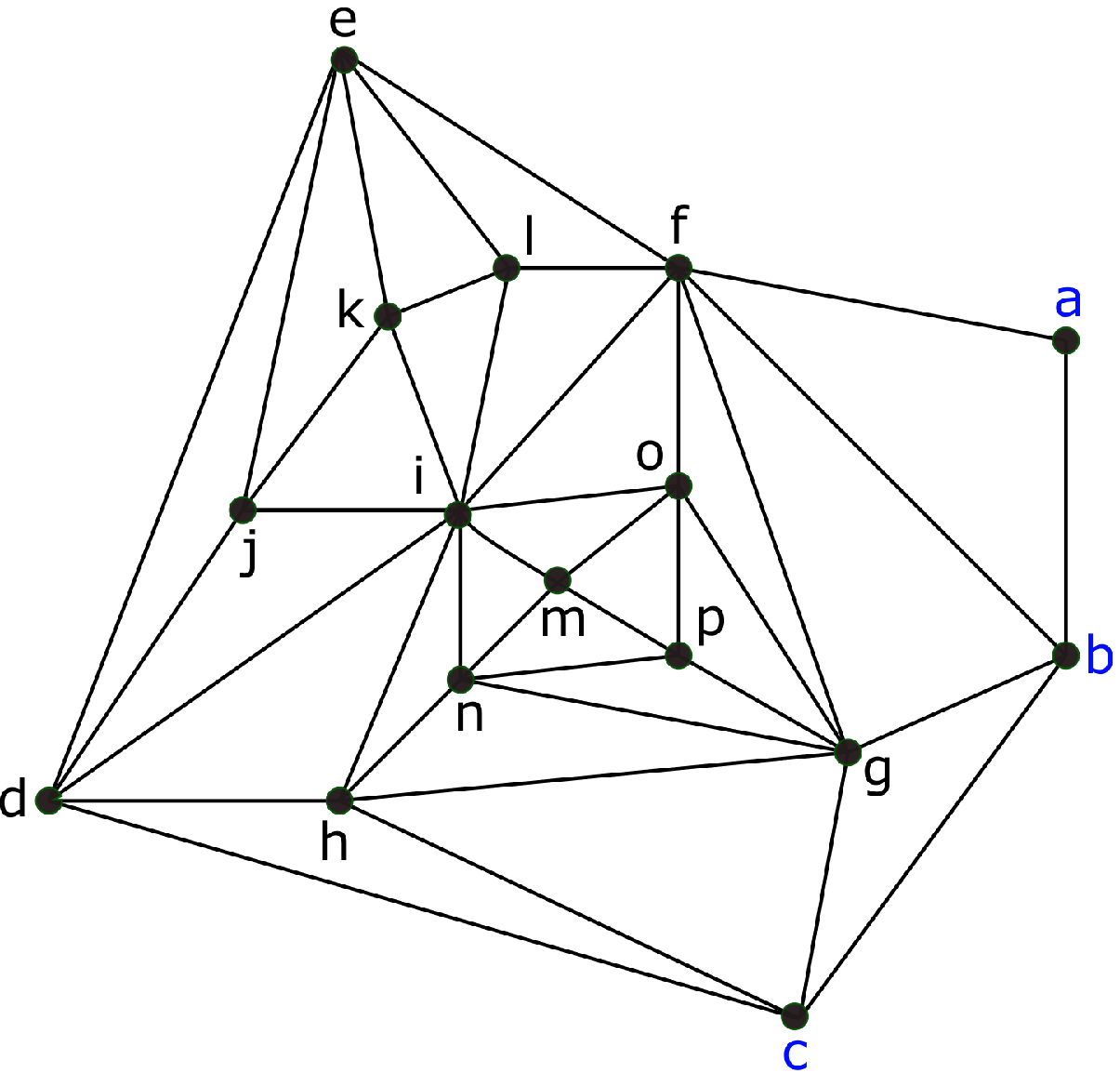}
\caption{
}
\label{eg0}
\end{subfigure}
\begin{subfigure}[b]{0.34\textwidth}
\includegraphics[width=.80\textwidth]{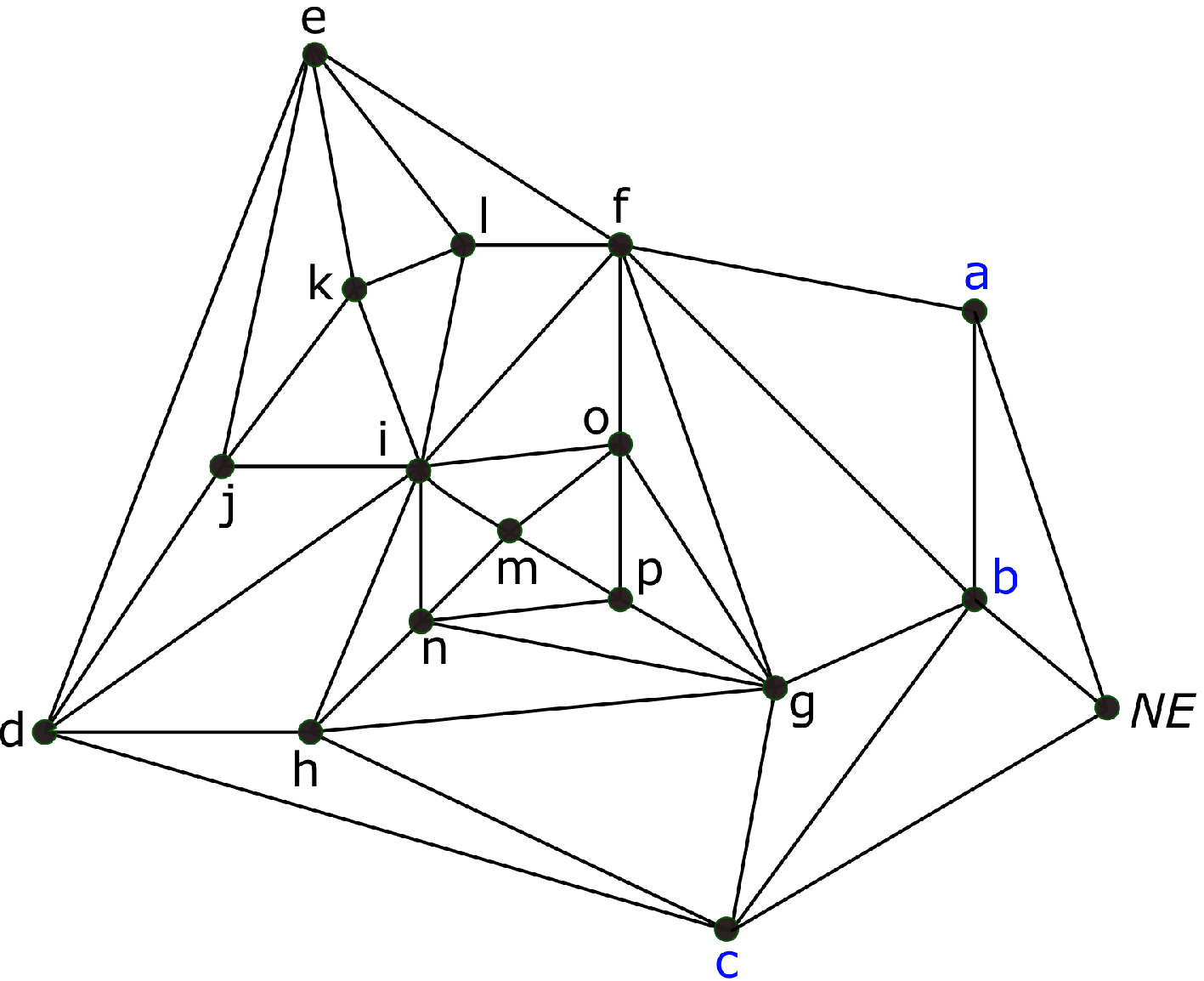}
\caption{
}
\label{Eeg3}
 \end{subfigure}
 \begin{subfigure}[b]{0.35\textwidth}
  \includegraphics[width=.82\textwidth]{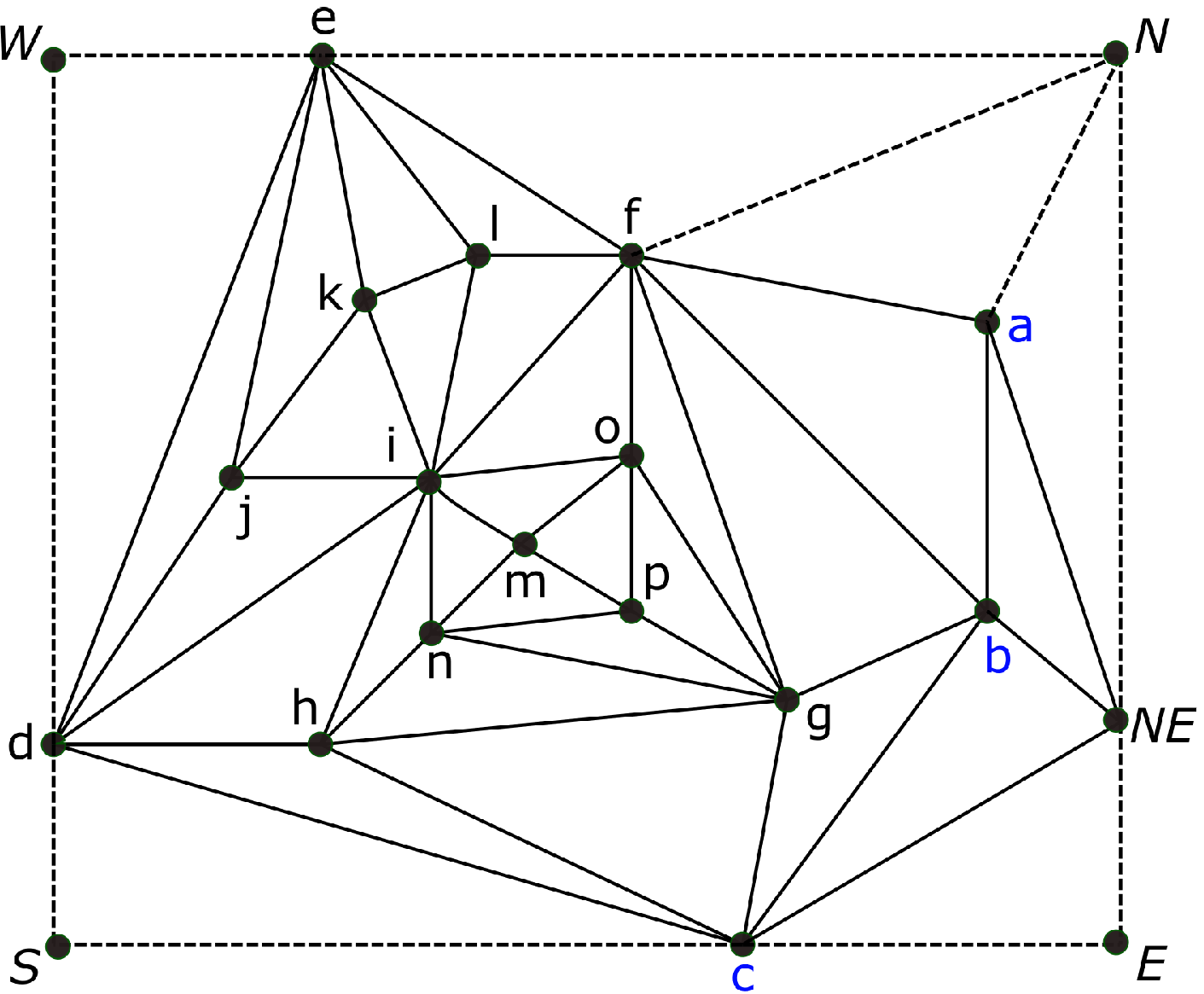}
  \caption{
  }
  \label{eg3}
  \end{subfigure}
  \caption{(a) A PTPG $G$, (b) Modified PTPG $G'$ with an extra vertex \textit{NE}, (c) Graph obtained after adding vertices $N, E, S$ and $W$}
 \end{figure} 
 
\begin{figure}[]
\includegraphics[width=7.5cm,height=6.55cm]{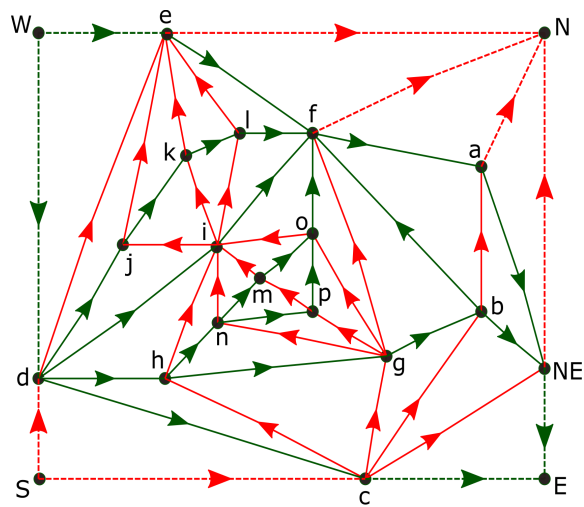}
\centering
\caption{REL of PTPG \textit{G'}}
\label{nontrivial}
\end{figure}

\begin{figure}[H]
\includegraphics[width=16cm,height=21cm]{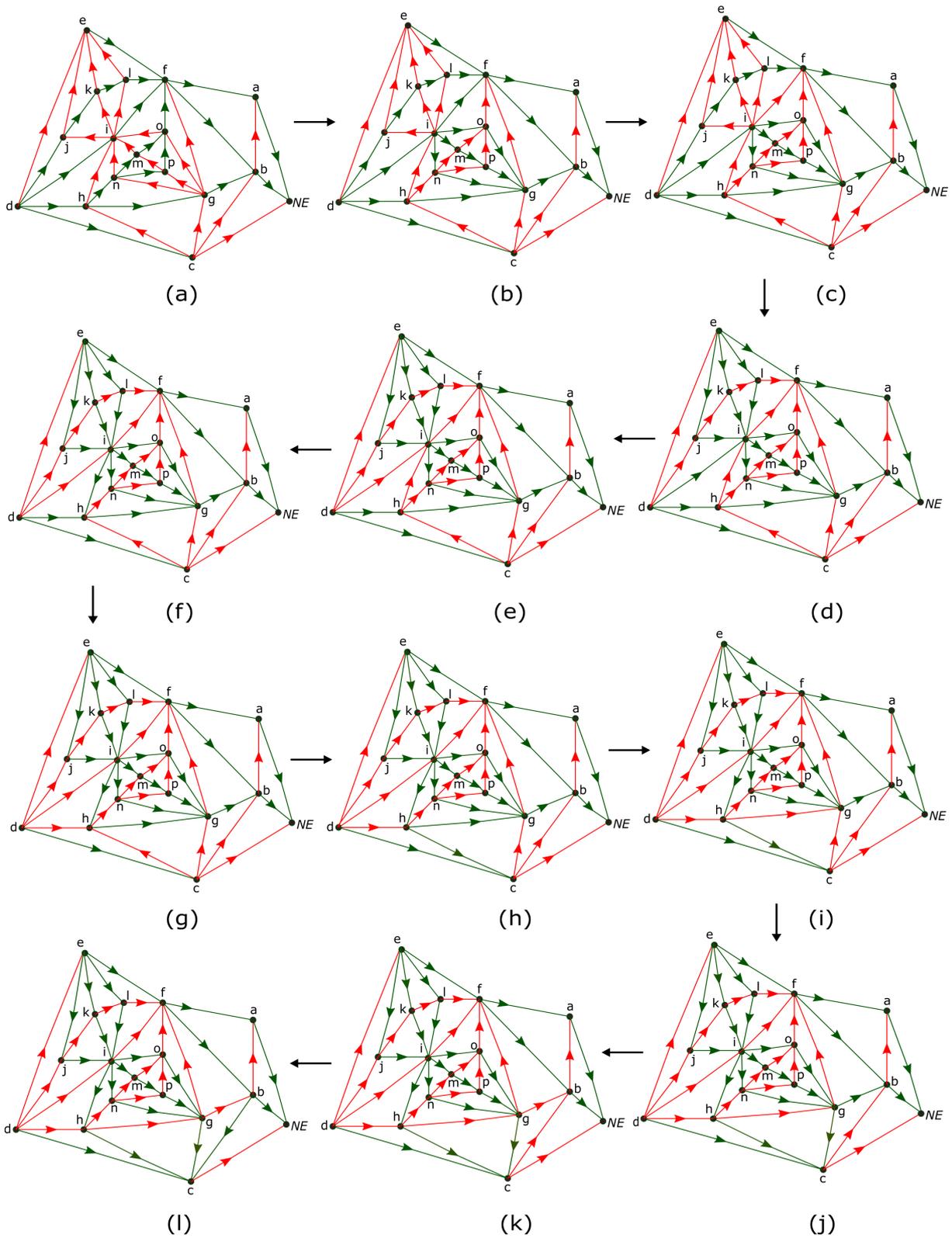}
\centering
\caption{(a) REL with an alternating four-cycle $g, h, i, f$ (at $19th$ level in the binary tree shown in Figure \ref{fm1}), (b - j) RELs obtained after flipping all the edges inside cycle $g, h, i, f$,  flipping edge $(f, i)$, 
flipping all the edges inside cycle $e, f, i, d$, flipping edge $(d, i)$, 
flipping edge $(i, h)$, 
flipping edge $(h, d)$, 
flipping edge $(c, h)$, 
flipping edge $(h, g)$, 
flipping edge $(c, g)$, 
flipping edge $(g, b)$,  
flipping edge $(b, c)$, respectively 
}
\label{fe1}
\end{figure}

\begin{figure}[]
 \centering
 \begin{subfigure}[b]{0.35\textwidth}
  \includegraphics[width=.83\textwidth]{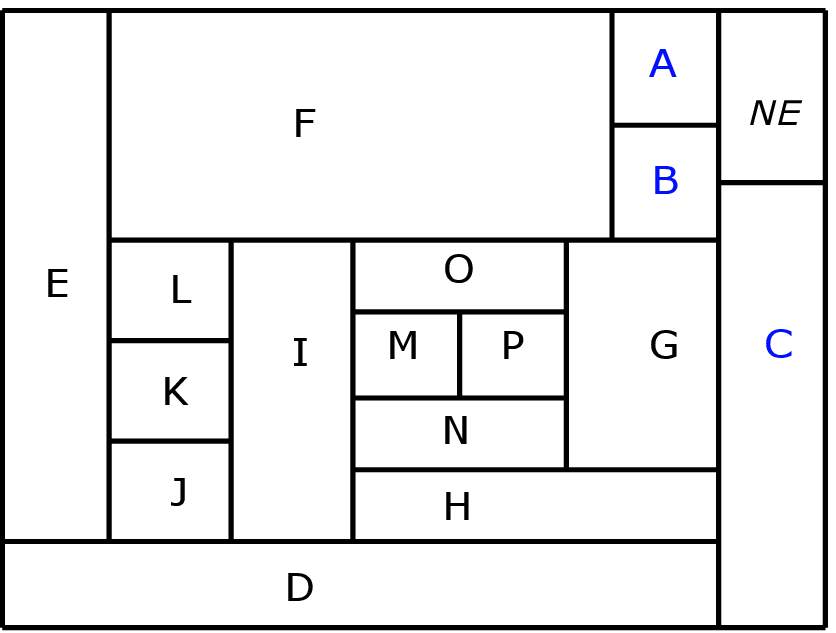}
  \caption{}
  \label{}
 \end{subfigure}
 \begin{subfigure}[b]{0.35\textwidth}
  \includegraphics[width=.83\textwidth]{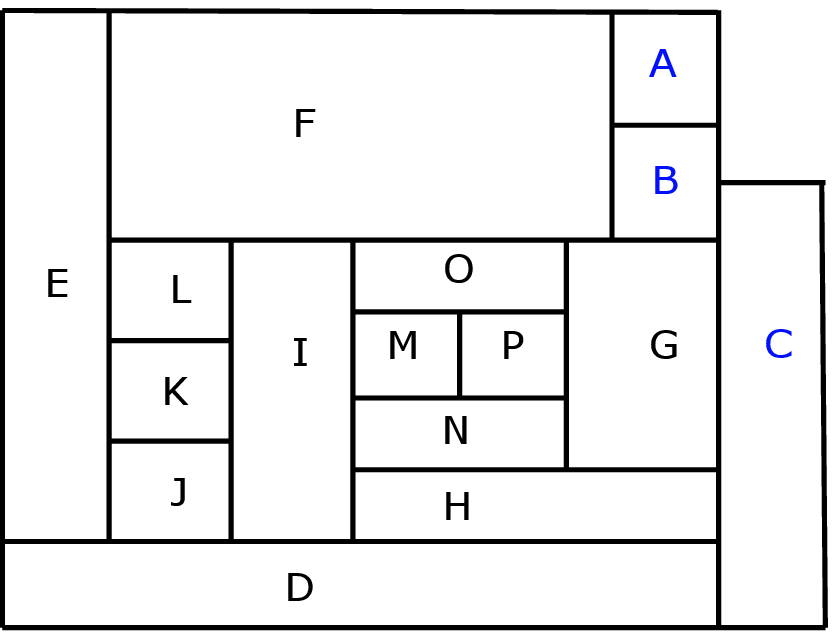}
  \caption{}
  \label{}
 \end{subfigure}
  \caption{(a) RFP, (b) Non-trivial $\mathbb{L}$ for the PTPG $G$ obtained from (a) by removing module \textit{NE} }
  \label{nfp}
 \end{figure}

\end{enumerate}

\subsection{Correctness of Algorithm \ref{EGDR}}
\label{correctness0}
\begin{figure}[]
 \centering
 \begin{subfigure}[b]{0.30\textwidth}
  \includegraphics[width=.68\textwidth]{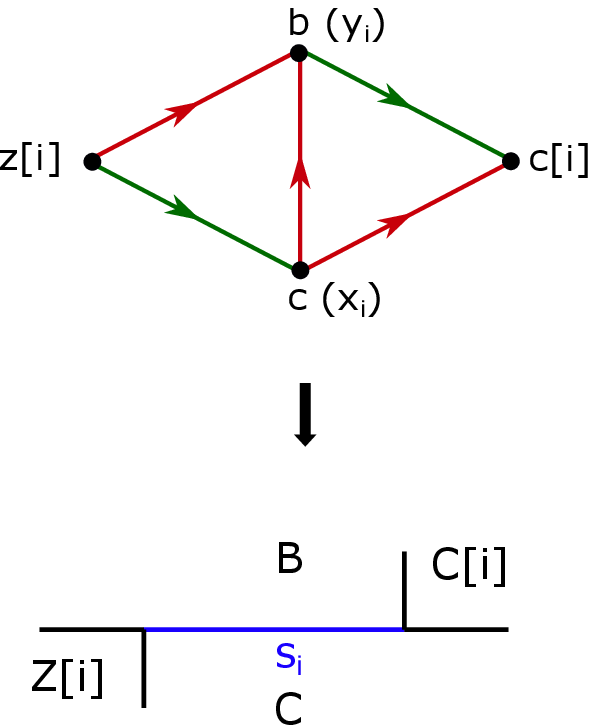}
  \caption{}
  \label{}
 \end{subfigure}
 \begin{subfigure}[b]{0.30\textwidth}
  \includegraphics[width=.68\textwidth]{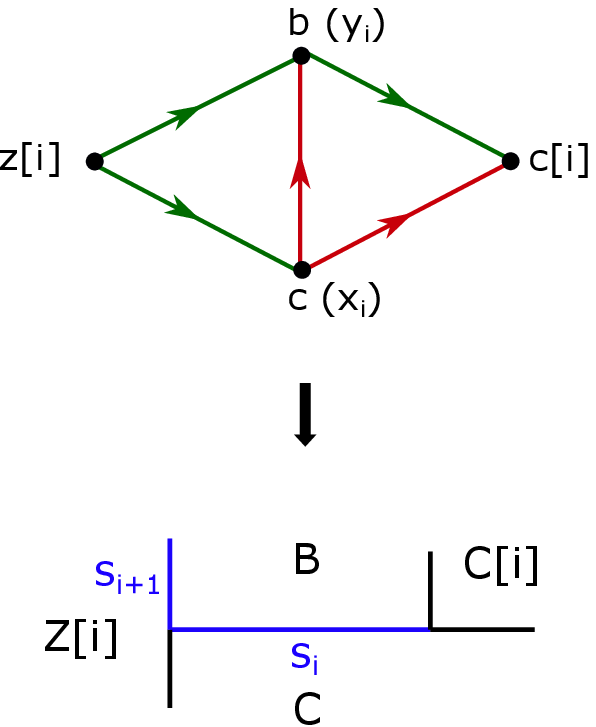}
  \caption{}
  \label{}
 \end{subfigure}
 \begin{subfigure}[b]{0.30\textwidth}
  \includegraphics[width=.68\textwidth]{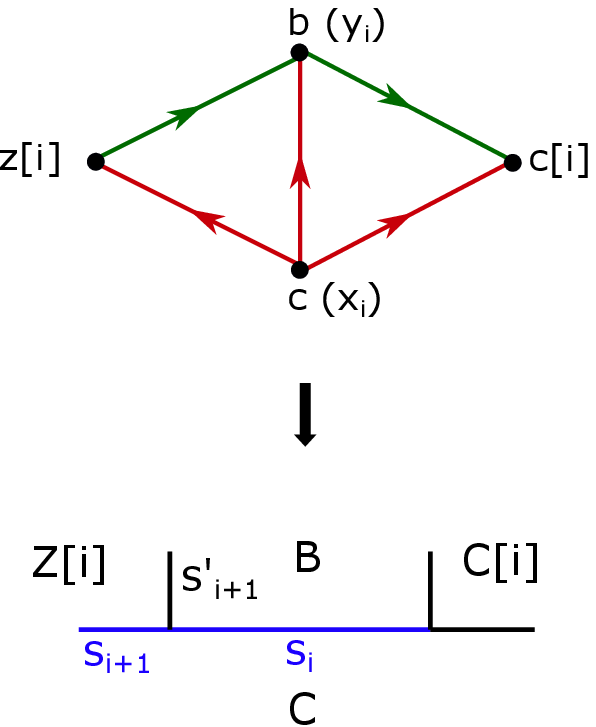}
  \caption{}
  \label{}
 \end{subfigure}
 \begin{subfigure}[b]{0.30\textwidth}
  \includegraphics[width=.68\textwidth]{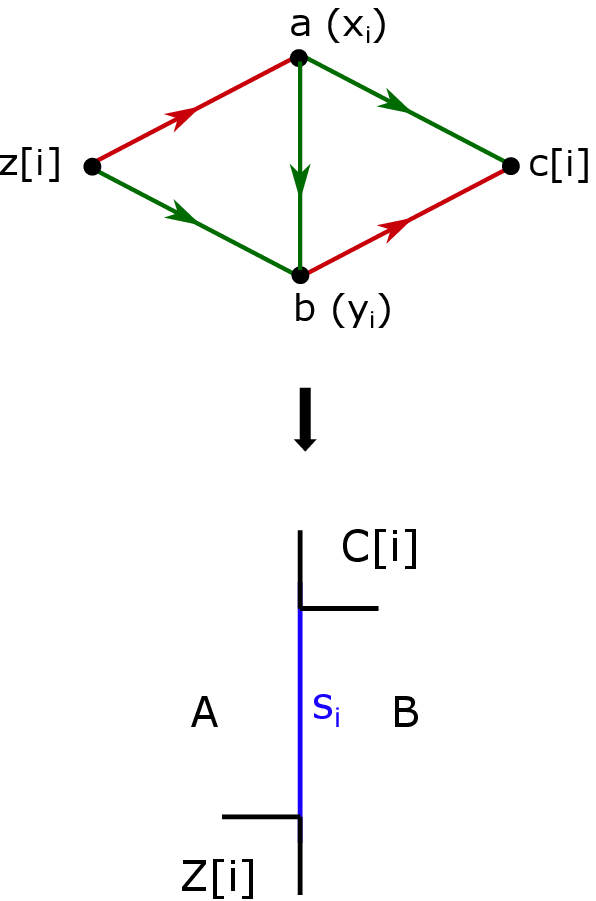}
  \caption{}
  \label{}
 \end{subfigure}
 \begin{subfigure}[b]{0.30\textwidth}
  \includegraphics[width=.68\textwidth]{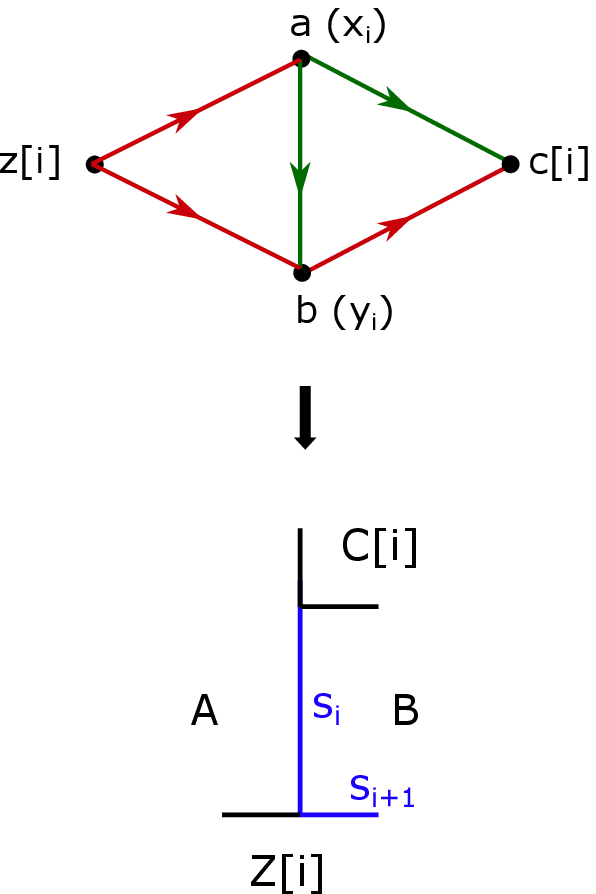}
  \caption{}
  \label{}
 \end{subfigure}
 \begin{subfigure}[b]{0.30\textwidth}
  \includegraphics[width=.68\textwidth]{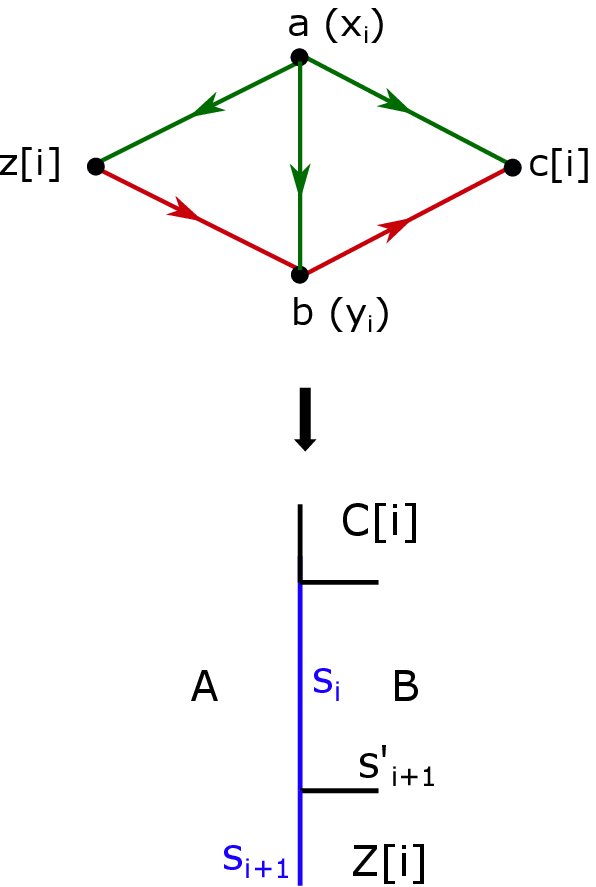}
  \caption{}
  \label{}
 \end{subfigure}
 \caption{Common line segments in the floor-plan corresponding to the different types of edges in the REL, where $A$, $B$, $C$, $Z[i]$ and $C[i]$ are the modules in the floor-plan corresponding to the vertices $a, b, c, z[i]$ and $c[i]$ of the REL (PTPG), respectively.  
 }
\label{lineseg}
\end{figure}

In Algorithm \ref{EGDR}, the input is a REL which is obtained from Algorithm \ref{EGDR1} corresponding to a PTPG $G'$, where $G'$ is a PTPG with an extra vertex $\textit{NE}$, adjacent to all the vertices of $P_1$ in PTPG $G$ and $(a, b, c)$ is a triplet. In the REL, edges $(a, b)$ and $(b, c)$ belong to the same set (either both are in $T_1$ or in $T_2$). For the correctness of Algorithm \ref{EGDR}, we have to prove that in the REL (obtained from Algorithm \ref{EGDR1}), one of the edges among $(a, b)$ or $(b, c)$ can always be flipped, and a new REL can be obtained in which edges $(a, b)$ and $(b, c)$ belong to  different sets.
To prove it, we claim the following results:
\begin{enumerate}[i.]
\item \textbf{For any interior edge \textbf{$(x_i, y_i)$} of a REL, there exist exactly two common neighbors of its end vertices $x_i$ and $y_i$.}
    
Each edge of a PTPG $G'$ is an interior edge in its REL because of the four-completion process. Therefore, for any edge of $G'$, there will be at least two common neighbors of its end vertices, but more than two common neighbors of the end vertices of any edge would create a separating triangle in the PTPG. As a result, for any edge there exist two common neighbours of its end vertices in the REL. Moreover, for any selected edge $A_j=(x_j, y_j)$ from the array list, a common neighbor of its end vertices ($x_j$ and $y_j$) at which we traverse first is called previous common neighbor $c[j]$. The another common neighbor of vertices $x_j$ and $y_j$ is a new common neighbor for edge $(x_j, y_j)$, denoted as $z[j]$.

\item \textbf{If both the edges $(a, b)$ and $(b, c)$ of the REL belong to set $T_1$, then $(b, c)$ must be flipped, otherwise $(a, b)$ must be flipped.}

Since we know if the concave corner is assumed to be in $\textit{NE}$ direction, then in the REL the labels of the edges $(a, b)$ and $(b, c)$ must be $T_1$ and $T_2$, respectively, in order to obtain a non-trivial $\mathbb{L}$. Hence, if both the labels of the edges $(a, b)$ and $(b, c)$ are in $T_1$, then we need to flip edge $(b, c)$ and if both the labels of the edges are in $T_2$ then we need to flip edge $(a, b)$.
In both the cases, we start from vertex $\textit{NE}$ as a previous common neighbor of the end vertices of edges $(a, b)$ and $(b, c)$. 

\item \textbf{For each selected edge $A_i=(x_i, y_i)$ in the array list, the previous common neighbor $c[i]$ of its end vertices will be adjacent to vertices $x_i$ and $y_i$ with edges belonging to different sets.}

We will prove this result using induction. For the initialization of the algorithm the base case will be obtained at $i=0$. The previous common neighbor of the end vertices of the first selected edge $(x_0, y_0)$ ($(a, b)$ or $(b, c)$), is taken as $c[0]=\textit{NE}$. Therefore, the vertex $\textit{NE}$ is adjacent to both of the vertices $x_0$ and $y_0$. Now, if edge $(x_0, y_0)$ belongs to set $X$ ($X$ can be $T_1$ or $T_2$), then in triangle $(x_0, \textit{NE}, y_0)$, there are two edges incident at vertex $\textit{NE}$, among which only one of the edges will belong to the same set as $(x_0, y_0)$ does (see Figure \ref{FLIPI3}), and that edge will be called $(x_0, \textit{NE})$. Another edge which belongs to different set (say $Y$) will be called $(y_0, \textit{NE})$. Hence, the base case is proved.
If $(x_0, y_0)$ is $(a, b)$, then $x_0 = a$ and $y_0 = b$ and if $(x_0, y_0)$ is $(b, c)$, then $x_0 = c$ and $y_0 = b$ (see Figure \ref{lineseg}).

Now for the inductive step, we assume that for a selected edge $A_j=(x_j, y_j)$ in the array list, the previous common neighbor $c[j]$ of its end vertices is adjacent to vertices $x_j$ and $y_j$ with edges belonging to different sets (edge $(x_j, c[j])$ belongs to set $X$ and edge $(y_j, c[j])$ belongs to set $Y$) and a new common neighbor of vertices $x_j$ and $y_j$ is denoted as $z[j]$. As we know that an edge can be flipped if it is surrounded by an alternating four-cycle of labelled edges in a REL. For edge $(x_j, y_j)$, the surrounding four-cycle is made up of four consecutive edges $(x_j, c[j])$, $(c[j], y_j)$, $(y_j, z[j])$, $(z[j], x_j)$. According to the label of the edges $(x_j, z[j])$, $(y_j, z[j])$, three cases are possible which are shown in Figure \ref{FLIPI3}:

\begin{enumerate}[(a)]

\item when edge $(x_j, z[j])$ belongs to set $Y$ and $(y_j, z[j])$ belongs to set $X$, then the four-cycle around edge $(x_j, y_j)$ is alternately labelled in the REL (refer to Figure \ref{flipi1}). Hence, edge $(x_j, y_j)$ is flippable and it is not required to select edge $A_{j+1}$ to flip it.

\item when both the edges $(x_j, z[j])$ and $(y_j, z[j])$ belong to set $Y$ (refer to Figure \ref{flipi2}), then we need to flip edge $(y_j, z[j])$ in order to flip edge $(x_j, y_j)$. Therefore, $(y_j, z[j])$ (say $(x_{j+1}, y_{j+1})$) is the next selected edge to flip and the previous common neighbor of the end vertices of edge $(x_{j+1}, y_{j+1})$ is $c[j+1]=x_j$. It can be observed that in triangle $(x_{j+1}, c[j+1], y_{j+1})$, edge $(x_{j+1}, c[j+1])$ belongs to the same set ($Y$) as $(x_{j+1}, y_{j+1})$ does and edge $(y_{j+1}, x_j)$ belongs to different set ($X$).

\item when edge $(x_j, z[j])$ belongs to set $X$ and $(y_j, z[j])$ belongs to set $Y$. Then, in order to obtain an alternating four-cycle around edge $(x_j, y_j)$, it needs to flip both the edges $(x_j, z[j])$ and $(y_j, z[j])$. To decide whether which edge to be flipped first, we select the edge which belong to the same set as edge $(x_j, y_j)$ belongs (since if we flip the edge first which belong to different set, then it would create a triangle with same labelled edges in the REL which is not possible). Hence, the next selected edge is $(x_j, z[j])$ (say $(x_{j+1}, y_{j+1})$) (refer to Figure \ref{flipi3}) and the previous common neighbor of its end vertices will be $c[j+1]=y_j$ which is adjacent to vertices $x_{j+1}$ and $y_{j+1}$ with differently labelled edges (edge $(x_{j+1}, c[j+1])$ belongs to set $X$ and edge $(y_{j+1}, c[j+1])$ belongs to set $Y$. 
\end{enumerate}
Hence, it can be seen that for each selected edge $(x_i, y_i)$, the previous common neighbor $c[i]$ of its end vertices is always adjacent to vertices $x_i$ and $y_i$ with differently labelled edges.
More precisely, in triangle $(x_i, c[i], y_i)$, two edges $(x_i, y_i)$ and $(x_i, c[i])$ belong to same set and edge $(y_i, c[i])$ belongs to a different set, for any selected edge $(x_i, y_i)$ and previous common neighbor $c[i]$ of its end vertices.

\item \textbf{The edges which are required to flip an edge $(a, b)$ (or $(b, c)$) forms a binary tree $B$, where the edges represent the vertices of $B$ and the root vertex corresponds to edge $(a, b)$ (or $(b, c)$).}

The array list is an ordered collection of the edges which are selected to flip. Whenever an edge is selected to flip from the REL, we insert it in the array list and the first edge in the array list will be $A_0$ = $(a, b)$ or $(b, c)$. Moreover, if a selected edge $A_i$ is flippable or becomes flippable as a consequence of flipping other edges in the REL then we delete $A_i$ from the array list.
If we represent an edge $A_i= (x_i, y_i)$ of the REL (which is required to flip) by a vertex of $B$ and the edges from the REL which are required to flip, to flip $A_i$, by its children, then a binary tree $B$ can be obtained. An edge in $B$ between two of its vertices shows that to flip an edge of the REL corresponding to the parent vertex, it is required to flip the edge which is corresponding to the child vertex.
As we have seen that to flip an edge $A_i= (x_i, y_i)$, it needs to flip at most two edges at a time in the REL. Hence, there can have at most two children of any vertex in $B$.
 To construct $B$, we start from edge $A_0$ ($(b, c)$ or $(a, b)$) by assuming it corresponds to the root vertex of $B$ and identify its children. To flip a selected edge $A_i$, if it is required to flip one edge only (say $A_{i+1}$), then in the array list, the next selected edge will be $A_{i+1}$ and the vertex $v_i$ corresponding to edge $A_{i}$ will have one child in $B$. Whereas, to flip a selected edge $A_i$, if it is required to flip two edges $A_{i+1}$ and $A'_{i+1}$ ($A_i$ and $A_{i+1}$ belong to same set (say $X$) and $A'_{i+1}$ belongs to a different set ($Y$) in the REL) then in $B$, vertex $v_i$ corresponding to edge $A_i$ will have two children namely, $v_{i+1}$ and $v'_{i+1}$ corresponding to edges $A_{i+1}$ and $A'_{i+1}$, respectively. Moreover in $B$, we assume a vertex as a left child if its corresponding edge in the REL is the next edge which is selected to flip after the edge corresponding to its parent. Therefore, vertex $v_{i+1}$ will be the left child of $v_i$ in $B$ and edge $A_{i+1}$ will be inserted as the next edge in the array list. In this way, we continue to select edges until we identify a leaf vertex in $B$.   
 We identify a vertex $v_i$ as a leaf in $B$, if for the corresponding edge $A_{i}$, the following conditions occur in the REL :
 
\begin{enumerate}[(a)]

\item Edge $A_{i}$ is flippable. 

\item Edge $A_i=(x_i, y_i)$ occurs for the second time in the array list (a vertex occurs for the second time in $B$), i.e., $A_i = A_j$ for some $j < i$.
    
\end{enumerate}

If $A_{i}$ is flippable then the corresponding vertex $v_i$ will be identified as a leaf in $B$, since to flip $A_{i}$, it is not required to flip any edge.
After identifying a leaf vertex $v_i$ in $B$ corresponding to an edge $A_i$ which is flippable, we flip $A_{i}$ in the REL and delete it from the array list while marking $v_i$ as visited in $B$. Then, for the parent vertex $v_{i-1}$ of $v_i$, we check whether $v_{i-1}$ has another child in $B$. If $v_i$ is the only child of $v_{i-1}$, then the edge $A_{i-1}$ in the REL corresponding to vertex $v_{i-1}$ of $B$, will also be flippable (since to flip edge $A_{i-1}$, it is required to flip one edge $A_{i}$, which is flipped). Again, by flipping edge $A_{i-1}$ in the REL, we delete it from the array list while marking vertex $v_{i-1}$ as visited in $B$. Whereas, if $v_i$ is not the only child of $v_{i-1}$, then we traverse at the another child of $v_{i-1}$ and select the corresponding edge from the REL as a next edge in the array list. Hence, when both the children of a vertex $v_i$ are marked as visited, we mark the vertex itself as visited in $B$. Moreover, whenever a vertex $v_i$ is marked as visited in $B$, the corresponding edge $A_i$ in the REL is flipped and deleted from the array list. However, if we traverse at a vertex $v_i$ in $B$ to identify its children, then the corresponding edge $A_i$ from the REL is inserted in the array list. 

If an edge $A_i=(x_i, y_i)$ occurs for the second time in the array list (a vertex occurs for the second time in $B$), i.e., $A_i$ = $A_j$ for some $j < i$ then we will prove that the vertex $v_i$ corresponding to edge $A_i$ will also be a leaf. 
Since, whenever there is an edge $A_i$ in the REL which is flippable, there is a four-cycle of alternately labelled edges around $A_i$. Then in the floor-plan, the common lines segments corresponding to the edge $A_i$ and the edges of surrounding four-cycle are depicted in Figures \ref{lineseg}a and \ref{lineseg}d. Whereas, if edge $A_i$ is not flippable and it is required to flip one edge to flip $A_i$, then the surrounding edges in the REL and the  common line segments in the corresponding floor-plan are shown in Figures \ref{lineseg}b and \ref{lineseg}e. In Figures \ref{lineseg}b and \ref{lineseg}e, the common line segments $s_i$ and $s_{i+1}$ correspond to the edges $A_i$ and $(y_i, z[i])$ (which is required to flip, to flip edge $A_i$) in the REL. Moreover, the common line segment $s_{i+1}$ is perpendicular to $s_i$, and the common line segment corresponding to edge $(z[i], x_i)$ (which is incident at the new common neighbor of the end vertices of edge $A_i$) is also perpendicular to $s_i$. Now, if it is required to flip two edges to flip $A_i$, then the four-cycle around edge $A_i$ in the REL and the corresponding line segments in the floor-plan are shown in Figures \ref{lineseg}c and \ref{lineseg}f. In Figures \ref{lineseg}c and \ref{lineseg}f, the common line segments $s_i$, $s_{i+1}$ and $s'_{i+1}$ are corresponding to the edge $A_i$, $(z_i, x_i)$ (which is selected first to flip) and $(z_i, y_i)$ (which is selected to flip after flipping edge $(z_i, x_i)$), respectively. Furthermore, the common line segment $s_{i+1}$ is aligned with the common line segment $s_i$ and $s'_{i+1}$ is perpendicular to $s_i$.

As we know, for a REL of the modified graph $G'$ (obtained after inserting a new vertex $\textit{NE}$ in PTPG $G$) with edges $(a, b)$ and $(b, c)$ belonging to the same set, a RFP can be obtained. After removing module $\textit{NE}$ from it, a trivial $\mathbb{L}$ can be obtained. In this trivial $\mathbb{L}$, corresponding to each common line segment between two modules, there exist an edge in the REL between the corresponding two vertices of the PTPG. Now, if we highlight the common line segments in trivial $\mathbb{L}$ corresponding to the edges from the REL which are inserted in the array list, in the same order in which they were inserted, then we obtain rectangular spiral structures (highlighted with blue lines) shown in Figure \ref{spirall}. These spiral structures are continuously changing as the edges are inserted or deleted in the array list. Whenever an edge is inserted or deleted in the array list, the corresponding common line segments are highlighted or removed from being highlighted, respectively, in the trivial $\mathbb{L}$. For a particular REL, these spiral structures move in clockwise or anti-clockwise direction depending on the first selected edge as $(b, c)$ or $(a, b)$, respectively (see Figures \ref{rsa} and \ref{rsb}). Whenever an edge $A_i$ is inserted in the array list then we highlight the corresponding line segment in the trivial $\mathbb{L}$. There can have three possibilities for the common line segment corresponding to the selected edge in the REL:

\begin{enumerate}[(a)]
 \item If $A_i$ is flippable then we delete it from the array list while removing the corresponding highlighted common line segment from the trivial $\mathbb{L}$.
 \item When $A_i$ is not flippable and to flip $A_i$ it is required to flip only one edge $A_{i+1}$, then the highlighted common line segment corresponding to edge $A_{i+1}$ will be perpendicular to the line segment corresponding to edge $A_i$ and forms a corner in the spiral structure (shown in Figures \ref{lineseg}b and \ref{lineseg}e). 
 \item When $A_i$ is not flippable and to flip $A_i$ it is required to flip two edges $A_{i+1}$ and $A'_{i+1}$, then the highlighted common line segment corresponding to edge $A_{i+1}$ will be aligned with the line segment corresponding to edge $A_i$ and forms a straight line (shown in Figure \ref{lineseg}c and \ref{lineseg}f).
\end{enumerate}

In any spiral structure, the outer side of any straight line (made up of the common line segments corresponding to the edges from the array list) is a part of a boundary wall of only one module in $\mathbb{L}$ and the end points of a straight line are the corners of the spiral structure. Since, if the outer side of any straight line is a part of boundary walls of two or more modules, then there will be a perpendicular line segment from the outer side of the straight line as well. Consequently, there will be a line segment which is not one-sided \cite{eppstein2009area}, which will be corresponding to a flippable edge in REL, which is a contradiction since the straight line represents that to flip the corresponding edges, it requires to flip one or two edges.

If an edge $A_i$ occurs for the second time in the array list such that $A_i = A_j$ for some $j < i$, then the common line segment corresponding to $A_i$ will be highlighted again, i.e, an inner round of the blue lines will intersect with its previous (outside) round of the spiral (refer to Figures \ref{struct}xix and \ref{struct}xxiv). Since the spiral structure moves in rectangular form, then there will a cycle (rectangle) made up of the highlighted common line segments corresponding to the edges $A_j, A_{j+1},\dots,A_{i}$ of the array list. From the cycle (rectangle) having four corners, we can conclude that among these edges, exactly four edges (say $A_k, A_l, A_m, A_n$) will be there, for which it is required to flip only one edge. The rectangle is made up of four straight lines and each of its outer side is a part of the boundary wall of a module, i.e., the rectangle is surrounded by four modules which are adjacent horizontally and vertically, alternatively. Moreover, the modules will be corresponding to the vertices (say $z[k], z[l], z[m]$ and $z[n]$) of the PTPG which are the new common neighbour of the end vertices of the edges $A_k, A_l, A_m, A_n$. Since, there are four corners in the rectangle and corners are formed when there is an edge in REL as shown in Figure \ref{lineseg}b. Hence, in the corresponding REL there will be an alternate four-cycle of vertices corresponding to these four alternately adjacent modules. The cycle (rectangle) of common line segments lies inside the corresponding alternate four-cycle, then all the edges $A_j, A_{j+1},\dots,A_{i}$, corresponding to these common line segments lie inside the four-cycle. As it is known from \cite{steven2021morphing} that we can flip all the edges together which are inside an alternate four-cycle in an REL. Hence we can flip all the edges coming inside the four-cycle $z[k], z[l], z[m], z[n]$, which includes edges $A_j, A_{j+1},\dots,A_{i}$ as well. After that, by deleting edges $A_j, A_{j+1},\dots,A_{i}$ from the array list, we select the last remaining edge as the next edge to select.
Then in $B$, all the descendants of the vertex $v_j$  will be marked as visited and $v_j$ itself be marked as visited. Meanwhile, in the trivial $\mathbb{L}$, the cycle (rectangle) made up of the common line segments corresponding to the edges $A_j, A_{j+1},\dots,A_{i}$ will also be removed from highlighted.
Therefore, vertex $v_i$ can be taken as a leaf vertex since it is of no use to show the descendants of $v_i$ over and over.

\item \textbf{The height of $B$ can be at most $e+1$, where $e$ is number of edges in the corresponding PTPG.}

To prove it we first show that the height of the $B$ will be finite. The height of a binary tree is the maximum number of vertices in a path from root vertex to leaf vertex. We know that the number of edges in the PTPG are finite, hence the number of vertices in $B$ will be finite because if any vertex occurs for the second time then that vertex will be taken as a leaf vertex. Hence, if the total number of edges in the corresponding PTPG are $e$, then in $B$ there can have at most $e+1$ vertices when a vertex occurs for the second time in any path of the $B$. 

\item \textbf{In a REL of a PTPG $G'$, with edges $(b, c)$ and $(a, b)$ belonging to the same set (obtained 
from Algorithm \ref{EGDR1}), one edge among $(b, c)$ and $(a, b)$ can always be flipped.}

If we assume that edge $(b, c)$ is the first selected edge to flip, then a rectangular spiral structure of the common line segments corresponding to the edges which are inserted in the array list is shown in Figure \ref{rsa} (blue lines). It starts from the common line segment between modules $B$ and $C$ and moves in a straight line towards walls $W_5$, $W_6$ and $W_1$ but can not intersect with these walls. Since, if the spiral intersect with wall $W_5$, then either vertex $c$ would be coming in four consecutive paths $P_1, P_2, P_3$, and $P_4$ or the PTPG would be $1$-connected. If the spiral intersect with wall $W_6$ (it can possible only in the first round of the spiral), then vertex $c$ will have a shortcut with a vertex of path $P_5$ (since module $C$ would be adjacent to a module sharing a wall segment with $W_6$). If the spiral intersect with wall $W_1$ (it can possible only in the first round of the spiral), then there will be a common neighbor of vertices $a_n$ and $c$ for some $n \geq 1$. Hence, for all the cases, we find that either the set of paths are not satisfying the characteristic given in Theorem \ref{th4} or the given PTPG is $1$-connected, which is a contradiction of our initial assumptions (since the input REL of the Algorithm \ref{EGDR} is obtained from the Algorithm \ref{EGDR1}, in which the input graph is a bi-connected PTPG and the set of paths are also chosen with the characteristics given in Theorem \ref{th4}). 

As we know, in $B$ there are finite number of vertices and in order to mark the root vertex (corresponding to the edge $(b, c)$ or $(a, b)$) as visited, it is required to mark all of its descendants as visited. Any leaf vertex in $B$, either will be corresponding to a flippable edge or will represent an edge which occurs for the second time in the array list. Therefore, either all the edges of array list will be deleted or any other edge will occur for the second time in the array list (the spiral can again intersect with its previous round). In this way at one time, one edge can occur for the second time in $B$, that can always be deleted. Hence, we will be able to flip all the edges that are required to flip edge $(b, c)$. 

Similarly, if $(a, b)$ is the first selected edge, then a rectangular spiral structure of the common line segments corresponding to the edges which are inserted in the array list is shown in Figure \ref{rsb} (blue lines). It starts from the common line segment between modules $A$ and $B$ and can move towards walls $W_4$, $W_3$ and $W_2$ but can not intersect with these walls. In the same manner, it can be proved that edge $(a, b)$ can be flipped as well. 

Hence, after combining all the results we conclude that by flipping edge $(a, b)$ or $(b, c)$, there can be obtained a new REL with edges $(a, b)$ and $(b, c)$ belonging to different sets.
$\square$
\end{enumerate}

\begin{figure}[H]
\includegraphics[width=17cm,height=22cm]{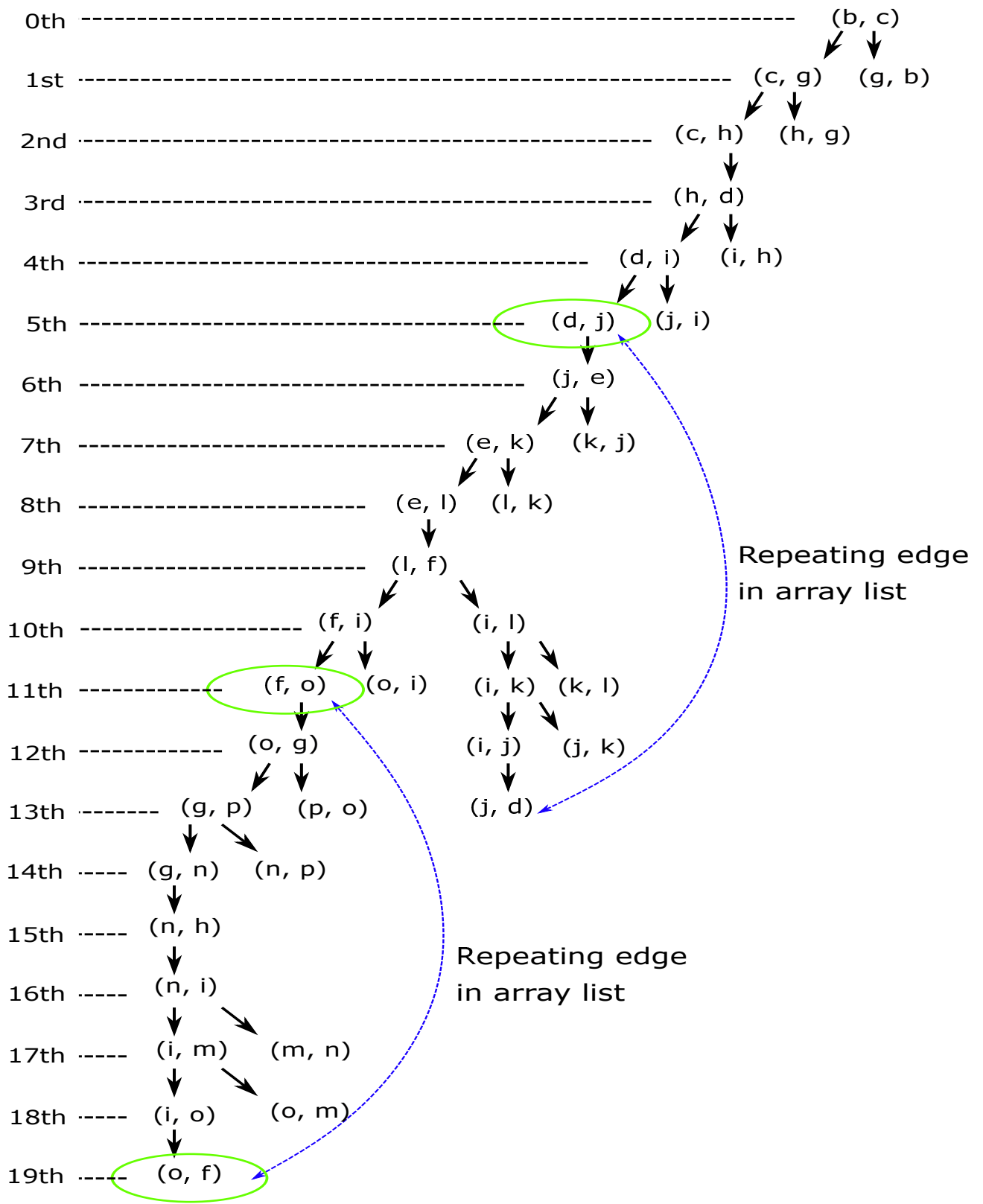}
\centering
\caption{Binary tree with root vertex corresponding to edge $(b, c)$ for the REL in Figure \ref{nontrivial}}
\label{fm1}
\end{figure} 

\begin{figure}[H]
 \centering
 \begin{subfigure}[b]{0.46\textwidth}
  \includegraphics[width=.90\textwidth]{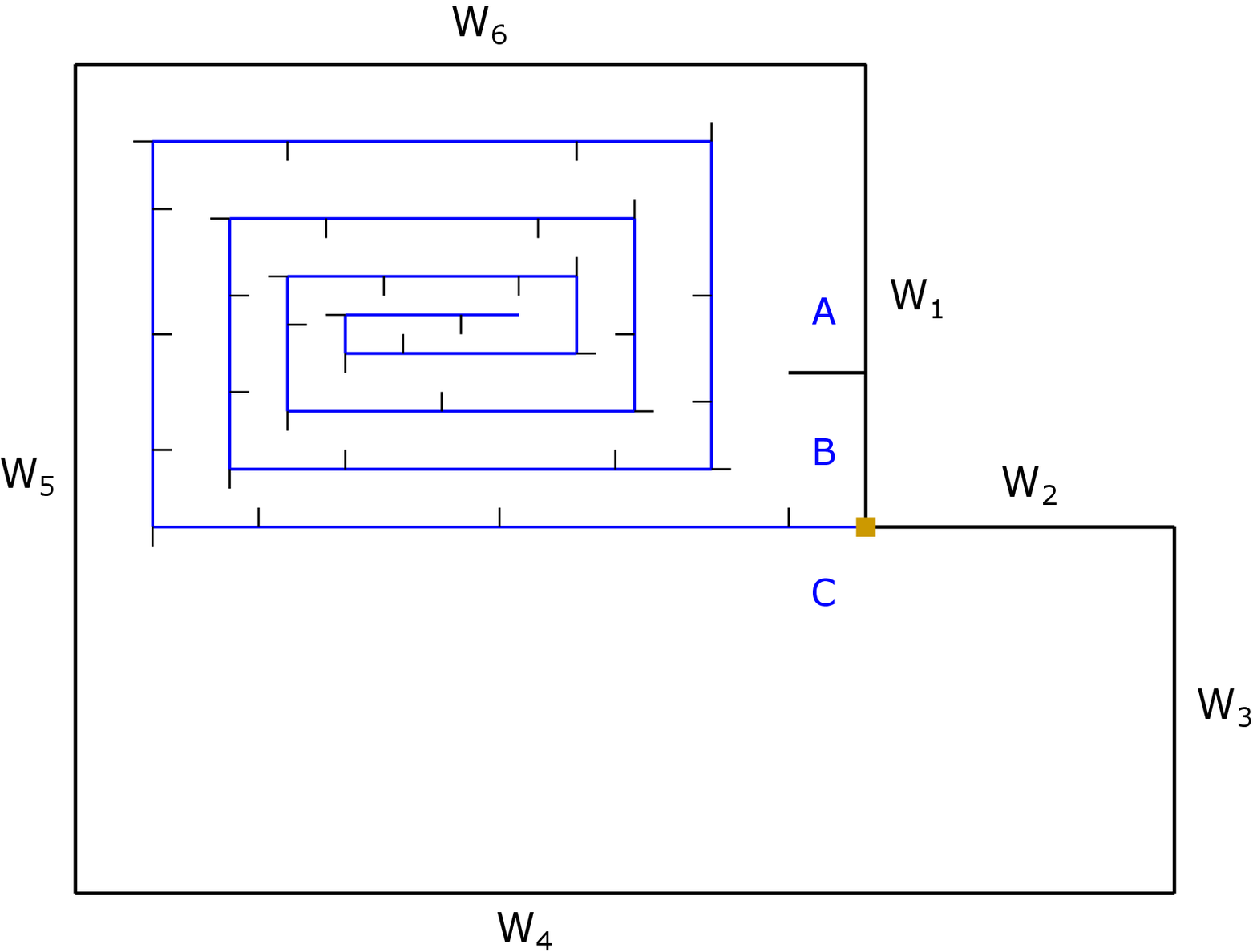}
  \caption{}
  \label{rsa}
 \end{subfigure}
 \begin{subfigure}[b]{0.46\textwidth}
  \includegraphics[width=.90\textwidth]{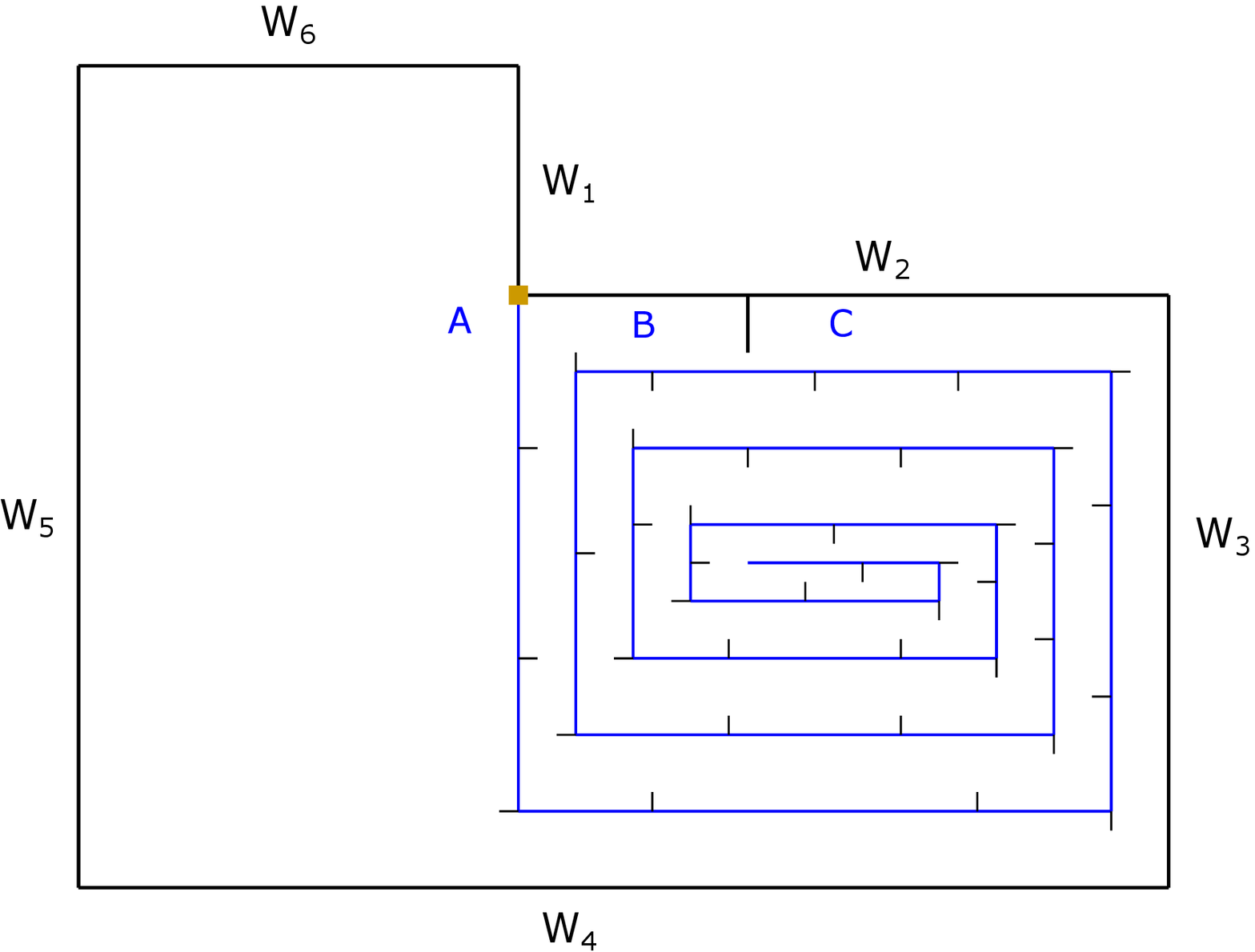}
  \caption{}
  \label{rsb}
 \end{subfigure}
 \caption{(a) Rectangular spiral when edge $(b, c)$ is to be flipped, (b) Rectangular spiral when edge $(a, b)$ is to be flipped}
 \label{spirall}
 \end{figure}

\begin{figure}[]
\includegraphics[width=16.8cm,height=21.5cm]{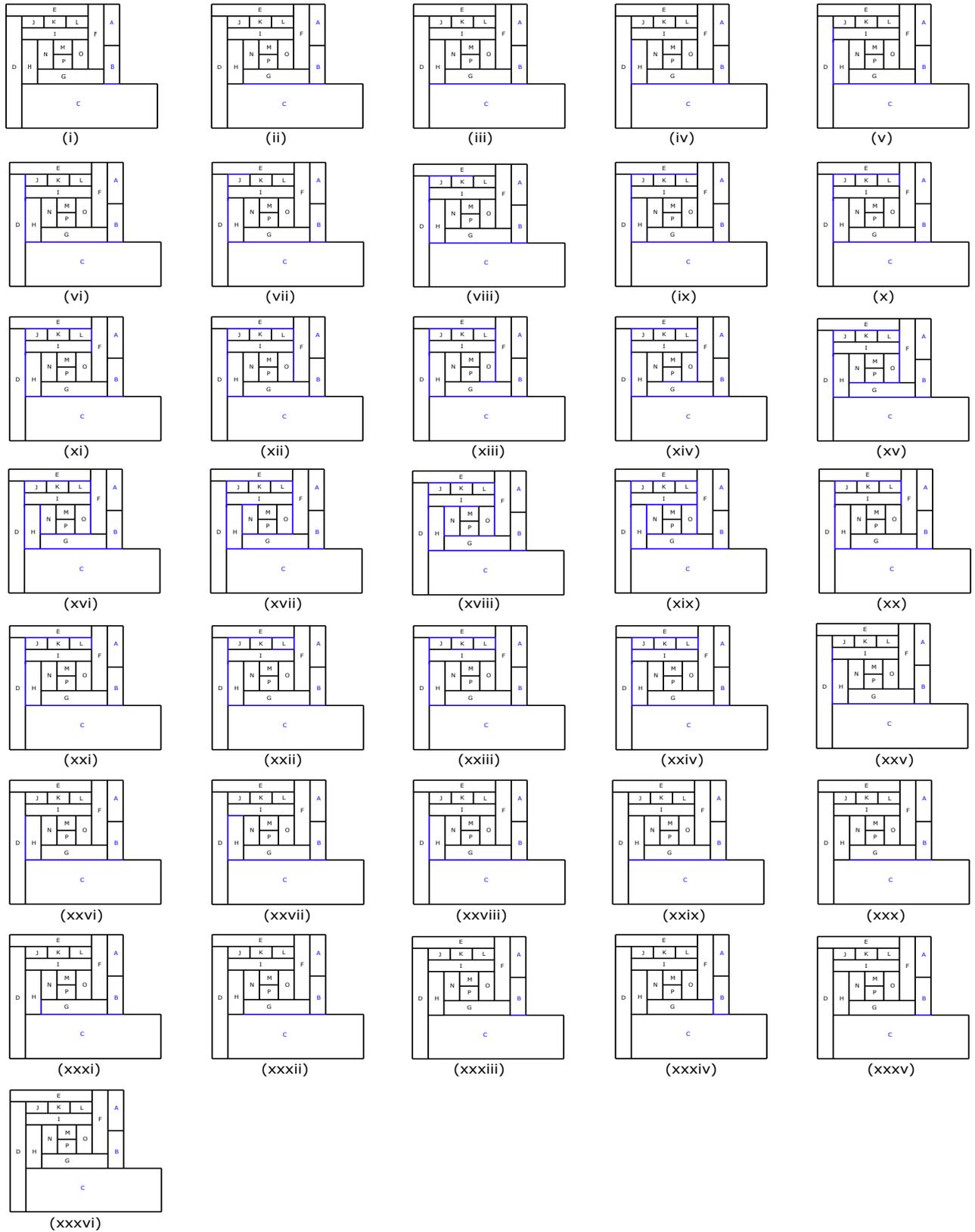}
\centering
\caption{Rectangular spiral structures (shown in blue lines) corresponding to the edges of the array list in the same order as inserted or deleted in Illustration \ref{illus2}}
\label{struct}
\end{figure}


\subsection{Correctness of Algorithm \ref{EGDR1}}
\label{correctness}

To prove the correctness of the Algorithm \ref{EGDR1}, we give explanation for each step.
In Algorithm \ref{EGDR1}, the input is taken a PTPG $G$. The first task is to identify a triplet of exterior vertices $(a, b, c)$ which satisfies the condition (ii) of Theorem \ref{th3}. After choosing a triplet $(a, b, c)$, we identify five paths $P_1, P_2, P_3, P_4, P_5$ for the outer boundary of $\mathbb{L}$ using the method given in the proof of Theorem \ref{th5}.
For any PTPG with four or less CIPs, we can always obtain five paths while assuming $(a, b, c)$ as a triplet, satisfying all the characteristic defined in Theorem \ref{th4}. Whereas, for PTPGs having five CIPs, we may or may not find such set of paths which satisfies all the characteristics.
Once we identify five paths which satisfies the characteristics given in Theorem \ref{th4} with respect to the triplet $(a, b, c)$, we insert a new vertex $\textit{NE}$ which is adjacent to all the vertices of path $P_1$ in PTPG $G$. The modified graph is called $G'$. Since, we are adding a new vertex at the outer boundary of $G$, then the number of CIPs in $G$ and $G'$ can differ. Moreover, the number of CIPs in PTPG $G'$ can never exceed the number of CIPs in PTPG $G$, because vertex $\textit{NE}$ is inserted at the outer boundary of $G$. Since if we assume that $P_1$ contains outer vertices $a_m, a_{m-1},\dots, a, b, c,\dots, c_{n-1}, c_n$ in clockwise order, then in $G'$, vertex $\textit{NE}$ is adjacent to each of the vertex of $P_1$ and all the vertices of $P_1$ except the two end vertices $a_m$ and $c_n$, are interior vertices in $G'$. Now, if the number of CIPs exceeds after adding vertex $\textit{NE}$, it implies that there was a shortcut between two vertices of path $P_1$, which contradicts that $P_1$ is a path. Hence, the number of CIPs can not exceed after inserting $\textit{NE}$ in $G$ but the number of CIPs in $G'$ can be less than to the number of CIPs in $G$. If a PTPG has five CIPs then each path ($P_1, P_2, P_3, P_4, P_5$) contains one of their end vertex from each of two different consecutive CIPs (refer to Figure \ref{drawing-2}). Therefore, the path $P_1$ also contain end vertices $a_2$ and $b_1$ from CIP$1$ and CIP$2$ respectively and after adding $\textit{NE}$, two CIPs will be reduced. Hence, after adding a new vertex $\textit{NE}$, there can be at most four CIPs and corresponding to the modified graph, a REL can be obtained. 
Now, in the REL if the label of both the edges $(a, b)$ and $(b, c)$ are different then we obtain the RFP and by removing the module $\textit{NE}$, a non-trivial $\mathbb{L}$ can be obtained using \cite{kant1997regular}. If both the edges $(a, b)$ and $(b, c)$ are same labelled in the REL, then we use Algorithm \ref{EGDR} to flip edge $(a, b)$ or $(b, c)$. After obtaining a REL with edges  $(a, b)$ and $(b, c)$ differently labelled, we obtain a non-trivial $\mathbb{L}$. $\square$

\begin{theorem}
{\rm Using Algorithm \ref{EGDR1}, a non-trivial $\mathbb{L}$ correspond to a PTPG $G$ can be obtained in $O(n^2)$ time, where $n$ and $m$ represents number of vertices and edges in any PTPG $G$, respectively.}
\end{theorem}
\proof We analyze step-wise the complexity of Algorithm \ref{EGDR1} , which includes Algorithm \ref{EGDR} as well.

In Step $1$, we search for a triplet $(a, b, c)$ (where neither vertices $a$ and $c$ are adjacent nor $a$ and $c$ have a common neighbour other than $b$) in $O(n)$ time (there can have a maximum of $n$ vertices at the outer boundary of any PTPG) while traversing at the outer boundary of the PTPG, if such a triplet exists. After choosing a triplet, in Step $6$, we look for five paths $P_1, P_2, P_3, P_4, P_5$ to create the outer boundary of $\mathbb{L}$. In case of five CIPs in a PTPG, we can choose five paths in $O(n)$ time by using the method given in Section \ref{set}, whereas in case of four of less CIPs we obtain five paths using the method described in the proof of Theorem \ref{th5}. In this method, initially we find $k$ ($k$ is the number of CIPs in the PTPG) paths dividing the outer boundary into $k$ parts using the same method given in Section \ref{set} and rest of the paths are obtained by splitting one or more paths. More precisely, in case of four CIPs to obtain the fifth path, we split the path containing vertices $(a, b, c)$ either from $a$ or from $c$ according to the cases described in the proof of Theorem \ref{th5}. This can be done $O(1)$ time. Similarly, for the PTPGs with less than four CIPs we can select five paths by splitting more than one paths in $O(n)$ time. In Step $7$, we add a vertex $\textit{NE}$ by making it adjacent to all the vertices of path $P_1$ which takes $O(1)$ time. Again for the modified graph $G'$ (obtained after adding a new vertex $\textit{NE}$ in $G$), four paths $P'_1, P'_2, P'_3, P'_4$ are computed using the method described in Section \ref{2.41}  which can be done in $O(1)$ time. After identifying four paths, we add four new vertices $N, E, S$ and $W$ while making them adjacent to all the vertices of paths $P'_1, P'_2, P'_3$ and $P'_4$, respectively. Hence the four-completion process also takes $O(1)$ time. For the next step, we can obtain an REL in $O(n)$ time for any PTPG (as proposed in \cite{kant1997regular}). In Step $10$, we can check the labels of the edges $(a, b)$ and $(b, c)$ in the REL in $O(1)$ time. If the labels belong to different sets, then we can get a RFP directly using the method given in \cite{bhasker1986linear} in linear ($O(n)$) time, otherwise we use Algorithm \ref{EGDR} to flip one of the edges among $(a, b)$ or $(b, c)$, which takes $O(n^2)$ time. Then by producing an RFP, we remove the corner module $\textit{NE}$ and find a non-trivial $\mathbb{L}$ correspond to the given PTPG. Now, we will compute the complexity of Algorithm \ref{EGDR}.
 
Since in Algorithm \ref{EGDR} the edges are selected using DFS to flip,  DFS time complexity will be $O(n+m)$. As we know, whenever an edge repeats in the array list there exist an alternating four-cycle. To detect an alternating four-cycle we check if some $A_i = (x_i, y_i)$ is already present in the array list, i.e., $A_i = A_j$ for some $j < i$ which takes $O(1)$ time. To find the four corner vertices of the four-cycle, we traverse through all the edges between $A_i$ and $A_j$ and choose the new common neighbor of the end vertices of the edges which are correspond to such vertices of $B$ which has a single child. This can be done in $O(m)$ time. Then we can identify all the edges inside the alternating four-cycle from the drawing of the REL and flip all these edges in $O(1)$ time. Moreover, the worst case time complexity will be $O(n.(n+m))$ $= n^2$, which occurs when for each edge which is added to array list, we detect a four-cycle (which is not possible).
Hence, the time complexity of Algorithm \ref{EGDR1} can be at most $O(n) + O(n^2)= O(n^2)$.

\section{Conclusion and future work}
\label{sec3}

In this work, we have analysed the behaviour of non-trivial NRFPs (specially a non-trivial $\mathbb{L}$), where the boundary of the floor-plan is inflexible and can not be transformed into a rectangular boundary, without disturbing the modules adjacencies. We have derived the existence conditions (including necessary and sufficient) of a non-trivial $\mathbb{L}$ correspond to any PTPG and developed an algorithm for the construction of a non-trivial $\mathbb{L}$, if it exists. For the future work, we will extend the present work on different types of other NRFPs (which are illustrated in Section \ref{1.1}) and investigate their graph theoretic properties. Moreover, flipping algorithm has an important role to ensure the non-triviality of the obtained floor-plan. It would be much simpler if we can demonstrate an algorithm where without using the flipping algorithm we can guarantee the non-triviality of the floor-plan.  

\bibliographystyle{alphaurl}
{\footnotesize\bibliography{main.bbl}}

\end{document}